\documentclass[a4paper,11pt]{article}
\usepackage{jcappub,macros,graphicx,float,mathtools,caption,subcaption,cancel}
\usepackage[linktocpage=true]{hyperref}
\hypersetup{colorlinks=true, linkcolor=blue, filecolor=magenta,  urlcolor=blue, citecolor=red,}
\usepackage{array}
\newcolumntype{M}[1]{>{\centering\arraybackslash}m{#1}}
\usepackage[normalem]{ulem}
\usepackage{bigints}
\arxivnumber{2501.05732} % Only if you have one

\graphicspath{{./figs/}}

\title{Relativistic magnetohydrodynamics in the early Universe}

\author[a]{Alberto Roper Pol,}
\author[a]{Antonino S.~Midiri\,}

\affiliation[a]{D\'epartement de Physique Th\'eorique, Universit\'e de Gen\`eve, Quai Ernest-Ansermet 24,
CH-1211 Gen\`eve, Switzerland}

\emailAdd{alberto.roperpol@unige.ch}
\emailAdd{antonino.midiri@unige.ch}

\abstract{
We review the conservation laws of magnetohydrodynamics (MHD)
in an expanding homogeneous and isotropic Universe that can be
applied to the study of early Universe physics
during the epoch of radiation domination.
The conservation laws for a conducting perfect
fluid with relativistic bulk velocities
in an expanding background are presented
(for the first time in their
non-conservation form, i.e., as dynamical equations for the velocity
and energy density fluid variables),
extending previous results that apply
in the limit of subrelativistic bulk motion.
Furthermore, it is shown that the subrelativistic limit presents new
corrections that have not been considered in previous work.
We discuss the conformal invariance of the MHD equations for
a radiation-dominated fluid and different types of
scaling of the fluid variables that are relevant
for other equations of state when the bulk velocity is subrelativistic.
In particular, we review the super-comoving coordinates
that have been proposed for matter-dominated fluids and present
this choice of coordinates
for any constant equation of state.
First-order fluid dynamics to include
imperfect relativistic fluids and the scaling of the transport coefficients with
temperature in the early Universe are presented.
We review the propagation of sound waves, Alfv\'en waves, and
magnetosonic waves in the early Universe plasma.
The Boris correction for relativistic Alfv\'en speeds is
presented and adapted for early Universe applications.
This review
is an extension, including new results,
of part of the lectures presented at
the minicourse ``Simulations of Early 
Universe Magnetohydrodynamics'' lectured by
A.~Roper Pol and J.~Schober at 
EPFL, as part of the six-week program ``Generation, evolution, and 
observations of cosmological magnetic fields'' at
the Bernoulli Center in May 2024.
}

\begin{document}

\maketitle

\tableofcontents

\section{Introduction}
\label{intro}

In this work, the conservation laws of magnetohydrodynamics (MHD)
in the early Universe are reviewed,
described in general relativity over the
Friedmann--Lema\^itre--Robertson--Walker (FLRW) background
metric tensor for a homogeneous and isotropic expanding Universe.
Since the focus of this work is the relativistic MHD description,
we only provide a brief introduction to
the FLRW model in \Sec{FLRW} with the objective to make this work self-contained.
We refer the reader to textbook references on cosmology,
for example \cite{Weinberg:1972kfs,Kolb:1990vq}, or \cite{Baumann_2022} for
a recent textbook.
For extended work on perfect fluids and electromagnetism
in general relativity, the literature is extensive; we recommend
the reader the following textbook references \cite{Misner:1973prb,Wald:1984rg,Schutz:1985jx,Landau1987Fluid,
Jackson:1998nia,Carroll:2004st,Rezzolla:2013dea}.

The MHD equations in a cosmological expanding background
have been studied following the
pioneer work of \cite{Brandenburg:1996fc,Jedamzik:1996wp,Brandenburg:1996sa,Subramanian:1997gi,Martel:1997hk}.
In particular, the MHD equations during the radiation-dominated
era in the early Universe have been considered, for example, to study the
evolution of primordial magnetic fields \cite{Brandenburg:1996fc,Jedamzik:1996wp,Brandenburg:1996sa,Subramanian:1997gi,Christensson:2000sp,Christensson:2002xu,Banerjee:2003xk,Banerjee:2004df,Kahniashvili:2010gp,Kahniashvili:2012vt,Kahniashvili:2012uj,Brandenburg:2014mwa,Kahniashvili:2015msa,Brandenburg:2016odr,Brandenburg:2017rnt,Brandenburg:2017neh,Brandenburg:2018ptt,Trivedi:2018ejz,Kahniashvili:2018mzl,Brandenburg:2020vwp,Perrone:2021srr,Dvornikov:2022cyz,Hosking:2022umv,Zhou:2022xhk,Uchida:2022vue,Armua:2022rvx,Brandenburg:2023rrd,RoperPol:2023bqa,Jedamzik:2023rfd,Brandenburg:2024tyi,Uchida:2024ude}, the production of gravitational
waves (GWs) from MHD turbulence \cite{RoperPol:2018sap,RoperPol:2019wvy,Kahniashvili:2020jgm,Brandenburg:2021aln,Brandenburg:2021tmp,Brandenburg:2021bvg,RoperPol:2021gjc,RoperPol:2021xnd,Kahniashvili:2021gym,RoperPol:2022iel,RoperPol:2022hxn,Sharma:2022ysf,RoperPol:2023bqa,Brandenburg:2023imm,EPTA:2023xxk,Caprini:2024hue,Caprini:2024gyk},
and the chiral magnetic effect \cite{Brandenburg:2017rcb,Rogachevskii:2017uyc,Schober:2017cdw,Schober:2018wlo,Schober:2020ogz,Brandenburg:2023rul,Brandenburg:2023aco,Schober:2024vtv}. 
Excellent reviews on the generation, evolution, and observational signatures of primordial magnetic fields exist
in the literature (cf.~\cite{Durrer:2013pga,Subramanian:2015lua,Vachaspati:2020blt} and the textbook \cite{Shukurov_Subramanian_2021}).

The aim of the present work is to review the MHD equations in
an expanding background, extending previous work to relativistic
bulk motion and presenting new results on the fully relativistic description
of perfect fluids and resistive MHD.
In particular, we also show that previous work assuming the $\gamma^2  \approx 1$
limit in the
subrelativistic regime (cf.~the original work \cite{Brandenburg:1996fc}, or the
reviews \cite{Durrer:2013pga,Subramanian:2015lua,Vachaspati:2020blt}
and references therein)
ignores time derivatives of the Lorentz factor, $\gamma$,
which can lead to additional non-linear terms of order
$U^2/\ell$, being $U$ and $\ell$ a characteristic velocity and length scale of the fluid, respectively, when the fluid is dominated by radiation.
Note that this term is of the same order
as the convective derivative, $(\uu \cdot \nab)\, \uu \sim U^2/\ell$, which is of pivotal
importance to describe, for example, non-linear fluid dynamics and turbulence.
Therefore,
the time derivative of $\gamma^2$ is not, in general,
negligible in a non-linear
description of subrelativistic fluid dynamics or MHD.

Many of the results in this work were presented by one of the
authors in the lectures of the EPFL course ``Simulations
of Early Universe Magnetohydrodynamics,'' in particular the theory lectures
on ``MHD in an expanding Universe,''
as a part of the six-week program ``Generation, evolution, and 
observations of cosmological magnetic fields'' at
the Bernoulli Center for Fundamental Studies.\footnote{\href{https://indico.cern.ch/e/cosmoMF}{``Generation, evolution, and 
observations of cosmological magnetic fields''} program at the Bernoulli Center at EPFL (Apr.\,28--Jun.\,7, 2024).}

With the growing interest in studying the evolution of primordial magnetic
fields, fluid perturbations, and GWs, among others, in the
early Universe, the results presented in this work can be used for
a broad range of cosmological applications.
Many of these studies have been performed in the subrelativistic limit
using the open-source
{\sc Pencil Code} \cite{PencilCode:2020eyn}, adapted to a radiation-dominated fluid with
a constant equation of state $p = \cs^2 \rho$, where $p$ is the pressure, $\rho$ is
the total energy density, and $\cs^2 = \third$ is the square of the speed of sound, following the
pioneer work of \cite{Brandenburg:1996fc}.
Current developments for including relativistic MHD in an
expanding background in {\sc Pencil Code}
are in progress by the authors and collaborators \cite{PC_relativistic}.
In addition, the open-source \CL \cite{Figueroa:2020rrl,Figueroa:2021yhd} is also being developed to
include an MHD description for early Universe studies \cite{Figueroa:2023xmq,CosmoLattice_MHD}.
In this work, we present the theoretical framework for
MHD studies in the early Universe
going beyond the currently studied subrelativistic
limit of bulk fluid motion, and present
new results that also lead to
corrections in the subrelativistic limit that should be addressed
in future studies.

In \Sec{FLRW}, the FLRW geometry of an isotropic
and homogeneous expanding Universe is reviewed.
This is considered to describe
the metric tensor over which the MHD conservation laws are studied.
The perturbations of the metric tensor induced by the velocity and magnetic
fields are assumed to be subdominant, such that the backreaction
of the metric tensor perturbations on the MHD equations can be neglected.
In \Sec{perfect_fluid_hydro}, the conservation laws of a fully relativistic
perfect fluid in the absence of electromagnetic fields
on an expanding Universe are presented.
Linear perturbations of these equations in the form of sound waves
are described in \Sec{sound_waves}.
The extension to imperfect fluids (with viscous and heat
transfer contributions) is then discussed in \Sec{viscosity},
and a review of the estimate of the shear and bulk viscosities,
and of the thermal conductivity of the primordial plasma is presented
in \Sec{transport_coeffs}.
In \Sec{Maxwell_sec}, Maxwell equations in an expanding background
are reviewed, introducing the definitions of
comoving electromagnetic fields and current density (\Sec{comoving_EM}),
together with the covariant formulation of the generalized Ohm's law (\Sec{cov_ohms}),
required to describe the current density for relativistic fluid
bulk motion.
We review the values of the conductivity in the primordial
Universe and show that its ratio to the Hubble time is in general
very large, justifying the usual MHD description where the
displacement current is neglected.
In \Sec{MHD_eom}, the MHD conservation laws in an expanding Universe for a relativistic charged
fluid are described, combining the equations of motion of the fluid with
Maxwell equations.
Linear MHD perturbations in an expanding Universe are studied
in \Secss{alfven_waves}{magnetosonic}.
In \Sec{boris_correction}, the Boris correction is introduced, which allows to correct
the Alfv\'en speed when it becomes relativistic, even in the subrelativistic
limit of bulk flow velocities.
We conclude and summarize the set of equations presented in this work in \Sec{conclusions}.

In the present work, a few assumptions are made to simplify the MHD 
description that can be relevant for some cosmological applications:
\begin{itemize}
    \item[{\em i)}] 
    Contributions from neutrino and electron free-streaming below
    the Silk damping scales are ignored.
    Their impact on MHD has been studied in the literature
    (cf.~\cite{Jedamzik:1996wp,Brandenburg:1996sa,Subramanian:1997gi} for pioneer work).
    The Silk damping scale becomes larger towards the onset of
    matter domination in the early Universe and, hence, it needs to be taken
    into account to study the correct evolution of large-scale
    primordial magnetic fields across the early Universe, especially towards
    the epoch of recombination.
    \item[{\em ii)}]
    The study of imperfect fluids
    in this work is only presented using a covariant
    description of Navier-Stokes viscosity
    and Fourier's thermal conductivity,
    following the classical irreversible thermodynamics (CIT)
    approach
    \cite{Weinberg:1971mx,Weinberg:1972kfs,Landau1987Fluid,Jedamzik:1996wp,Subramanian:1997gi,Romatschke:2009im,Rezzolla:2013dea}.
    It is known that this description leads to acausal fluid perturbations
    and a relativistic description of viscous fluid dynamics is an active
    field of research (cf.~the review \cite{Romatschke:2009im} and references therein).
    \item[{\em iii)}]
    At high energies, asymmetries between left and right-handed particles
    can lead to a chiral induced current \cite{Joyce:1997uy} that can
    produce vorticity and drive magnetic
    fields in processes known as the chiral vortical and magnetic effects.
    The latter effect has been studied in the context of primordial magnetic fields
    using MHD simulations
    (cf.~\cite{Brandenburg:2017rcb,Rogachevskii:2017uyc,Schober:2017cdw} for pioneer work).
\end{itemize}

\paragraph{Conventions and notation.}

Natural units with $c = \hbar = 1$ and
Heaviside-Lorentz units with $\mu_0 = 1$ for electromagnetic fields are considered along
the text.
In these units, the Maxwell equations in Minkowski space-time are
\begin{equation}
    \partial_t \EE = \nab \times \BB - \JJ\,, \qquad \nab \cdot \EE = J^0\,,
    \qquad \partial_t \BB = - \nab \times \EE\,, \qquad \nab \cdot \BB = 0\,.
\end{equation}

In general, $t$ and $\tau$ are used for cosmic and conformal time, respectively.
Derivatives with respect to $t$ and $\tau$ are denoted by a dot (e.g., $\dot{a}$) and a prime (e.g., $a'$), respectively.
The early Universe geometry is described by the
homogeneous, isotropic, and spatially flat Friedmann-Lema\^itre-Robertson-Walker (FLRW) metric tensor, $g_\munu = a^2 \eta_\munu$
with $\eta_\munu = m \, {\rm diag}
\{-1, 1, 1, 1\}$ being the Minkowski metric tensor.
We will always assume
that the MHD perturbations are small compared to the background,
such that their effect on the metric does not feedback into
the MHD equations.
The signature is $m \, (-\, +\, +\, +)$, such that $m = \pm 1$ allows
to keep track of the signature-dependent quantities.
Taking $m = 1$, space-time intervals are spacelike and it is more commonly used
in general relativity, for example in \cite{Misner:1973prb,Wald:1984rg,Carroll:2004st},
while $m = -1$ yields timelike intervals and it is more common in
particle physics and cosmology, for example in \cite{Baumann_2022,Landau1987Fluid}.
In \Sec{FLRW}, a generic signature will be considered,
while in the remaining of the
text, when the equations of motion are computed, $m = 1$ will be used for
compactness of the calculations as the
resulting equations of motion are not affected by the signature choice.

In most of the text, the coordinates are
$X^0 = \tau$ and $X^i = x^i$ with $x^i$ being the comoving space coordinates.
Only in \Sec{FLRW}, $X^0$ is allowed
to be a generic $\alpha$-time, such that
$a^\alpha \dd \tau_\alpha = \dd t$, which reduces to cosmic or conformal
when $\alpha = 0$ or 1, respectively.
When we describe spatial vectors in flat space-time,
we indistinguishably use upper or lower indices as the indices can be risen or lowered
with Minkowski metric tensor $\eta^{ij} = \delta^{ij}$ in the $m = 1$ signature,
e.g., $u^i = u_i$ or $\partial^i = \partial_i$.
We will still keep both upper and lower indices because
Einstein summation is only assumed over one lower and one upper index, e.g., $u^2 = u^i u_i$.

\subsection{Relativistic MHD equations in an expanding Universe}

For the impatient reader, let us present the set of relativistic MHD equations (conservation of energy and momentum) in an expanding Universe describing the evolution of the total energy density
$\rho$ and the peculiar velocity field $\uu$
that we find in this work
(see \Sec{MHD_eom}).
Up to our knowledge, the fluid equations of motion
have not been fully considered in this form\footnote{The set of relativistic MHD equations
are usually considered in their conservation form (cf.~\cite{Font:2008fka,Rezzolla:2013dea}), describing the evolution of the stress-energy
tensor components, or using different combinations of the fluid variables.
In this work, we also present these equations in terms of the primitive fluid variables $\rho$ and $\uu$ in the so-called non-conservation form (see details in \Sec{perfect_fluid_hydro}).} in previous work:
\begin{subequations}
\label{summary_rel}
\begin{align}
    \partial_{\tau} \ln \tilde \rho \, = &\, -
    \frac{1 + \cs^2}{1 - \cs^2 u^2}
    \nab \cdot \uu
    -
    \frac{1 - \cs^2}{1 - \cs^2 u^2} \, (\uu \cdot \nab) \ln \tilde \rho +
    \frac{1}{1 - \cs^2 u^2} \frac{1}{\tilde \rho} \bigl[\tilde f_\tot^0
    (1 + u^2) - 2 \, \uu \cdot \tff_\tot \bigr]
    \nonumber \\
    &\, + 
    \frac{1 - 3 \cs^2}{1 - \cs^2 u^2} \,  (1 + u^2) \, \HH
    \,, \label{cont_summary_rel}
    \\
    D_{\tau} \uu  = &\,
    \frac{\uu}{(1 - \cs^2 u^2)\gamma^2} \biggl[
    \cs^2 \, \nab \cdot \uu + 
    \cs^2 \frac{1 - \cs^2}{1 + \cs^2} (\uu \cdot \nab) \ln \tilde \rho - \frac{1}{\tilde \rho}
    \biggl(
    \tilde f_\tot^0 - \frac{2 \cs^2}{1 + \cs^2} \uu \cdot \tff_\tot
    \biggr) +
    (3 \cs^2 - 1) \, \HH
    \biggr]  \nonumber \\ 
    &\,  -   \frac{\cs^2}{1 + \cs^2}  \frac{\nab \ln \tilde \rho}{\gamma^2}
     + \frac{1}{1 + \cs^2} \,
    \frac{\tff_\tot}{\tilde \rho \gamma^2}\,,   \label{mom_summary_rel}
\end{align}
\end{subequations}
where $D_\tau = \partial_\tau + u_i \partial^i$ is the material derivative,
$\tau$ corresponds to conformal time, $a$ is the scale factor, and $\HH = a'/a$ is the conformal Hubble rate.
The velocity field $\uu$
corresponds to the peculiar velocity with
respect to the Hubble observer $\uu = \dd \xx (\tau)/\dd \tau$, i.e., comoving with the Universe expansion (see \Sec{four_velocity}).
The comoving energy density and pressure are $\tilde \rho = a^4 \rho$ and $\tilde p = a^4 p$, respectively, where a constant
equation of state is considered to describe their ratio, $\tilde p = \cs^2 \tilde \rho$.
Comoving variables and spatial coordinates $\xx$,
together with conformal time, are chosen to exploit
the conformal invariance of the equations for a radiation-dominated fluid with
$\cs^2 = \third$ \cite{Brandenburg:1996fc,Subramanian1997} (see \Sec{perfect_fluid_hydro} and, in particular,
\Sec{cons_perf_fluid}),
as can be seen by the fact that no
dependence on $a$ is left in \Eqq{summary_rel}
when $\cs^2 = \third$.

We note that when $\cs^2 \sim {\cal O} (1)$,
an additional term proportional
to $\uu$ appears in the momentum equation, which is not present in
usual MHD (i.e., in fluids predominantly composed by massive particles,
with $\cs^2 \ll 1$), as well as modifications in some coefficients
of the different terms that depend on $\cs^2$.
The appearance of this additional term is a clear indication that,
even when the equations become conformally flat for $\cs^2 = \third$,
the MHD dynamics in the early Universe could be modified with respect
to usual MHD \cite{Brandenburg:1996fc}.
An important aspect of these modifications is that they
can, for example, lead to the production of
vorticity, $\oom = \nab \times \uu$, from an initial curl-free configuration of the velocity field,
even in the absence of
external forcing (i.e., for the modified Euler equations).
Therefore, vorticity, defined as $\nab \times \uu$, is no longer a topological invariant, as for the
usual Euler equations when $\cs^2 \ll 1$ \cite{Rezzolla:2013dea}.
This has been studied in the subrelativistic limit in \cite{Dahl:2021wyk,Dahl:2024eup,vorticity} and in the relativistic limit in \cite{vorticity}.
The equation describing the conservation of vorticity can be found by taking
the curl of \Eq{mom_summary_rel} (see details in \App{vorticity}),
\begin{align}
    D_\tau \oom + \oom \, \nab \cdot \uu - (\oom \cdot \nab)
    \uu =  &\, \frac{\oom \Psi}{(1 - \cs^2 u^2) \gamma^2} +
    \uu \times \nab \Biggl[
    \frac{\Psi}{(1 - \cs^2 u^2)\gamma^2}\Biggr] \nonumber \\ 
    &\, +  \frac{(\nab \tilde p - \tff_{\rm tot}) \times \nab(\tilde \rho \gamma^2)}
    {(1 + \cs^2) \tilde \rho^2 \gamma^4} + \frac{\nab \times \tff_{\rm tot}}{(1 + \cs^2)
    \tilde \rho \gamma^2}\,, \label{vort_summary}
\end{align}
where $\Psi$ is given in \Eq{Psi},
\begin{equation}
    \Psi = \cs^2 \nab \cdot \uu + \cs^2 \frac{1 - \cs^2}{1 + \cs^2} (\uu \cdot \nab)
    \ln \tilde \rho - \frac{\tilde f_\tot^0}{\tilde \rho} + \frac{2 \, \cs^2}{1 + \cs^2} \frac{\uu \cdot \tff_\tot}{\tilde \rho} + (3 \cs^2 - 1) \HH \,.
\end{equation}
Indeed, from \Eq{vort_summary} one can see that the second term of the right-hand side can lead
to the production of vorticity when the initial vorticity is zero in the absence of
external forces and baroclinic terms, i.e., when $\tff_\tot = 0$ and
$\nab \tilde p \times \nab(
\tilde \rho \gamma^2) = 0$ \cite{vorticity}.

The comoving four-force $\tilde f^\mu_\tot = a^6 f^\mu_\tot$ can correspond to any
force applied to the fluid and it comes from additional
contributions to the stress-energy tensor on top of those corresponding
to a perfect fluid $T^\munu_{\rm pf}$.
In particular, we study in \Sec{viscosity} the inclusion of out-of-local-thermal-equilibrium effects like viscosity and heat transfer
by incorporating a deviatoric tensor $\Pi^\munu$, such that
$T^\munu = T^\munu_{\rm pf} - \Pi^\munu$, yielding
an imperfect (viscous) force $f^\nu_{\ipf} = \partial_\mu \Pi^\munu$.
The inclusion of electromagnetic forces to the fluid is required for the
study of MHD (see \Sec{MHD_eom}).
It can be described by incorporating the electromagnetic stress-energy tensor, $T^\munu \to T^\munu + T^\munu_{\rm EM}$, which leads to the
inclusion of
the Lorentz force $f^\nu_{\rm Lor} = - \partial_\mu T^\munu_{\rm EM}$ (see \Sec{four_lorentz}).
When electromagnetic fields are included, \Eqq{summary_rel}
become coupled to Maxwell equations in an expanding background (see \Sec{Maxwell_sec}),
which can be expressed for the Hubble observer in
the following way (using the temporal or Weyl gauge $A_0 = 0$ and defining the magnetic
vector potential $\AAA$, such that $\tBB = \nab \times \AAA$),
\begin{equation}
    \partial_\tau \tEE = \nab \times \tBB - \tJJ\,, \qquad
    \nab \cdot \tEE = \tilde J^0\,, \qquad
    \partial_\tau \AAA = - \tEE\,,
\end{equation}
where $\tEE$ and $\tBB$ are the comoving electric and magnetic fields, related to the covariant components of the
Faraday tensor, $\tilde E_i = F_{i0}$ and $\tilde B^i = \half \varepsilon^{ijk} F_{jk}$ with
\begin{equation}
    F_{\mu \nu} = \partial_\mu A_\nu - \partial_\nu A_\mu\,.
\end{equation}
The comoving four-current $\tilde J^\mu = a^4 J^\mu$
is described by
the generalized Ohm's law [see \Sec{cov_ohms} and, in particular, \Eq{Ohms_law}].

\subsection{MHD equations in the limit of subrelativistic bulk motion}

In the limit of subrelativistic bulk motion, $u^2 \ll 1$,
\Eqq{summary_rel} reduce to
\begin{subequations}
\begin{align}
    \lim_{u^2 \ll 1} \partial_\tau \ln \tilde \rho \, = &\, -  (1 + \cs^2) \,\nab \cdot \uu
    - (1 - \cs^2) (\uu \cdot \nab) \ln \tilde \rho +
    \frac{1}{\tilde \rho} (\tilde f^0_\tot - 2 \, \uu \cdot \tff_\tot)
    + (1 - 3 \cs^2)\, \HH\,, \label{cont_summary_nonrel} \\ 
    \lim_{u^2 \ll 1} D_\tau \uu = &\, \uu \, \cs^2 \, \biggl[\nab \cdot \uu + \frac{1 - \cs^2}{1 + \cs^2} (\uu \cdot \nab) \ln \tilde \rho \biggr] - \frac{\uu}{\tilde \rho}
    \biggl(\tilde f^0_\tot - \frac{2 \cs^2}
    {1 + \cs^2} \uu \cdot \tff_\tot \biggr)
    \nonumber \\ 
    &\, - \frac{\cs^2}{1 + \cs^2} \nab \ln \tilde \rho + \frac{1}{1 + \cs^2} \frac{\tff_\tot}{\tilde \rho}
    + (3 \cs^2 - 1) \, \uu \, \HH\,, 
    \label{mom_summary_nonrel}
\end{align}
\label{summary_nonrel}
\end{subequations}
and \Eq{vort_summary} reduces to
\begin{align}
    \lim_{u^2 \ll 1} D_\tau \oom + \oom \, \nab \cdot \uu - (\oom \cdot \nab)
    \uu  = \oom \cs^2 \Psi  + \cs^2 \,
    \uu \times \nab \Psi 
    + \frac{(\nab \tilde p - \tff_{\rm tot}) \times \nab \ln \tilde \rho + \nab \times \tff_{\rm tot}}{(1 + \cs^2)
    \tilde \rho}\,.
\end{align}

A Hubble friction term proportional to $(3 \cs^2 - 1) \, \HH$ appears
when $\cs^2 \neq \third$
in both the momentum and energy conservation equations as a consequence of the Universe
expansion, breaking the conformal invariance of the equations.
Note that in the subrelativistic limit,
the Hubble-dependent term in the energy equation
vanishes if one rescales $\tilde \rho = a^{3\,(1 + \cs^2)} \rho$, as it is
shown in \Sec{generic_scaling}.
The terms $\tilde f^0_{\rm tot}$ and $\uu \cdot \tff_\tot$ correspond to
energy dissipation and power exerted by the different forces on the fluid.
In addition, we find that $\partial_\tau \gamma^2$, omitted in previous
work,
presents terms that are non-negligible even in the subrelativistic
limit (see \Sec{perfect_fluid_hydro} and, in particular, \Sec{rel_hydrodynamics}), yielding a modification
in the coefficient of $(\uu \cdot \nab) \ln \tilde \rho$ in the
energy equation\footnote{We note that the corresponding correction to \Eq{cont_summary_nonrel}
was recently
pointed out in \cite{Dahl:2024eup} in Minkowski space-time
before this work was published. \label{Dahlspaper}} [cf.~\Eq{cont_summary_nonrel}] and in the coefficient of $\uu\, (\uu \cdot \nab)
\ln \tilde \rho$ in the momentum equation [cf.~\Eq{mom_summary_nonrel}].

\section{Friedmann--Lema\^itre--Robertson--Walker background}
\label{FLRW}

At large distances in our Universe, the distribution of galaxies becomes homogeneous and
isotropic as we look farther away from us.
This is known as the cosmological principle and its strongest observation evidence
is the uniformity of the cosmic microwave background (CMB), with observed
anisotropies in the CMB temperature being only of a
few parts in $10^5$ \cite{Durrer:2008eom}.
Under the assumptions of homogeneity and isotropy, Einstein field
equations have an exact solution, corresponding to the
Friedmann--Lema\^itre--Robertson--Walker (FLRW) metric tensor.

In this section, we give a brief review of the FLRW model that will be
used in the rest of this work to describe the background metric over which
the equations of motion for fluids and electromagnetic fields are
described in an expanding Universe.
For further details on cosmology and general relativity,
the reader is referred to the extensive textbook literature in
these fields, for example \cite{Weinberg:1972kfs,Kolb:1990vq,Baumann_2022,Misner:1973prb,Wald:1984rg,Schutz:1985jx,Jackson:1998nia,Carroll:2004st,Rezzolla:2013dea}.

\subsection{Geometry with cosmic time as $X^0$}

We can start choosing our space-time coordinates as $X^\mu = (t, x^i)$ being
$t$ the cosmic time and $\xx = x^i$ the comoving spatial coordinates.
The line element is described by the metric tensor $g^\munu$,
\begin{equation}
    \dd s^2 = g_\munu \dd X^\mu \dd X^\nu\,,
\end{equation}
where $g_\munu$ corresponds to the Minkowski metric tensor
$\eta_\munu = m \, {\rm diag} \{-1, 1, 1, 1\}$ in special
relativity.
In general relativity, the metric tensor is a dynamical variable,
coupled to the distribution of matter and energy via
the Einstein equations that relate the space-time
geometry and the stress-energy distribution.
However, exploiting the homogeneous and isotropic geometry of the Universe at large scales,
the dependence of $g_\munu$ is reduced to the time dependence of a scale factor $a$.
Space-time can then be foliated in spatial hypersurfaces that can have positive, negative,
or zero curvature, depending on whether the space-line element can be given as a 3-sphere
embedded in 4-dimensional Euclidean space, as a hyperboloid embedded in 4-dimensional Lorentzian space,
or directly as the 3-dimensional Euclidean space.
Unifying all cases, the spatial contribution to the
line element is
\begin{equation}
    \dd {\ell}^2 = g_{ij} \dd x^i \dd x^j = a^2 \, \gamma_{ij} \dd x^i \dd x^j\,, \qquad {\rm with \ }
    \gamma_{ij} = \delta_{ij} + k \frac{x_i x_j}{1 - k x^2}\,,
\end{equation}
where $k \in \mathbb{R}$ is the curvature of the Universe and $k^{-1/2}$
has dimension of length.
Note that spatial homogeneity and isotropy allow us to reduce
the ten independent components
of $g_\munu$ to the scale factor $a$ and the curvature $k$.
Based on CMB observations, the curvature
is very close to zero \cite{Planck:2013pxb}.
We will show in \Sec{Friedmann} that the curvature contribution to the
total energy of the Universe reduces at earlier times, thus justifying
to neglect it in the early Universe.
Then, the line element can be expressed as
\begin{equation}
\dd s^2 = - m (\dd t^2 -a^2 \dd \xx^2) = -m \, a^2 (\dd \tau^2 - \dd \xx^2)\,,
\end{equation}
where $\tau$ is the conformal time, defined such that
$a \dd \tau = \dd t$, and $a \, x^i = r^i$ relates the comoving, $x^i$, and physical, $r^i$, spatial coordinates.
The physical velocity of an object in an FLRW Universe is
\begin{equation}
    u^i_{\rm phys} \equiv \frac{\dd r^i (t)}{\dd t} = a \frac{\dd x^i (t)}{\dd t}
    + r^i H = \frac{\dd x^i (\tau)}
    {\dd \tau} + x^i \, \HH \,, \label{u_phys}
\end{equation}
where 
$x^i(\tau) \equiv x^i\bigl(t(\tau)\bigr)$ are the comoving coordinates expressed in terms of conformal time,
$H \equiv \dot a/a$ and $\HH = a'/a$ are the Hubble and conformal Hubble rates of expansion of the Universe.
The first term is the peculiar velocity, $\uu = a \dot \xx (t) = \xx' (\tau)$,
measured by a comoving
observer, and $\rr  H = \xx \HH$ is the Hubble flow.
We note that the velocities are described taking the time
derivative of the positions $\xx (\tau)$ or $\rr (t)$ describing the trajectory of the observer.
In general, a dot is used to denote derivatives with respect to
cosmic time and a prime for derivatives with respect to conformal time
in the following.

Then, the FLRW metric tensor and its inverse when one chooses $X^0 = t$
are
\begin{equation}
    g_\munu = m \, {\rm diag}\{-1, a^2, a^2, a^2\}\,, \qquad
    g^\munu =
    m \, {\rm diag}\{-1, a^{-2}, a^{-2}, a^{-2}\}\,,
\end{equation}
with determinant $g = {\rm det} \, g_\munu = -a^6$.
We can use $g_\munu$ to lower indices, such that $X_\mu = g_\munu X^\nu = m (- t, a^2 \xx)$.
The non-vanishing Christoffel symbols in the Levi-Civita connection
are
\begin{equation}
    \Gamma^0_{ij} = a^2 \, H \delta_{ij} = a \, \HH \delta_{ij} \,,
    \qquad 
    \Gamma^i_{0j} = H \delta^i_j =  \frac{\HH}{a} \delta^i_j
    \,. \label{Christoffel_t}
\end{equation}
As a reminder, the connection coefficients can be computed from the
metric tensor as
\begin{equation}
    \Gamma^\mu_{\nu\sigma} = \half \, g^{\mu\lambda}(\partial_\sigma g_{\lambda \nu} + \partial_\nu g_{\lambda \sigma} - \partial_\lambda g_{\nu \sigma})\,.
\end{equation}

\subsection{Conformal time as $X^0$}

Alternatively, one can consider conformal time as the $X^0$ variable
(this will be the usual choice in the following sections),
such that $X^\mu = (\tau, \xx)$, $X_\mu = m \, a^2 (-\tau, \xx)$,
and the metric tensor components are
\begin{equation}
    g_\munu = a^2 \, m \, {\rm diag}\{-1, 1, 1, 1\}\, = a^2 \, \eta_\munu\,, \qquad g^{\mu\nu} =
    a^{-2} \eta^{\mu \nu}\,.
    \label{gmunu}
\end{equation}
Using conformal time as $X^0$, the non-zero Christoffel symbols are
\begin{equation}
    \Gamma^0_{00} = \HH\,,\qquad 
    \Gamma^0_{ij} = \HH \, \delta_{ij}\,,\qquad
    \Gamma^i_{0j} = \HH\, \delta^i_{j}\,. \label{Christoff}
\end{equation}
One can indistinguishably use cosmic or conformal time 
as $X^0$ to find the
conservation laws.
However, note that the metric and connection tensor components depend on this choice
and, hence, it is crucial to be consistent
with modifications of the choice of geometric variables.

\EEq{gmunu} shows explicitly that $g_\munu$ under FLRW geometry
can be expressed as a Weyl transformation
of the Minkowski metric tensor.
This property will be crucial to show how conservation laws are invariant
under conformal transformations in \Sec{perfect_fluid_hydro}
for a radiation-dominated
perfect fluid, in \Sec{Maxwell_sec} for electromagnetism, and in \Sec{MHD_eom} for MHD.

\subsection{Generic $\alpha$-time $\tau_\alpha$ as $X^0$}

In specific situations,
it might be useful to allow for a generic definition of the
time variable, an $\alpha$-time, as
used, for example, in \CL \cite{Figueroa:2020rrl,Figueroa:2021yhd},
which is defined such  that $a^\alpha \dd \tau_\alpha = \dd t$.
The $\alpha$-time reduces to cosmic and conformal times when $\alpha = 0$ and $1$,
respectively.
For the choice
$X^0 = \tau_\alpha$,
$g^{00} = -m \, a^{-2\alpha}$ and $g_{00} = - m \,a^{2\alpha}$,
and
the line element is
\begin{equation}
    \dd s^2 = - m ( a^{2\alpha} \dd \tau^2_\alpha - a^2 \dd \xx^2 )\,.
\end{equation}
In the present work, the equations of motion will be described in a
generic $\alpha$-time.
However, we will keep $X^0 = \tau$ in most of the text and will then
transform the equations of motion to the generic $\tau_\alpha$.
Of course, the different choices of $X^0$
do not affect the resulting equations of motion.
When one chooses $X^0 = \tau_\alpha$, the non-zero
components of the Christoffel symbols are
\begin{equation}
    \Gamma^0_{00} = \alpha \HH_\alpha\,, \qquad \Gamma_{ij}^0 = a^{-2(\alpha - 1)} \HH_\alpha \, \delta_{ij}\,, \qquad \Gamma^i_{0j} = \HH_\alpha \, \delta^i_j\,,
\end{equation}
where
$\HH_\alpha = (\partial_{\tau_\alpha} a)/a = a^{\alpha - 1} \HH$.
Note that our $\HH_\alpha$ is denoted by $\HH$ in \cite{Figueroa:2020rrl}, while we restrict $\HH \equiv a'/a$ to the conformal
Hubble rate in this work.

\subsection{Four-velocity}
\label{four_velocity}

The four-velocity of a particle following a path $X^\mu (s)$ is
\begin{equation}
    U^\mu = \frac{\dd X^{\mu} (\tilde s)}{\dd \tilde s}\,,
\end{equation}
where $\tilde s$ is the proper time, such that
$\dd \tilde s^2 = - m \dd s^2$.
To find its exact form in units of the peculiar velocity $u^i \equiv \dd x^i (\tau)/\dd \tau$,
where $\xx (\tau)$ describes the trajectory of the fluid particles,
we note that the proper time
can be expressed in the following way
\begin{equation}
    \dd \tilde s^2 = - m \, g_{\munu} \dd X^{\mu} \dd X^{\nu} =
    a^2 (\dd \tau^2 - \dd \xx^2) = \,
    \frac{a^2 \dd \tau^2}{\gamma^2}\,,
\end{equation}
where
$\gamma = (1 - u^2)^{-1/2}$ is
the usual Lorentz factor in special relativity.
Hence, the four-velocity can be expressed as
\begin{equation}
    U^\mu = \frac{\dd X^\mu (\tilde s)}{\dd \tilde s} =
    \frac{\dd X^\mu (\tau)}{\dd \tau}
    \frac{\dd \tau}{\dd \tilde s} = \gamma (1, \uu)/a\,, \label{four-v}
\end{equation}
where 
$X^\mu(\tau) \equiv X^\mu \bigl( \tilde{s}(\tau) \bigr)$ and
$X^0 = \tau$ is chosen to define the four-velocity.
For a generic choice of coordinates, one can express the
four-velocity as 
\begin{equation}
    U^\mu = \frac{\dd X^\mu (\tilde s)}{\dd \tilde s} =
    \frac{\dd X^\mu (\tau_\alpha)}{\dd
    \tau_\alpha} \frac{\dd \tau_\alpha}{\dd t} \frac{\dd t}{\dd \tilde s} = 
    \gamma \, a^{-\alpha} \frac{\dd X^\mu 
    (\tau_\alpha)}{\dd \tau_\alpha}\,,
\end{equation}
with $X^\mu(\tau_{\alpha}) \equiv X^\mu \bigl( \tilde{s}(\tau_{\alpha}) \bigr)$, and define a velocity field $\uu_\alpha = 
\dd \xx_\alpha (\tau_\alpha)/\dd \tau_\alpha$, with $a^\alpha \xx_\alpha = \rr$,
where $\alpha = 0$ yields the physical spatial coordinates $r^i$
and the physical velocity $\uu_{\rm phys}$ [cf.~\Eq{u_phys}].
However, such a choice of the velocity
would require a redefinition of the Lorentz factor in FLRW as the
peculiar velocity $\uu$ becomes
\begin{equation}
    \uu = \uu_\alpha + (\alpha - 1) \, \HH_\alpha \, \xx_\alpha\,.
\end{equation}
Therefore, only the choice $\alpha = 1$ to define $\uu$ allows
to maintain the special relativity definition of the Lorentz factor.
For this reason, we will use this convention and the four-velocity
components given in \Eq{four-v} in the following sections.
If one alternatively chooses a different $X^0 = \tau_\alpha$, the four-velocity is expressed
as $U^\mu = \gamma \, (a^{- \alpha}, \uu/a)$.
In particular, for $X^0 = t$,
the four-velocity is $U^\mu = \gamma (1, \uu/a)$, which has been used
in previous work (cf. \cite{Brandenburg:1996fc}).

Noting that the line element can be expressed as
\begin{equation}
    \dd s^2
    = g_{\mu \nu} \dd X^{\mu} \dd X^{\nu}
    =
    \dd X^\mu \dd X_\mu\,,
\end{equation}
$U^\mu$ is required to satisfy the normalization condition,
\begin{equation}
    U^\mu U_\mu = -m\, \frac{\dd X^\mu \dd X_\mu}{\!\dd s^2} = -m\,.
\end{equation}

The four-acceleration $a^\mu$ is computed by parallel transport of the derivative of the
four-velocity along itself,
\begin{equation}
    a^\mu = U^\nu U^\mu\covderV{\nu} \,,
\end{equation}
where
the subscript $; \nu$ indicates the gravitational covariant derivative (see \Sec{sec:covder}).
Particles in free fall follow trajectories that are determined by $a^\mu = 0$ (in analogy
to no acceleration in the absence of forces in Newtonian physics), yielding the geodesic equation
\begin{equation}
    \frac{\dd U^\mu}{\dd \tilde s} + \Gamma^{\mu}_{\nu \lambda} U^\nu U^\lambda = 0\,.
\end{equation}

\subsection{Covariant derivative}
\label{sec:covder}

The covariant derivative has been introduced to define the four-acceleration, which can be obtained by parallel transport of a vector
along the curvature of space-time.
The reader is referred to excellent textbooks on general relativity
that are available in the literature
for a
geometrical description of covariant derivatives and general relativity \cite{Weinberg:1972kfs,Baumann_2022,Misner:1973prb,Wald:1984rg,Schutz:1985jx,Carroll:2004st,Rezzolla:2013dea}.
We restrict this section to simply provide useful expressions
of the covariant derivative that will be used to find the conservation
laws in the following sections:
\begin{itemize}
    \item For a scalar field $\phi$,
    \begin{equation}
        \phi\covderS{\mu} = \partial_\mu \phi\,.
    \end{equation}
    \item For a rank-1 tensor $U^\nu$,
    \begin{equation}
    U^\nu\covderV{\mu} = \partial_\mu U^\nu + U^\lambda \Gamma^{\nu}_{\lambda \mu}\,,
\end{equation}
where its divergence can be expressed as $U^\mu\covderV{\mu} = \tfrac{1}{\sqrt{-g}} \partial_\mu(\sqrt{-g} \, U^\mu)$.
\item For a rank-2 tensor $T^\munu$,
\begin{align}
    T^\munu\covderT{\mu} = &\, \partial_\mu T^\munu
    + \Gamma^\mu_{\mu \sigma} T^{\sigma \nu}
    + \Gamma^\nu_{\mu \sigma}
    T^{\mu \sigma}
    = \frac{1}{\sqrt{-g}} \partial_\mu
    (\sqrt{-g} \, T^\munu)
    + \Gamma^\nu_{\mu\sigma} T^{\mu \sigma}\,.
    \label{Fmunu}
\end{align}
\end{itemize}
For the FLRW metric tensor with no curvature, 
the Christoffel symbols are given in \Eq{Christoff}
using conformal time as the $X^0$ coordinate.
We observe from \Eq{Fmunu} that, for an antisymmetric tensor $F^\munu$ (for
example, the Faraday tensor), the term $\Gamma^{\nu}_{\mu \sigma} F^{\mu \sigma}$ vanishes due to the symmetry of the connection coefficients in the
lower indices.

\subsection{Friedmann equations}
\label{Friedmann}

The dynamics of the metric tensor and the stress-energy distribution is determined
by the Einstein field equation in general relativity,
\begin{equation}
    G_\munu + \Lambda g_\munu = \Mpl^{-2} \, T_\munu\,, \label{Einstein}
\end{equation}
where the
reduced Planck mass is $\Mpl = (8\pi G)^{-1/2} \simeq 2.4 \times 10^{18}$ GeV.
The cosmological constant $\Lambda$ can alternatively be
included as a vacuum energy density $\rho_{\rm vac}$ contributing to
the stress-energy tensor $T_\munu \to T_\munu - \Lambda g_\munu \Mpl^2$ and then consider
\begin{equation}
    G_\munu = \Mpl^{-2} T_\munu\,.
\end{equation}
Let us start focusing on the Einstein tensor,
\begin{equation}
    G_\munu = R_\munu - \half R g_\munu\,, \label{Gmunu}
\end{equation}
where $R_\munu$ is the Ricci tensor
\begin{equation}
    R_\munu = \partial_\lambda \Gamma^{\lambda}_{\munu} - \partial_\nu \Gamma^{\lambda}_{\mu \lambda} + \Gamma^{\lambda}_{\lambda \rho} \Gamma^{\rho}_{\munu} - \Gamma^{\rho}_{\mu \lambda}
    \Gamma^{\lambda}_{\nu \rho}\,,
\end{equation}
and its trace $R = R^\mu_{\ \mu}$ is the Ricci scalar.
From this expression, and using the Christoffel symbols of the FLRW
metric tensor (cf.~\Eq{Christoffel_t} or \Eq{Christoff} for $X^0=t$ or $\tau$), one finds the non-zero components of the Ricci tensor,
\begin{subequations}
\begin{align}
    R^0_{\ \, 0} = \,  &\,  3 m \, \frac{\ddot a}{a} = \frac{3 m}{a^2} \biggl(
    \frac{a''}{a} - \HH^2 \biggr)\,, \\
    R^i_{\ \, j} = \,  &\,  m\, \biggl(\frac{\ddot a}{a} + 2 H^2
    + 2 \frac{k}{a^2}\biggr) \delta^i_{j} = \frac{m}{a^2} \biggl(\frac{a''}{a} + \HH^2 + 2 k \biggr) \delta^i_{j} \,,
\end{align}
\end{subequations}
where we have recovered the curvature $k$ in the metric tensor
(cf.~\cite{Weinberg:1972kfs,Kolb:1990vq,Baumann_2022,Rezzolla:2013dea} for details).
The Ricci scalar is
\begin{equation}
    R = R^0_{\ 0} + R^i_{\ i} = 6 m \, \biggl(
    \frac{\ddot a}{a} + H^2 + \frac{k}{a^2} \biggr) =
    \frac{6 m}{a^2} \, \biggl(
    \frac{a''}{a}  + k \biggr)\,.
\end{equation}
We note that
$R_{ij} \propto g_{ij}$ and $R_{i0} = 0$ are expected based on homogeneity and isotropy.
The non-zero terms of the Einstein tensor are
\begin{subequations}
\label{Gmunu_comp}
\begin{align}
     G^0_{\ 0} = &  -3 m \biggl(H^2 + \frac{k}{a^2}\biggr) = - \frac{3m}{a^2} (\HH^2 + k )\,, \label{G00_comp} \\
     G^i_{\ j} = & - m \biggl(2 \frac{\ddot a}{a} + H^2 + \frac{k}{a^2}\biggr) \delta^i_j = - \frac{m}{a^2} \biggl(2 \frac{a''}{a} - \HH^2
     + k\biggr) \delta^i_j\,. \label{Gij_comp}
\end{align}
\end{subequations}

Finally, the stress-energy tensor of the Universe can be obtained taking into account homogeneity
and isotropy at large scales.
Hence, $T_{0i} = 0$, and $T_{ij} = \bar p \, g_{ij}$
is the isotropic pressure tensor.
Furthermore, $T_{00} = \bar \rho \, g_{00}$ is the background energy density.
In the reference frame of the Hubble observer $U^\mu = (1, {\pmb 0})/a$, the covariant
stress-energy tensor is
\begin{equation}
    T^{\mu}_{\ \ \nu} = (\bar p + \bar \rho) \, U^\mu U_\nu + m
    \, \bar p \, g^\mu_{\ \, \nu} =  m\,
    {\rm diag}\{-\bar \rho, \bar p, \bar p, \bar p\} \,.
    \label{perf_fluid}
\end{equation}
This stress-energy tensor corresponds to the one for a perfect fluid\footnote{A perfect fluid is a fluid composed by particles in
local thermal equilibrium described, for classical subrelativistic systems, by a Maxwell-Boltzmann distribution
function $f_0$ (see \Sec{viscosity}), and it is the required description to satisfy the homogeneity
and isotropy of the Universe at large scales.} with $\bar p$ and $\bar \rho$
denoting the average pressure and energy density in the Universe as a function of time, which
will be determined by the particle content at each time.
For an equation of state with a constant ratio of pressure to
density, $\bar p = w \bar \rho$,
the Universe can be considered to be dominated by massive dust particles
(matter domination) with $w = 0$, by massless particles (radiation domination) with
$w = \onethird$, or by dark energy with $w = - 1$, which is equivalent to introducing a cosmological
constant $\Lambda$.
A dark energy contribution is necessary to explain the accelerated rate of expansion of the Universe observed
at present time, corresponding
to a homogeneous, negative pressure
in the stress-energy tensor
$T^\munu = - m \, \rho_{\rm vac} \, g^\munu$.
A vacuum energy has been estimated in quantum field theory (QFT), however,
its estimate differs by
many orders of magnitude with respect to
the observed critical energy density of the Universe at present time,
$\rho_{\rm vac}^{\rm QFT}/\rho_\crit^0
\sim 10^{56}$ \cite{Weinberg:2008zzc}.
Hence, the origin of dark energy is still an unsolved problem.

Einstein field equations automatically imply the conservation of the stress-energy tensor as
$G^\munu\covderT{\mu} = 0$ and, hence,
\begin{equation}
    T^\munu\covderT{\mu} = 0\,. \label{cons_Tmunu}
\end{equation}
These conservation laws are consistent with the relativistic Boltzmann
equation, which describes the classical conservation of the distribution
function of particles $f$ in a fluid.
They can be obtained from the first moment of the Boltzmann equation
(cf.~\cite{Cercignani2002RelativisticBoltzmann,Rezzolla:2013dea} for details).
For the perfect fluid description and a comoving observer $U^\mu = (1, {\pmb 0})/a$,
the energy equation is found from the temporal component, $T^{\mu 0}\covderT{\mu} = 0$,
\begin{equation}
    \partial_\tau \bar \rho +
    3 \, (1 + w) \, \HH \, \bar \rho = 0\,.
\end{equation}
The resulting continuity
equation is of first-order, hence we can transform $\partial_\tau \to \partial_{\tau_\alpha}$ and $\HH \to \HH_\alpha$ to describe it for any generic
$\alpha$-time.
This equation can be integrated to find
\begin{equation}
    \int \frac{\dd \bar \rho}{\bar \rho} = -
    3 \int \frac{\dd a}{a} (1 + w)\,,
\end{equation}
which, for a constant $w$, gives a solution
$\bar \rho \sim a^{-3 (1 + w)}$.
In particular,
$\bar \rho \sim a^{-3}$, $a^{-4}$, $a^0$ for matter, radiation, and vacuum dominations.
We note that the introduction of a cosmological constant does not modify the energy equation
as $(\Lambda g^\munu)\covderT{\mu} = 0$.

Finally, Friedmann equations are obtained introducing \Eqs{Gmunu_comp}{perf_fluid} into \Eq{Einstein},
\begin{equation}
    \HH^2 = a^2 \frac{\bar \rho}{3 \Mpl^2} - k\,, \qquad \frac{a''}{a} =
    \frac{a^2}{6 \Mpl^2}
    (\bar \rho - 3 \bar p) - k\,. \label{HH_friedmann}
\end{equation}
Note that the curvature can be absorbed in the total energy density defining $\rho_k = - 3 \Mpl^2 k/a^2$,
\begin{equation}
\label{eq:Friedmann}
     \HH^2 = \frac{a^2}{3 \Mpl^2}  \bar \rho\,,
    \qquad \frac{a''}{a} = \frac{a^2}{6 \Mpl^2}(
    \bar \rho - 3 \bar p)\,.
\end{equation}
These equations allow us to evolve the scale factor $a$ as a function of $\tau$ once the
background values $\bar \rho$ and $\bar p$ are known.
Each of the energy contributions can be normalized to the
present-day critical energy density $\rho_\crit = 3 \Mpl^2 H_0^2$, corresponding
to a closed universe with $k = 0$,
\begin{align}
    \Omega_\rad = \frac{\rho_\rad}{\rho_\crit} = \Omega_{r, 0}  \, (a/a_0)^{-4}\,, \qquad \Omega_\mat = 
    \Omega_{m, 0} \, (a/a_0)^{-3}\,, \qquad \Omega_k =
    \Omega_{k, 0} \, (a/a_0)^{-2}\,,
\end{align}
where the values at present time have been inferred from CMB 
observations as described by the $\Lambda$CDM
model (with a cosmological constant $\Lambda$ describing
dark energy, cold dark matter, CDM,
and baryons contributing to the matter content of the Universe) \cite{Planck:2013pxb,Planck:2014ylh,Planck:2018vyg},
\begin{equation}
    \Omega_{r, 0} \simeq 9.4 \times 10^{-5}\,, \qquad \Omega_{m, 0} \simeq 0.32\,, \qquad
    |\Omega_{k, 0}| \leq 0.01\,,
    \qquad \Omega_\Lambda \simeq 0.68\,.
\end{equation}
Then, the Hubble rate as a function of the scale factor $a$ can be expressed as a  ratio to its value at present time,
\begin{equation}
    \frac{H^2 (a)}{H_0^2} = \Omega_{r, 0} \, (a/a_0)^{-4} + \Omega_{m, 0} \, (a/a_0)^{-3} + \Omega_{k, 0} \,
    (a/a_0)^{-2} + \Omega_{\Lambda} = \Omega (a)\,.
\end{equation}

As curvature decreases slower than matter with a ratio $\Omega_k/\Omega_m \sim a/a_0$, the relevance of curvature in the total energy budget
decreases at earlier times.
Hence, it is safe to set $\Omega_k = 0$ in the early Universe.
We note that going back in time, we find an era of matter domination before $\Lambda$ would become
the dominant contribution, and an even earlier period of radiation domination, since radiation
decreases faster than matter, prior to matter-radiation equality.
The latter corresponds to the era of the Universe history that we
focus on in the present work.

Assuming that at any time, the Universe is dominated by a single component,
\Eq{HH_friedmann} can be solved to find the evolution of the scale factor with conformal time as
$a\sim \tau^{2/(1 + 3w)}$ for $w \neq -\third$.
In particular,
$a \sim \tau^2$, $\tau$, and $-1/\tau$ for matter, radiation, and vacuum\footnote{The evolution of the scale factor when $w = - 1$ is $a_0^{-1} - a^{-1} = \sqrt{\bar \rho/3} \, (\tau - \tau_0)\, \Mpl^{-1}$.} (dark-energy) dominations.
In terms of cosmic time, we find $a \sim t^{2/[3(1 + w)]}$ for $w \neq -1$,
i.e., 
$a \sim t^{2/3}$ and $t^{1/2}$ for $w = 0$ and $\third$ respectively,
and $a \sim e^{Ht}$ for $w = -1$.
The exponential expansion
in vacuum domination corresponds to a de Sitter Universe,
observed at present time, and considered during
the period of inflation in the early Universe.

Considering both matter and radiation near equality (eq), such that $\bar \rho = \bar \rho_m + \bar \rho_r = \half \bar \rho_{\rm eq} [(a_{\rm eq}/a)^3 + (a_{\rm eq}/a)^{4}]$, we can find a solution
\begin{equation}
    a(\tau) = a_{\rm eq} \bigl[(\tau/\tau_\ast)^2 + 2 (\tau/\tau_\ast)\bigr]\,,
    \qquad \tau_\ast = \frac{\tau_{\rm eq}}{\sqrt{2} - 1}\,,
\end{equation}
valid at epochs when mass and radiation contributions are almost equal, like, for example,
around the epoch of recombination, $a_{\rm rec} \sim 9 \times 10^{-4}$, which is close to $a_{\rm eq} \sim 3 \times 10^{-4}$.

\section{Relativistic fluid dynamics in an expanding Universe}
\label{perfect_fluid_hydro}

In this section, the equations of motion describing
the energy density $\rho$ and the velocity field $\uu$
in an expanding background are computed for perfect fluids, i.e.,
fluids that are in local thermal equilibrium (LTE).
Deviations with respect to LTE will be considered
in \Sec{viscosity}
to include viscous effects and heat fluxes,
corresponding to the first-order description of imperfect fluids.
In first-order fluid dynamics,
the viscous terms
follow a Navier-Stokes description and heat fluxes correspond to
those described by Fourier's law of thermal conduction.
As both corrections lead to acausal fluids, they are not a satisfactory
theory for relativistic imperfect fluids.
We briefly discuss this issue in \Sec{viscosity} and refer the reader to
the review \cite{Romatschke:2009im} and the textbook \cite{Rezzolla:2013dea} for more extensive descriptions.
The conservation laws for charged fluids in presence of electromagnetic
fields are described in \Sec{MHD_eom}, leading
to the equations of motion that, combined with Maxwell equations (see \Sec{Maxwell_sec}), describe magnetohydrodynamics (MHD).

Simulations of the early Universe describing the dynamics
of a radiation-dominated fluid
have been performed in the subrelativistic regime,
following the pioneer work of \cite{Brandenburg:1996fc}, in the
context of decaying MHD turbulence (e.g., \cite{Brandenburg:2017neh}),
production of GWs (e.g., \cite{RoperPol:2019wvy}), or chiral decaying MHD (e.g., \cite{Schober:2017cdw}).
For a more comprehensive literature, see references in the introduction.
In their work, the equations of motion for a perfect fluid are found from the conservation
of the stress-energy tensor,
$T^\munu\covderT{\mu} = 0$ after setting $\gamma^2 \to 1$,
\begin{subequations}
\label{eqs_old}
\begin{align}
     \partial_\tau  \ln \tilde \rho =  - &\, \frac{4}{3}
    \bigl[\nab \cdot \uu + (\uu \cdot \nab) \ln \tilde \rho\bigr]\,, 
    \label{continuity_old}\\ 
    \partial_\tau \uu + (\uu \cdot \nab) \, \uu = &\,
    \frac{\uu}{3} \bigl[\nab \cdot \uu + (\uu \cdot \nab) \ln \tilde
    \rho \bigr] - \frac{1}{4} \nab \ln \tilde \rho\,, \label{momentum_old}
\end{align}
\end{subequations}
where $\tilde \rho = a^4 \rho$ is the comoving energy density, and
the comoving pressure is described using the ultrarelativistic
equation of state $\tilde p = \cs^2 \tilde \rho$ with  $\cs^2 = \third$ (see \Sec{eos}).
In this case, the fluid equations of motion are conformally invariant \cite{Brandenburg:1996fc} (see \Sec{cons_perf_fluid}).
In the following, we will present
the derivation of these equations for a generic constant 
equation of state
and will extend them
to include relativistic effects, as already presented in \Eqq{summary_rel}.
Furthermore, we will show in \Secs{conservation_rel_perf}{rel_hydrodynamics} [see, in particular, \Eqq{eqs_subrelativistic_1}]
that these equations are required
to be corrected (by the terms indicated below in red) to:
\begin{subequations}
\label{eqs_new}
\begin{align}
    \partial_\tau \ln \tilde \rho = - &\, \frac{4}{3}
    \bigl[\nab \cdot \uu + \red{\bf \half} (\uu \cdot \nab) \ln \tilde \rho\bigr]\,, \label{continuity_new} \\
    \partial_\tau \uu + (\uu \cdot \nab) \, \uu = &\, \frac{\uu}{3} \bigl[\nab \cdot \uu + \red{\bf \half} (\uu \cdot \nab) \ln \tilde
    \rho \bigr] - \frac{1}{4} \nab \ln \tilde \rho\,,
    \label{momentum_new}
\end{align}
\end{subequations}
due to the fact that the term
$\partial_\tau u^2$ presents non-negligible contributions in the
subrelativistic limit, which had been previously
ignored, as a consequence of setting $\gamma^2 \to 1$.
Indeed, note that $\partial_\tau \gamma^2 = \gamma^4 \partial_\tau u^2$ is related to the conservation
of the kinetic energy contribution to the energy density $T^{00} = \rho + \rho_\kin$,
where $\rho_\kin = (p + \rho) \gamma^2 u^2$ (see \Sec{Tmunu_perf}). 
Therefore, the time derivative of $\rho_\kin$, $\partial_\tau \rho_\kin =
\gamma^2 (1 + \cs^2) (u^2 \partial_\tau \rho + \gamma^2 \rho \, \partial_\tau u^2)$ balances
with the power exerted by the forces to the fluid, e.g., $\uu \cdot 
\nab p$.
These forces include terms of leading order
also in the subrelativistic limit when $\cs^2 \sim {\cal O} (1)$
that lead to the corrections in \Eqq{eqs_new}.

\subsection{Equation of state}
\label{eos}

We are in general interested in the situations when the Universe
is dominated by radiation, such that $\cs^2 \approx \third$.
In general, the energy density can be decomposed into the energy
density of massive ($\rho_\mat$) and massless radiation ($\rho_\rad$) particles,
\begin{equation}
    \rho = \rho_\mat + \rho_\rad = \rho_m (1 + \varepsilon)
    + \rho_\rad\,,
\end{equation}
where $\rho_m$ is the rest-mass density
and $\rho_m \, \varepsilon$ is
the internal energy density.

When $\rho_\mat \ll \rho_\rad$, then the fluid is dominated by
radiation $\rho \simeq \rho_\rad$, and the pressure is described by the ultrarelativistic equation
of state $p_\rad = \cs^2 \rho_\rad \simeq \third \rho_\rad$.
When massive particles are present but still the contribution to the pressure is dominated by radiation
particles, we model the plasma with an equation of
state described by a constant $\cs^2$, such that
the total pressure is $p = \cs^2 \rho$, i.e.,
\begin{equation}
    p = p_\mat + p_\rad \simeq p_\rad = \cs^2 (\rho_\mat + \rho_\rad) = \third \rho_\rad\,.
\end{equation}
In this case, one can approximate the squared
speed of sound as
\begin{equation}
    3 \cs^2 = \frac{1}{1 + \rho_\mat/\rho_\rad}\,.
\end{equation}
For such an equation of state, and under the assumption that the
particles in the fluid are coupled, such that the
fluid can be described with a common four-velocity $U^\mu$ (for interacting multifluid
systems, see \cite{Rezzolla:2013dea} and references therein),
the system of momentum and energy
conservation equations, $T^\munu\covderT{\mu} = 0$,
is closed.
In this work, we will restrict our MHD description to this type of systems: fluids
with radiation particles corresponding to the dominant contribution to the
pressure, such that $p/\rho$ is a constant $\cs^2 \sim {\cal O} (1)$.

On the other hand, when $\rho_\rad \ll \rho_\mat$, the fluid is
dominated by massive particles, and $p \ll \rho$.
In this case, the ideal gas equation of state relates
the pressure and the internal energy density, $p =
\cs^2 \, \rho_m \, \varepsilon$ with a constant $\cs^2$ when
one ignores entropy variations in LTE.
Therefore, as the equation of state does not directly relate
$p$ to the total $\rho = \rho_m(1 + \varepsilon)$,
the system obtained from
$T^\munu\covderT{\mu} = 0$ is no longer closed.
Hence,
it is needed to introduce the
rest-mass conservation law for coupled fluids,
${J^\mu_\mat}\covderT{\mu}= 0$,
with $J^\mu_\mat = \rho_m U^\mu$.
In the subrelativistic regime and ignoring the expansion of the
Universe, the system of equations would reduce to
the usual Newtonian limit of mass, momentum, and energy conservation
(Euler equations)
\begin{align}
    D_t \rho_m + \rho_m \nab \cdot \uu = 0\,, \qquad 
    \rho_m D_t \uu + \nab p = 0\,,
    \qquad 
    \rho_m D_t \varepsilon +
    p \nab \cdot \uu = 0 \,.
\end{align}

In the following, we will always assume that the pressure and total
energy density can be described by an equation of state $p = \cs^2 \rho$,
i.e., that the radiation pressure is the dominant
contribution to the total one, such that
the $T^\munu\covderT{\mu} = 0$
conservation laws
already represent a closed system.
Furthermore, we will assume that the fluid perturbations are smaller
than the background, such that they do not feedback on the metric
tensor.
Hence, we will consider the fluid perturbations over the FLRW
background metric tensor described in the previous section.
For generality, we will
allow the background energy density and
pressure, $\bar \rho$ and $\bar p$, to
contain contributions in addition to those from the fluid, $\rho$
and $p$, such that the background constant describing the equation of state, $w$,
does not necessarily need to be equal to $\cs^2$ (see \Sec{Friedmann}).

\subsection{Stress-energy tensor of a perfect fluid}
\label{Tmunu_perf}

The stress-energy tensor of a perfect fluid can be described as
\begin{subequations}
\begin{equation}
    T^\munu = (p + \rho) \, U^\mu U^\nu + p g^\munu \,.
    \label{Tmunu} 
\end{equation}
\end{subequations}
In the following, we consider the signature
$m = 1$ and $X^0 = \tau$ for compactness
as this choice does not affect the equations of motion.
The stress-energy tensor can also be expressed in terms
of the projection tensor $h^\munu = g^\munu + U^\mu U^\nu$,
\begin{equation}
    T^{\mu \nu} = \rho \, U^\mu U^\nu + p h^\munu\,,
\end{equation}
where $h^\munu$ satisfies $h^\munu U_\nu = 0$.
The different components of the stress-energy tensor are:
\begin{equation}
    a^2 T^{00} = (p + \rho) \, \gamma^2 - p\,, \qquad
    a^2 T^{0i} = (p + \rho) \, \gamma^2 \, u^i\,, \qquad
    a^2 T^{ij} = (p + \rho) \, \gamma^2 \, u^i u^j + p \delta^{ij}\,.
    \label{Tij_perf_fluid}
\end{equation}
Note that, in the following,
we do not distinguish between $u^i$ and $u_i$, as $\uu$ corresponds
to the peculiar velocity, and not to the spatial components of the
four-velocity.
The covariant
energy density in the Hubble observer frame can be rearranged to
\begin{equation}
   - T^{0}_{\ \, 0} = (p + \rho) \, \gamma^2 u^2 + \rho = \rho_{\rm kin} + \rho\,,
\end{equation}
using the identity $\gamma^2 u^2 = \gamma^2 - 1$,
where $\rho_{\rm kin} = (p + \rho) \gamma^2 u^2$ is the relativistic kinetic
energy density.

The trace of the stress-energy tensor plays an important role in the
conservation laws, as we will see in the following.
We first express the trace of the spatial components as
\begin{equation}
    T^{i}_{\ \, i} = a^2 T^{ij} \delta_{ij}
    = \rho_{\rm kin} + 3 p\,, 
\end{equation}
such that the trace of the stress-energy tensor becomes
\begin{equation}
    T = T^{\mu}_{\ \mu} = T^{0}_{\ 0} + T^i_{\ i} =
    3 p - \rho\,.
    \label{trace_Tmumu}
\end{equation}
Then, for a radiation-dominated fluid, $p = \third \rho$,
the trace of the stress-energy tensor vanishes.
This condition implies that the
equations of motion are conformally invariant,
i.e., the equations in an expanding background reduce to those in flat Minkowski
space-time after a conformal transformation \cite{Brandenburg:1996fc,Subramanian:1997gi}, as
we show in the following section.

\subsection{Conservation laws of a perfect fluid}
\label{cons_perf_fluid}

The conservation laws are obtained from \Eq{cons_Tmunu} as a consequence of Bianchi identities
$G^\munu\covderT{\mu} = 0$,
\begin{equation}
    T^\munu\covderT{\mu} = 
    \frac{1}{\sqrt{-g}} \partial_\mu (\sqrt{-g}\, T^\munu)
    + \Gamma^\nu_{\mu\sigma} T^{\mu \sigma}
    = 0\,,
    \label{cons}
\end{equation}
where $; \mu$ is the gravitational covariant derivative given in \Eq{Fmunu},
$\partial_\mu$ the partial
derivative, $\sqrt{-g} = a^4$ for the FLRW metric tensor using $X^0 = \tau$, and $\Gamma^\nu_{\musig}$ are the FLRW Christoffel symbols, given in
\Eq{Christoff}.
The conservation laws correspond to the conservation of energy
and momentum and can be found setting $\mu = 0$ and $i$, respectively, in \Eq{cons}.
They characterize
a closed system of equations (for the energy density $\rho$ and the
peculiar velocity $\uu$) when the equation of state relating the
total energy density and the pressure is known, e.g., $p = \cs^2 \rho$.
This allows us to study coupled fluids composed by radiation and
massive particles when $\cs^2 \sim {\cal O} (1)$ is a known constant.
Alternatively, an equivalent set of the equations of
motion can be found by projecting
$T^\munu\covderT{\mu} = 0$ in the parallel, $U_\nu T^\munu\covderT{\mu} = 0$,
and perpendicular directions to $U_\nu$, $h_{\nu \lambda} T^\munu\covderT{\mu} = 0$,
leading to the energy and momentum equations, respectively \cite{Rezzolla:2013dea}.

\subsubsection*{3.3.1 \ Conformal invariance and conservation laws}

An important aspect of the conservation laws of perfect fluids
is that they are conformally invariant when the trace of the
stress-energy tensor vanishes \cite{1977ApJ...211..361L,Brandenburg:1996fc,Subramanian:1997gi}.
To show this result, let us consider two metrics that are related by a conformal transformation as
$g^{\munu} = \Omega^2 (x^\mu) \, \tilde g^\munu$.
This corresponds, for example, to the FLRW metric
tensor $g^\munu$ when
we use  conformal time as the time coordinate, where
$\tilde g^\munu = \eta^\munu$ would correspond to flat Minkowskian space-time and $\Omega = a^{-1}
(\tau)$ is the inverse scale factor.
The transformation of the covariant derivative of a symmetric $T^\munu$ in a metric tensor
$g^\munu$ is
\begin{align}
    T^\munu\covderT{\mu} = &\, \frac{1}{\sqrt{-g}} \partial_\mu
    (\sqrt{-g} \, T^\munu)
    + \Gamma^\nu_{\musig} T^\musig \nonumber \\ = &\,
    \frac{\Omega^4}{\sqrt{-\tilde g}} \partial_\mu
    (\sqrt{-\tilde g} \, \Omega^{-4} \, T^\munu) +
    \tilde \Gamma^\nu_{\musig} T^\musig - 2 \, T^{\munu} \partial_\mu \ln \Omega
    + T^\mu_{\ \mu} \, \partial^\nu \ln \Omega\,,
\end{align}
where $\tilde \Gamma^\nu_{\musig}$ are the Christoffel symbols
of the metric tensor $\tilde g^\munu$.
Introducing the comoving stress-energy tensor $\tilde T^\munu = \Omega^{-6}\, T^\munu$, we find
\begin{equation}
    T^\munu\covderT{\mu} = \Omega^6 \bigl(\tilde T^\munu\covderT{\tilde \mu} +
    \tilde T^\sigma_{\ \, \sigma} \,
    \tilde g^\munu \partial_\mu \ln \Omega\bigr) = 0 \Leftrightarrow
    \tilde T^\munu\covderT{\tilde \mu}
    + 
    \tilde T^{\sigma}_{\ \, \sigma} \,
    \tilde g^\munu \partial_\mu \ln \Omega =0\,,
    \label{cons_expand}
\end{equation}
where $ ; \tilde \mu$
is the covariant derivative in the
metric tensor $\tilde g^\munu$.

As we have shown in \Eq{trace_Tmumu}, the stress-energy tensor of a perfect fluid is traceless when $p = \third \rho$.
Then, in this case, the equations of motion become invariant under conformal
transformations,
\begin{equation}
    T^\munu\covderT{\mu} = 0 \Leftrightarrow
    \tilde T^{\munu}\covderT{\tilde \mu} = 0 \,. \label{cov_notrace}
\end{equation}
In particular, for a homogeneous and isotropic
expanding Universe, $\tilde g^\munu = \eta^\munu$ is the Minkowski
metric tensor, such that the equations of motion are conformally flat,
$\partial_{\mu} \tilde T^\munu = 0$.
For non-zero trace,
the conservation laws \Eq{cons_expand} in an expanding background with $\Omega = a^{-1}$ become
\begin{equation}
    \partial_\mu
    \tilde T^{\mu 0}  + \tilde T \, \HH = 0\,, \qquad
    \partial_\mu \tilde T^{\mu i} = 0\,.\label{cons_FLRW}
\end{equation}
The comoving trace
is $\tilde T = a^4 \, T = 3 \tilde p - \tilde \rho$,
where we have also defined $\tilde \rho = a^4 \, \rho$ and $\tilde p = a^4 \, p$ as the comoving energy density and pressure.
Therefore, the energy and momentum equations are 
directly found in conformal time using \Eq{cons_FLRW},
\begin{equation}
\partial_\tau \tilde T^{00}  + \partial_i \tilde T^{0i} =  (\tilde \rho - 3 \tilde p)
\, \HH \,,  \qquad  
    \partial_\tau \tilde T^{0i} + \partial_j \tilde T^{ij} = 0\,.
    \label{momentum1}
\end{equation}
We notice that this system of equations can be generalized to different geometries in general relativity, following the so-called Valencia formulation (see \cite{Rezzolla:2013dea} and references therein),
where the $- \tilde T\, \HH$ term
in the right-hand side of the energy equation corresponds to the extrinsic curvature
$K_{ij}$ contracted with the perfect fluid
stress-energy tensor $\tilde T^{ij}$ in FLRW geometry \cite{Rezzolla:2013dea}.

Moreover, these equations can be described in a generic $\alpha$-time,
by rescaling $\partial_\tau \to a^{1 - \alpha} \partial_{\tau_\alpha}$.
Similarly, we can also rescale $\HH \to a^{1 - \alpha} \HH_\alpha$ with $\HH_\alpha = (\partial_{\tau_\alpha} a)/a$, such that the equations of motion in $\alpha$-time become
\begin{align}
    \partial_{\tau_\alpha} \tilde T^{00} + a^{\alpha - 1} \partial_i \tilde T^{0i} =    (\tilde \rho - 3 \tilde p)\, \HH_\alpha \,, \qquad 
    \partial_{\tau_\alpha} \tilde T^{0i} + a^{\alpha - 1} \partial_j \tilde T^{ij} = 0\,. \label{conformal}
\end{align}
Note that for cosmic time, $a^{-1} \partial_j$ corresponds to a derivative
with respect to the physical coordinate $r_i = a\, x_i$ and $\HH/a = H$.
In general, the factor $a^{\alpha - 1}$ can be absorbed in a
space-coordinate $\xx_\lambda = a^{-\lambda} \, \xx$ with $\lambda = \alpha - 1$.
This rescaling is possible and reduces to a conformally flat system of equations
when $p = \third \rho$ as long as the mapping $g^\munu = a^{-2} \eta^\munu$ is satisfied.

From now on, unless otherwise stated, we will consider $\alpha = 1$ for compactness of the resulting
equations.
To recover the expression for any $\alpha$-time, it is enough to include
the $a^{\alpha - 1}$ term in the divergence term
$\partial_i \tilde T^{i \mu}$
and to substitute $\HH \to \HH_\alpha$.

\subsubsection*{3.3.2 \ Conservation laws for the physical stress-energy components}

Using the comoving stress-energy tensor components, $\tilde T^\munu = a^6 T^\munu$, allows us
to find conformally flat equations of motion for a perfect fluid
when $\rho = 3 p$ and the coordinates $X^\mu$ are chosen such that
the FLRW metric tensor maps to Minkowski space-time as $g^\munu = a^{-2} \eta^\munu$
(i.e., taking conformal time and comoving space coordinates).
Let us now consider the equations of motion
of the physical covariant
stress-energy components $T^0_{\ \, \mu} = T^{0\nu} g_\munu = a^2 \,
T^{0\nu} \eta_\munu$ in an expanding background,
\begin{subequations}
\label{generic_eom}
\begin{align}
    \partial_{\tau_\alpha} T^0_{\ \, 0} \, + &\,
    a^{\alpha - 1} \partial_i  T^{i}_{\ \, 0} = (T^{i}_{\ \, i} - 3\, T^0_{\ \, 0})
    \, \HH_\alpha  = f^{H}_{0} \,, \label{generic_eom1} \\ 
    \partial_{\tau_\alpha} T^0_{\ \, i} \, + &\,  a^{\alpha - 1} \partial_j T^{j}_{\ \, i} =
    - 4 \, T^{0}_{\ \, i}\, \HH_\alpha = f^{H}_{i}  \,, \label{generic_eom2}
\end{align}
\end{subequations}
where $f^{H}_{\mu}$
corresponds to an effective Hubble ``force'' (Hubble friction) that appears
due to the expansion of the Universe \cite{Jedamzik:1996wp},
\begin{equation}
    f^{H}_{0} = (4 \gamma^2 - 1)
    (p + \rho) \, \HH_\alpha \,, \qquad
    f^{H}_i = - 4 \,  (p + \rho) \, \HH_\alpha \gamma^2 u^i\,.
    \label{fH_zero}
\end{equation}

In the subrelativistic regime, when $\gamma^2 \simeq 1$, the stress-energy
terms in the trace simplify to $T^0_{\ \, 0} = -\rho$ and $T^i_{\ \, i} = 3 p$,
such that the equations of motion
(\ref{generic_eom})
become
\begin{subequations}
\begin{align}
    \partial_{\tau_\alpha} (\rho \gamma^2) 
    + a^{\alpha - 1} \partial_i \bigl[(p + \rho) u^i \bigr]
   =  - & 3\, (p + \rho) \, \HH_\alpha  \,, \label{cont_noscaling} \\
    \partial_{\tau_\alpha} \bigl[( p + \rho) \gamma^2 \, u^i\bigr]  + a^{\alpha - 1}
    \partial_j \bigl[(p + \rho) \, u^i u^j + p \delta^{ij}\bigr] = - & 4 \,
    (p + \rho) \, \HH_\alpha  \, u^i\,.
 \label{mom_noscaling}
\end{align}
\end{subequations}
We have kept the $\gamma^2$ terms in the time derivatives
of $T^{0}_{\ \, 0}$ and $T^{0}_{\ \, i}$ because, as anticipated at
the beginning of the section, $\partial_\tau \gamma^2$
leads to additional subrelativistic terms for a generic value
of $p/\rho = \cs^2$
that are only negligible for dust, when $\cs^2 \ll 1$.

\subsubsection*{3.3.3 \ Generic scaling and super-comoving coordinates}
\label{generic_scaling}

Let us now consider a generic scaling of the fluid variables,
\begin{equation}
    \tilde p = a^\beta p\,, \qquad \tilde \rho = a^\beta \rho\,,
    \qquad \tilde u^i = a^\delta u^i\,,
\end{equation}
such that $\beta = 4$ and $\delta = 0$ recover the comoving variables
used in \Eq{momentum1}.
The rescaled stress-energy components include the scaling
of the fluid variables and an additional $a^2$ to take into
account the prefactor in \Eq{Tij_perf_fluid},
\begin{subequations}
\label{Tmunu_aux}
\begin{align}
    \tilde T^{00} = &\, a^{2 + \beta}
    T^{00} = (\tilde p + \tilde \rho)\, \gamma^2 - \tilde p\,, \label{T00_aux} \\
    \tilde T^{0i} = &\, a^{2  + \beta + \delta} T^{0i}
    = (\tilde p + \tilde \rho ) \, \gamma^2 
    \, \tilde u^i\,, \\ \tilde T^{ij} = &\, a^{2 + \beta + 2 \delta}
    \, (\tilde p + \tilde \rho) \,
    \gamma^2 \, \tilde u^i \tilde u^j + a^{2 + \beta} \,
    \tilde p \, \delta^{ij}\,. \label{Tij_aux}
\end{align}
\end{subequations}
The energy and momentum conservation
equations for this generic scaling become
\begin{subequations}
\label{aux_eqs}
\begin{align}
    \partial_{\tau_\alpha} \tilde T^{00} \, + &\, a^{\alpha - 1 - \delta} \partial_i \tilde T^{0i} = \tilde f^0_H\,, \label{aux_cons}
    \\
    \partial_{\tau_\alpha} \tilde T^{0i} \, + &\, a^{\alpha - 1 - \delta}
    \partial_j \bigl[(\tilde p + \tilde \rho) \, \gamma^2 \, \tilde u^i \tilde u^j + a^{2\delta}
    \tilde p \, \delta^{ij} \bigr]
    =  \tilde f^i_H \,, \label{aux_mom}
\end{align}
\end{subequations}
where we find Hubble forcing components\footnote{The
Hubble forcing terms $\tilde f_H^\mu$ resultant from the rescaling cannot be
directly rescaled to the components in \Eqq{generic_eom}, since they also
incorporate the contribution from the time
derivatives of the scale factor when rescaling the
stress-energy tensor components $\tilde T^{0\mu}$.} proportional to $\HH_\alpha$
in both the energy and momentum equations,
\begin{equation}
    \tilde f^0_H =  -  
    \bigl[(3 - \beta + a^{-2\delta} \tilde u^2)
    (\tilde p + \tilde \rho) \gamma^2 + 
    \beta \, \tilde p
    \bigr] \, \HH_\alpha  \,, \qquad \tilde f^i_H = -
    (4 - \beta - \delta)
    \, \tilde T^{0i}\, \HH_\alpha \,. \label{Hubble_force}
\end{equation}
Note that these terms reduce to those in \Eq{fH_zero} when $\beta = \delta = 0$, and to those in \Eq{conformal} when $\beta = 4$ and $\delta = 0$.
The latter choice appears naturally when we try to get rid of the Hubble
friction terms.
In first place, to make $\tilde f^0_H$ vanish
for a relativistic bulk velocity with $u^2 \sim {\cal O}(1)$,
the only possible choice is $\delta = 0$.
Then, we observe that only for $\beta = 4$, the dependence on $\gamma^2$ vanishes.

In the following, we will consider a
Hubble friction $\tilde f_H^\mu$ to allow a generic
choice of the rescaling.
However, note that when $\delta \neq 0$, the relation
between the Lorentz factor and the scaled velocity
$\tilde u^i$ is not the same as in special relativity,
\begin{equation}
    \gamma^2 = \frac{1}{1 - a^{-2 \delta} \tilde u^2}\,.
    \label{Lorentz_alpha}
\end{equation}

\subsubsection*{3.3.4 \ Subrelativistic limit}

It seems like $\beta = 4$ and $\delta = 0$ is, in general,
the best choice for the scaling of the fluid variables as it allows to obtain
conformally flat equations of motion when the fluid particles are relativistic.
However, when the fluid bulk motion is subrelativistic,
a different scaling to get rid of the Hubble term in the energy
equation, $\tilde f_H^0$, can be found and can be
a better choice in some cases.
Taking the limit $u^2 \ll 1$ in \Eq{Hubble_force}, we find
\begin{equation}
    \tilde f_H^0 = -  \bigl[(3 - \beta) \, \tilde \rho + 3 \tilde p \bigr]\, \HH_\alpha\,, \qquad
    \tilde f_H^i = - (4 - \beta - \delta )
    \, \tilde T^{0i} \, \HH_\alpha\,.
\end{equation}
Taking a constant equation of state $\tilde p = \cs^2 \tilde \rho$,
$\tilde f_H^0$ vanishes when $\beta = 3 \, (1 + \cs^2)$ for any value of $\cs^2$,
and the remaining Hubble term in the momentum equation,  $\tilde f_H^i$,
becomes
\begin{equation}
    \tilde f_H^i =(3 \cs^2 + \delta - 1)\, \tilde T^{0i}\, \HH_\alpha\,.
\end{equation}
Then, the energy equation becomes conformally flat for
any choice $\alpha = \delta + 1$,
which allows us to get rid of the scale-factor-dependent term in \Eq{aux_cons}.
The energy and momentum equations become
\begin{subequations}
\label{eqs_subrel_aux}
\begin{align}
    \partial_{\tau} (\tilde \rho \gamma^2) \, + &\, (1 + \cs^2) \, \partial_i (\tilde \rho u^i)
    = 0\,, \label{cont_subrel_aux} \\
    \partial_\tau (\tilde \rho \gamma^2 u^i) \, + &\,
    \partial_j (\tilde \rho u^i u^j) + a^{2 \delta}  \frac{\partial_i \tilde p}{1 + \cs^2}
     = (3\cs^2 + \delta - 1) \, \tilde \rho u^i \,  \HH_\alpha  \,,\label{mom_subrel_aux}
\end{align}
\end{subequations}
where we are only left with a dependence on $\HH_\alpha$ in the momentum equation,
corresponding to a Hubble friction term that will affect the velocity field evolution.

This will be a useful choice when $\tilde p \neq \third \tilde \rho$ in the subrelativistic regime.
Indeed,
this choice minimizes the appearance
of Hubble-dependent terms in the equations of conservation of 
energy and momentum due to the fact that a
non-vanishing $\tilde f_H^0$ in the energy equation
will propagate to the momentum equation, as we will see in \Sec{rel_hydrodynamics}.
We remind the reader that a choice yielding a vanishing $\tilde f_H^0$
was not in general possible for relativistic flows.

\subsubsection*{3.3.5 \ Super-comoving coordinates when $\cs^2 \ll 1$}

In the previous section, we have shown that $\alpha = \delta + 1$ is
a useful scaling for subrelativistic bulk flows with $u^2 \ll 1$.
A particular choice of $\delta = 1$ and $\alpha = 2$ (known as super-comoving
coordinates) is common
in cosmological simulations during matter domination as it allows to get rid of the friction terms as originally proposed in \cite{Martel:1997hk},
based on previous work in \cite{1980Afz....16..769S}.
An alternative choice for the super-comoving coordinates was proposed
in \cite{Banerjee:2004df}.
The latter choice adapts the $\alpha$-time to the Universe expansion during
matter domination, setting
$\alpha = \threehalf$ and $\delta = \half$, and allows to keep a constant Hubble rate
in $\alpha$-time, being a useful choice to compute the evolution of primordial magnetic
fields across recombination \cite{Jedamzik:2018itu,Trivedi:2018ejz,Jedamzik:2020krr,Jedamzik:2023rfd,Jedamzik:2025cax}.

For the case of a matter-dominated fluid ($\cs^2 \ll 1$),
it is possible to recover conformally flat fluid equations
in the subrelativistic regime, following the super-comoving variables choice
proposed in \cite{Martel:1997hk}.
We can see from \Eq{mom_subrel_aux} that the choice $\delta = 1 - 3 \cs^2$
allows us to get rid of the Hubble friction. However,
as a consequence of $\delta \neq 0$, an additional dependence with
$a$ appears in the pressure gradient.
We can also get rid of this term
when $\cs^2 \ll 1$, since then the enthalpy is dominated by the energy density,
$\tilde p + \tilde \rho \approx \tilde \rho$ in
\Eqq{Tmunu_aux}.
Hence, the only dependence on the pressure
remains in the $\tilde T^{ij}$ component, such that 
$\tilde \rho$ and $\tilde p$ become decoupled in the equations,
allowing us to separately rescale
\begin{equation}
    \tilde p = a^\chi \, p\,, \qquad {\rm with}  \quad \chi = \beta +
     2 \delta\,.
\end{equation}

Finally, to get rid of the Hubble friction, we can choose $\delta =  1$ and, hence, $\alpha = 2$,
$\beta = 3$, and $\chi = 5$.
The resulting conservation laws then become conformally flat \cite{Martel:1997hk},
\begin{align}
    \partial_{\tau_\alpha} (\tilde \rho \gamma^2) + \partial_i (\tilde \rho \tilde u^i)
    =  0\,, \qquad 
    \partial_{\tau_\alpha} (\tilde \rho \gamma^2 \tilde u^i) + \partial_j
    (\tilde \rho \tilde u^i \tilde u^j) + \partial_i \tilde p = 0\,.
\end{align}
We note that the relation between the Lorentz factor and $\tilde u^i$ is not
the same as in Minkowski space-time when $\delta \neq 0$ [cf.~\Eq{Lorentz_alpha}].
However, as we deal with subrelativistic flows in this limit, this correction
becomes negligible.

An alternative choice of the
super-comoving coordinates is found when one chooses
$\alpha = \threehalf$ and  $\delta = \half$, which allows to absorb the
Universe expansion in the time coordinate when the Universe
is matter-dominated, $a \sim t^{2/3}$, such that $\ln a \sim \tau_\alpha$ and,
hence, $\HH_\alpha$ is constant in $\alpha$-time \cite{Banerjee:2004df}.
For a generic background equation of state $w \neq -1$, the scale factor
evolves as $a \sim t^{2/[3 (1 + w)]}$ (see \Sec{Friedmann}).
Hence, the choice $\alpha = \threehalf (1 + w)$ and $\delta = \half (1 + 3w)$ allows to
compensate for the
evolution of the scale factor.
The resulting momentum equation (for $\cs^2 \ll 1$) is
\begin{equation}
    \partial_{\tau_\alpha} (\tilde \rho \gamma^2 \tilde u^i) +
    \partial_j (\tilde \rho \gamma^2 \tilde u^i \tilde u^j)
    + a^{\beta - \chi + 1 + 3w} \, \frac{\partial_i \tilde p}{1 + \cs^2} =
    \half (3 w - 1) \tilde \rho \tilde u^i\, \HH_\alpha \,.
\end{equation}
Then, the choice $\chi = \beta + 2 \delta = 4 + 3 w$ leads
to the following energy and momentum equations,
\begin{subequations}
\begin{align}
    \partial_{\tau_\alpha} (\tilde \rho \gamma^2) \, + &\, \partial_i (\tilde \rho
    \tilde u^i) = 0\,, \\
    \partial_{\tau_\alpha} (\tilde \rho \gamma^2 \tilde u^i) \, +
    &\,
    \partial_j (\tilde \rho \tilde u^i \tilde u^j) + \partial_i \tilde p
    = \half (3w - 1) \, \HH_\alpha \,
    \tilde \rho \tilde u^i\,,
\end{align}
\end{subequations}
where the scale-factor dependence only remains on the right-hand
side of the momentum equation, proportional to $\half \HH_\alpha$, which is
a constant due to the particular choice of $\tau_\alpha$.
These equations allow to use an $\alpha$-time that follows the Universe
expansion with a minimal dependence on the scale factor
to study baryon evolution in a generic background.
They generalize the super-comoving coordinates
introduced in \cite{Banerjee:2004df} for $w = 0$.
Furthermore, when the background is dominated by radiation $ w = \third$, even though the
particles of the fluid are dust, the friction term vanishes.

\begin{table}[t]
    \centering
\begin{tabular}{|M{4.5cm}|M{2.0cm}|M{2.0cm}|M{2.0cm}|M{2.0cm}|M{2.0cm}|} \hline
& $\beta$ & $\chi$ & $\alpha$ & $\delta$ \\ \hline
    \shortstack[c]{ \rule{0pt}{12pt}
    Relativistic $u^2\sim {\cal O} (1)$ \\ or $\cs^2 = \onethird$
    \cite{1977ApJ...211..361L,Brandenburg:1996fc}
    }
   & 4 & 4 & 1 & 0 \\ \hline 
   Subrelativistic $u^2 \ll 1$  & $3 \, (1 + \cs^2)$ &
   $3 \, (1 + \cs^2)$ & 1 + $\delta$ & $\delta$ \\ \hline
   \shortstack{\rule{0pt}{12pt} Subrelativistic dust $\cs^2 \ll 1$ \\ (super-comoving \cite{Banerjee:2004df})} &  3 & $4 + 3 w$ & $\threehalf (1 + w)$ & 
   $\half (1 + 3 w) $ \\ \hline
   \shortstack{\rule{0pt}{12pt} Subrelativistic dust $\cs^2 \ll 1$  \\
   (super-comoving \cite{Martel:1997hk})} & 3 & 5 & 2 & 1 \\ 
   \hline
\end{tabular}
    \caption{Summary of the fluid variables choice of
    scaling $\tilde \rho = a^\beta \rho$, $\tilde p = a^\chi \, p$,
    $a^{\alpha - 1} \dd \tau_\alpha = \dd \tau$, and
    $\tilde u^i = a^\delta u^i$.
    The dependence on $w$ of the super-comoving variables are
    a generalization proposed in this work with respect to the $w = 0$ case
    \cite{Martel:1997hk}.
    }
    \label{tab:my_label}
\end{table}

Table~\ref{tab:my_label} summarizes the different choices of the
scaling of the
fluid variables discussed.
For relativistic bulk fluid velocities, it will in general be useful
to consider $\beta = \chi = 4$, $\alpha = 1$ (conformal time), and $\delta = 0$,
corresponding to the conformal transformation discussed at the beginning of the section, as 
this choice leads to conformally flat equations when $\cs^2 = \third$.
On the other hand,
for subrelativistic bulk velocities, it is possible to find a
conformally flat energy equation even when $\cs^2 \neq \third$ with the
choice $\beta = \chi = 3 \, (1 + \cs^2)$ and $\alpha = 1 + \delta$.
The resulting
momentum equation presents an $a^{2\delta}$ dependence in the pressure gradient and a Hubble friction [cf.~\Eq{mom_subrel_aux}].
For dust, when $\cs^2 \ll 1$, $\tilde p$ and $\tilde \rho$ become
decoupled and it is possible to choose $\chi = \beta + 2 \delta \neq \beta$ to get rid of scale-factor dependence in the pressure gradient.
Two useful choices then either make the momentum equation conformally
flat or compensate the time coordinate with
the Universe expansion.
The former choice corresponds to the super-comoving
coordinates $\alpha = 2$ and $\delta = 1$, proposed in \cite{Martel:1997hk},
which
provides conformally flat fluid equations.
The latter choice generalizes
the super-comoving coordinates suggested for $w = 0$ in \cite{Banerjee:2004df}.
It corresponds to $\alpha = \threehalf (1 + w)$ and $\delta = \half 
(1 + 3w)$,
resulting in a Hubble friction $\tilde f_H^i = \half \, (3w - 1) \,
\tilde \rho \tilde u^i \, \HH_\alpha$ with a constant $\HH_\alpha$ in $\tau_\alpha$.
The two different choices might be useful in different circumstances,
depending on whether we want to use results from usual fluid dynamics
exploiting the conformal invariance of the equations, or whether
we prefer to use a time variable that evolves compensating for the
evolution of the scale factor, yielding then an additional Hubble
friction term in the usual fluid dynamic equations.

\subsection{Conservation form of relativistic fluid dynamics}
\label{conservation_rel_perf}

The equations of motion found in \Sec{cons_perf_fluid}, together
with an equation of state $p = \cs^2 \rho$, already
constitute a closed system of equations that can be solved
to compute the stress-energy tensor components $\tilde T^{0 \mu}$:
\begin{align}
    \partial_{\tau_\alpha} \tilde T^{0 \mu} + &\,
    a^{\alpha - 1} \partial_j \tilde T^{j \mu} =
    \tilde f_H^\mu\,,
    \label{mom_aux}
\end{align}
corresponding to \Eqq{aux_eqs} for $\delta = 0$.
For this choice, $\tilde T^{ij} = a^{2 + \beta} \, T^{ij}$,
and the Hubble friction due to the expansion of the Universe,
$\tilde f^\mu_H$, is taken from \Eq{Hubble_force},
\begin{equation} \label{fH_nu}
    \tilde f_{H}^0 =
    \bigl[(\beta - 4) \, \tilde T^{00} - \tilde T\bigr] \HH_\alpha\,,
     \qquad \tilde f_H^i =
    \, (\beta - 4)\, \tilde T^{0i} \,  \HH_\alpha\,.
\end{equation}

These equations are expressed in the so-called conservation form, which is more compact than the non-conservation form
of relativistic fluid dynamics that we will discuss in \Sec{rel_hydrodynamics}
[already presented in \Eqq{summary_rel}].
The latter system has the advantage that it directly
solves for the primitive fluid variables $\tilde \rho$ and $u^i$, while in
the conservation form the primitive variables need to be
non-linearly reconstructed from the stress-energy
components $\tilde T^{0\mu}$.
Note that this terminology only refers to the structure of the equations
and it is related to the conservation of fluxes at the discrete cells in numerical
simulations \cite{Rezzolla:2013dea}, with particular relevance
for finite volume methods \cite{Anderson:1992}.
However,
both forms describe the same
conservation laws of the fluid and, hence, they are equivalent in the continuum
limit.
For a particular application, it is not always obvious which form is
more suitable. Hence, it is useful to test and compare the numerical
solutions using both methodologies.

In general, a system of partial differential
equations is said to be in the conservation form if it
is expressed in the following way
\begin{equation}
    \partial_{\tau_\alpha} {\cal U}^a + \partial_j {\cal S}^{aj} = {\cal F}^a\,, \label{cons_form}
\end{equation}
for an array of variables ${\cal U}^a$, fluxes ${\cal S}^{aj}$, and forces ${\cal F}^a$.
In the case of perfect-fluid relativistic fluid dynamics in an expanding
background, taking ${\cal U}^\mu = \tilde T^{0\mu}$,
${\cal S}^{\mu j} = a^{\alpha - 1} \tilde T^{\mu j}$, and
${\cal F}^\mu = \tilde f^\mu_H$, \Eq{mom_aux} is recovered.

\begin{figure}
    \centering
    \includegraphics[width=0.65\linewidth]{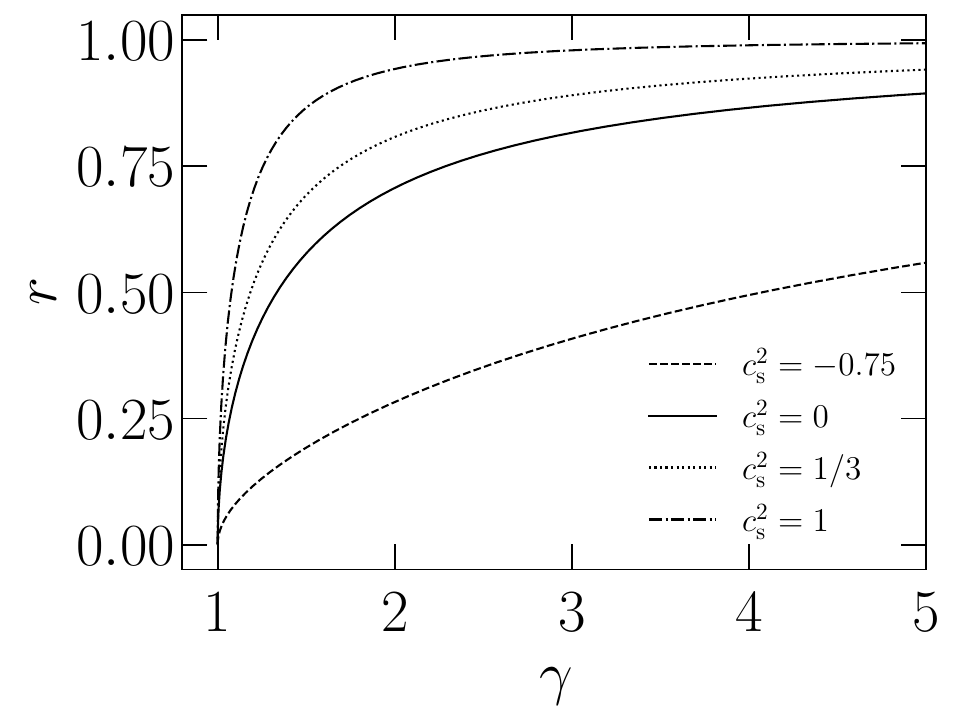}
    \caption{Ratio $r = \sqrt{\tilde T^{0i} \tilde T^{0i}}/\tilde T^{00}$ of a perfect fluid as a function of the Lorentz factor
    $\gamma$ for different constant values of $\cs^2 = \tilde p/\tilde \rho$.}
    \label{r_vs_gamma}
\end{figure}

The conservation form requires to express the primitive variables $\tilde \rho$ and $u^i$
in terms of the variables $\tilde T^{0\mu}$ to
compute $\tilde T^{ij}$.
Once we have a closed form for the fluxes $\tilde T^{\mu j}$ and forces $\tilde f^\mu_H$
in terms of $\tilde T^{0\mu}$, the system in \Eq{mom_aux} can be
numerically computed.
Assuming a generic but constant $\cs^2 = \tilde p/\tilde \rho$,
we first compute the following variables,
\begin{equation}
    \tilde T^{00} = (1 + \cs^2) \, \tilde \rho \, \biggl( \gamma^2 - \frac{\cs^2}{1 + \cs^2}\biggr)\,, \qquad \tilde T^{0i} \tilde T^{0i} = (1 + \cs^2)^2 \tilde \rho^2 \gamma^2 (\gamma^2 - 1)\,,
\end{equation}
and define the ratio
\begin{equation}
    r^2 = \frac{\tilde T^{0i} \tilde T^{0i}}{({\tilde T^{00}})^2} =
    \frac{\gamma^2(\gamma^2 - 1)}{\Bigl(\gamma^2 - \frac{\cs^2}
    {1 + \cs^2}\Bigr)^2}\,, \label{r2_vs_gamma}
\end{equation}
which is shown in \Fig{r_vs_gamma} as a function of $\gamma$ for different
values of $\cs^2$.
We note that for this relation to be valid we require $\delta = 0$, as any rescaling
of $u^i$ would affect the relation between $\gamma^2$ and $u^2$.
The previous equation can be solved for $\gamma^2$,
\begin{equation}
    \gamma^2 = \frac{1}{2(1 - r^2)} \left[ 1 - 2 r^2 \frac{\cs^2}{1 + \cs^2} + \sqrt{1 - 4 r^2 \frac{\cs^2}{(1 + \cs^2)^2}}
    \ \right] \,, \label{gamma2_r2}
\end{equation}
where we take the positive root of the quadratic equation to ensure
$\gamma^2 \geq 1$ when $\cs^2 \leq 1$.
Then, the fluid primitive variables can be expressed in terms of $\tilde T^{0\mu}$
and $\gamma^2$ as
\begin{equation}
    \tilde \rho = \frac{\tilde T^{00}}{(1 + \cs^2) \gamma^2 - \cs^2}
    \,, \qquad 
    u^i = \frac{\tilde T^{0i}}{(1 + \cs^2) \tilde \rho \gamma^2} 
    = \frac{\tilde T^{0i}}{\tilde T^{00}} \biggl[1 - \frac{\cs^2}{(1 + \cs^2)\gamma^2} \biggr]\,,\label{u_i_conservative}
\end{equation}
such that the flux $\tilde T^{ij}$ is
\begin{equation}
    \tilde T^{ij} = \bigl[(1 + \cs^2) \, \gamma^2 u^i u^j + \cs^2 \delta^{ij} \bigr]\, \tilde \rho = \frac{\tilde T^{0i} \tilde T^{0j}}{(1 + \cs^2) \tilde \rho \gamma^2} + \cs^2\, \tilde \rho \, \delta^{ij}\,.
\end{equation}
Using this formalism,
the non-linear relativistic system of equations can be solved
evolving the $\tilde T^{0\mu}$ components
and then the primitive fluid variables $\tilde \rho$
and $u^i$ can be reconstructed from $T^{0\mu}$.
A similar procedure has been used, e.g., in the Higgsless formulation,
to numerically compute the fluid perturbations induced in the
primordial plasma by first-order phase transitions in \cite{Caprini:2024gyk,Jinno:2020eqg,Jinno:2022mie}.

\subsubsection*{Conservation form in the subrelativistic limit}

In the subrelativistic limit, $u^2 \ll 1$,
the components of the
stress-energy tensor become
\begin{equation}
    \tilde T^{00} = \tilde \rho \,, \qquad \tilde T^{0i} = (1 + \cs^2)
    \, \tilde \rho u^i\,, \qquad \tilde T^{ij} =
    \frac{1}{1 + \cs^2}
    \frac{\tilde T^{0i} \tilde T^{0j}}{\tilde T^{00}}
    + \cs^2 \, \tilde T^{00} \, \delta^{ij}\,.
\end{equation}
These relations are tempting due to their simplicity,
as they already provide a straightforward relation between
$\tilde T^{ij}$ and $\tilde T^{0\mu}$.
However, as we have already mentioned,
time derivatives of $\gamma^2$ lead to additional
terms in the energy conservation equation that are not negligible
in the subrelativistic regime, as we will prove in the following.
Therefore, if one sets $\gamma^2 \to 1$ to express $\tilde T^{ij}$ in terms
of $\tilde T^{0\mu}$, indirectly $\partial_\tau \gamma^2 \to 0$ is being assumed
and, hence, subrelativistic terms are ignored.
To show this and recover all relevant terms, let us expand
\Eq{gamma2_r2} up to first order in $r^2 \equiv \tilde T^{0i} \,
\tilde T^{0i}/(\tilde T^{00})^2 \sim {\cal O} (u^2)$,
\begin{equation}
    \gamma^2 = 1 + \frac{r^2}{(1 + \cs^2)^2} + {\cal O} (u^4)\,.
\end{equation}
The $r^2 \to 0$ limit is equivalent to the subrelativistic limit, as can
be seen in \Fig{r_vs_gamma}.
In fact, the expansion up to order $r^{2n}$ is equivalent
to an expansion up to order $u^{2n}$.
Then, substituting $\gamma^2$ up to first order in $r^2$ into
$\tilde T^{0\mu}$, we find
\begin{equation}
    \tilde T^{00} =  \biggl(1 + \frac{r^2}{1 + \cs^2}\biggr)
    \tilde \rho + {\cal O} (u^4) \,, \qquad \tilde T^{0i} =  
    \biggl(1 + \cs^2 + \frac{r^2}{1 + \cs^2} \biggr)\tilde \rho \, u^i + {\cal O} (u^4) \,, \label{T0mu_r2}
\end{equation}
which allows us to express $\tilde T^{ij}$ up to first order in $u^2$,
\begin{equation}
    \tilde T^{ij} = \frac{1}{1 + \cs^2} \frac{\tilde T^{0i} \tilde T^{0j}}{\tilde T^{00}} \biggl(1 + \frac{r^2 \cs^2}{(1 + \cs^2)^2}
    \biggr) + \cs^2 \tilde T^{00} \biggl(1 - \frac{r^2}{1 + \cs^2}
    \biggr)\, \delta^{ij} + {\cal O} (u^4)\,.
\end{equation}
This relation then
takes into account the corresponding corrections in $\tilde T^{0\mu}$ up to first order in $u^2$
and leads to the correct subrelativistic limit, as we show in the following.
Therefore, this is an appropriate way to close the system of 
subrelativistic equations in their conservation form.

To explicitly show the subrelativistic corrections due to the inclusion
of $r^2$ in the time derivatives, let us take the conformal time derivatives of \Eq{T0mu_r2}
\begin{subequations}
\begin{align}
    \partial_\tau \tilde T^{00} =  \, &\, \partial_\tau \tilde \rho + 
    \frac{\tilde \rho}{1 + \cs^2} \partial_\tau \bigl[r^2 + {\cal O} (r^4)\bigr]
    \,, \label{partialt_T00} \\
    \partial_\tau \tilde T^{0i} = \, &\, (1 + \cs^2) \, \partial_\tau (\tilde \rho u^i) + \frac{\tilde \rho u^i }{1 + \cs^2}
    \partial_\tau \bigl[r^2 + {\cal O} (r^4)\bigr]\,. \label{partialt_T0mu}
\end{align}
\label{partialt_T0muu}
\end{subequations}
Note that the next-to-leading-order term is proportional to
$\partial_\tau r^2$, which can be expressed as
\begin{equation}
    \partial_\tau r^2 = (1 + \cs^2)^2\, \partial_\tau \gamma^2 = (1 + \cs^2)^2 \, \gamma^2  \, \partial_\tau u^2 = (1 + \cs^2)^2\, \partial_\tau u^2 + {\cal O}(u^4) \,.
\end{equation}
Therefore, this term is proportional to $\partial_\tau u^2$.
To estimate $\partial_\tau u^2$, let us first expand the momentum equation,
setting $\mu = i$ in \Eq{mom_aux},
as we will do for the full
relativistic system in \Sec{rel_hydrodynamics}, and using \Eq{partialt_T0mu} to find
\begin{equation}
    u^i \, \partial_\tau \tilde \rho + \tilde \rho \, \partial_\tau u^i
    + \frac{\tilde \rho u^i}{(1 + \cs^2)^2} \partial_\tau r^2 + \, \partial_j \bigl( \tilde \rho u^i u^j  \bigr) + \frac{\cs^2}{1 + \cs^2}
    \partial_i \tilde  \rho
    =  \frac{\tilde f_H^i}{1 + \cs^2} + {\cal O} (u^2)\,.
\end{equation}
Then,
taking the product of $u^i$ with the momentum equation,
and keeping only leading-order terms, we find
\begin{equation}
    \half \tilde \rho \, (1 + \cs^2) \, \partial_\tau u^2 = - \cs^2 \, \uu \cdot \nab \tilde \rho + \uu \cdot \tff_H + {\cal O} (u^2)\,,
    \label{dt_u2}
\end{equation}
where the leading-order terms in the
right-hand side correspond to the work done by the
pressure gradient, $\uu \cdot \nab \tilde p$,
and the Hubble friction, $\uu \cdot \tff_H$.
As expected for perfect fluids,
the work done by these forces is a reversible process and,
hence, it does not produce entropy \cite{2008Kundu}.
Plugging \Eq{dt_u2} back into \Eqq{partialt_T0muu},
the correct subrelativistic limit of
$\partial_\tau \tilde T^{0\mu}$ can be found,
taking into account the contribution from $\partial_\tau u^2$,
\begin{subequations}
\begin{align}
   \lim_{u^2 \ll 1} \partial_\tau \tilde T^{00} = \, &\, \partial_\tau \tilde \rho -
    2 \, \cs^2 \, \uu \cdot \nab \tilde \rho + 2\, \uu \cdot \tff_H\,,
    \\
   \lim_{u^2 \ll 1} \partial_\tau \tilde T^{0i} = \,  &\,
     (1 + \cs^2) \, \partial_\tau (\tilde \rho
    u^i) - 2\, u^i \, \cs^2 \, \uu \cdot \nab \tilde \rho + 2 \, u^i \,
    \uu \cdot \tff_H\,. \label{T0i_dt}
\end{align}
\label{T0mu_dt}
\end{subequations}

\EEqs{dt_u2}{T0mu_dt} show that neglecting the next-to-leading 
order term $r^2$ in the time derivatives of the stress-energy tensor components in 
\Eq{T0mu_r2} is only justified in the subrelativistic limit when $\cs^2 \ll 1$ but not
for a generic $\cs^2 \sim {\cal O} (1)$.
Substituting this term back in the energy and momentum
conservation equations,
we find
\begin{subequations}
\label{eqs_subrelativistic_1}
\begin{align}
    \lim_{u^2 \ll 1}  \partial_\tau \ln \tilde \rho \, =  &\, - (1 + \cs^2)\,
    \nab \cdot \uu  - \red{(1 - \cs^2)}\, (\uu \cdot \nab) \ln \tilde \rho
    +
    \frac{1}{\tilde \rho} (\tilde f_H^0 - 2\, \uu \cdot \tff_H)\,,
    \label{cont_subrelativistic_1}
    \\ 
    \lim_{u^2 \ll 1}  D_\tau \uu = &\, \uu \,\cs^2 \biggl[\nab \cdot \uu +
    \red{\frac{1 - \cs^2}{1 + \cs^2}} \, (\uu \cdot \nab) \ln \tilde \rho\biggr] -
    \frac{\uu}{\tilde \rho} \biggl(
    \tilde f_H^0 - \frac{2 \cs^2}{1 + \cs^2} \, \uu \cdot \tff_H
    \biggr) \nonumber \\  &\, - \frac{\cs^2}{1 + \cs^2} \nab \ln \tilde \rho +
     \frac{\tff_H}{(1 + \cs^2) \tilde \rho}\,,
     \label{mom_subrelativistic_1}
\end{align}
\end{subequations}
with the corrections found indicated in red.
This set of equations is equivalent to the one presented
in the introduction; cf.~\Eqq{summary_nonrel}, in the absence of external forces
(only the Hubble friction has been included so far).
They generalize the set of equations used in previous work to
values of the speed of sound different than $\cs^2 = \third$ but still relativistic,
i.e., $\cs^2 \sim {\cal O} (1)$,
introducing terms that break conformal invariance, i.e., Hubble friction
$\tilde f^\mu_H$.
For all values $\cs^2 \neq 0$,
these expressions include
corrections to the energy
and momentum equations with respect to previous work in the 
subrelativistic limit (cf.~\Eqq{eqs_new} for a radiation-dominated fluid).

Using an appropriate subrelativistic limit of the stress-energy
tensor components, the
subrelativistic Euler equations in their non-conservation form, 
explicitly describing the
dynamics of the comoving energy density $\tilde \rho$
and the velocity field $\uu$,
have been found.
In the next section, we compute the relativistic version of the
conservation laws in the non-conservation form.

\subsection{Non-conservation form of relativistic fluid dynamics}
\label{rel_hydrodynamics}

The non-conservation form of the fluid equations
is obtained expressing explicitly
the stress-energy tensor components in terms of the primitive
fluid variables, $\tilde \rho$ and $\uu$.
In the following, the extension of the system of
equations given in \Eqq{eqs_subrelativistic_1}
to the fully relativistic regime is done,
up to our knowledge,
for the first time for a perfect fluid in an expanding background.
For compactness, we set $\alpha = 1$ and $\delta = 0$ in the following, but
the equations can be generalized to any rescaling choice.

\subsubsection*{3.5.1 \ Relativistic energy equation}

Let us start with the equation of energy conservation taking $\mu = 0$
in \Eq{mom_aux}, and dividing the equation
by $(\tilde p + \tilde \rho) \gamma^2$,
\begin{equation}
    D_{\tau} \ln (\tilde p + \tilde \rho) + D_\tau \ln \gamma^2 + \nab \cdot \uu =
    \frac{\partial_{\tau} \tilde p + \tilde f_H^0}{(\tilde p + \tilde \rho) \gamma^2}\,,
    \label{cont1}
\end{equation}
where $D_\tau = \partial_\tau + u_i \partial^i$ is the material
derivative.
In terms of a constant
speed of sound $\tilde p = \cs^2 \tilde \rho$, the energy equation becomes
\begin{equation}
    D_\tau \ln \tilde \rho + D_\tau \ln \gamma^2 + \nab \cdot \uu = \frac{\cs^2}{(1 + \cs^2)\gamma^2}
    \partial_\tau 
    \ln \tilde \rho + \frac{\tilde f_H^0}{(1 + \cs^2)\tilde \rho \gamma^2}\,. \label{cont0}
\end{equation}
As discussed,
previous work considered this expression directly taking the
subrelativistic limit $u^2 \ll 1$ for $\cs^2 = \third$
and $\beta = 4$,
such that $\tilde f_H^0 = 0$.
However, as shown in the previous section, $\partial_\tau \gamma^2 = \gamma^4 \partial_\tau u^2 = \gamma^2 \partial_\tau r^2/(1 + \cs^2)$ cannot be
neglected in the subrelativistic regime.
In the following, we will keep all terms to find the fully relativistic
conservation laws of perfect fluids.

An alternative (but, of course, equivalent) version of the energy conservation
equation in \Eq{cont0} can be found projecting the conservation laws in \Eq{cons_expand} with the
comoving four-velocity $\tilde U^\mu = a U^\mu = \gamma(1, \uu)$,
\begin{equation}
    \tilde U_\nu \bigl(\partial_\mu \tilde T^\munu - \tilde f_H^\nu \bigr) = 
    \tilde U_\nu (\tilde p + \tilde \rho) \tilde U^\mu \partial_\mu \tilde U^\nu +
    \tilde U_\nu \tilde U^\nu \partial_\mu [(\tilde p + \tilde \rho) \tilde U^\mu]
    + \tilde U^\mu (\partial_\mu \tilde p - \tilde f^H_\mu) \,.
\end{equation}
Taking into account the normalization condition of $\tilde U^\mu$, i.e., $\tilde U^\mu \tilde U_\mu = -1$,
it follows that $\tilde U_\nu \partial_\mu \tilde U^\nu = 0$.
Note that here
we use the Minkowski metric to lower/rise indices for comoving tensors, such that
$\tilde U_\nu = \eta_\munu \tilde U^\mu$, since we can study the equations
in Minkowski space-time after a conformal transformation.
Hence,
\begin{align} 
    \tilde U_\nu \bigl(\partial_\mu \tilde T^\munu - \tilde f_H^\nu \bigr)
    = &\, - \partial_\mu[(\tilde p + \tilde \rho) \tilde U^\mu]
    + \tilde U^\mu (\partial_\mu \tilde p - \tilde f_\mu^H) = 0 \nonumber \\
    \Rightarrow &\ \tilde U^\mu \partial_\mu 
    \tilde \rho + (\tilde p + \tilde \rho)\,
    \partial_\mu \tilde U^\mu
    + \tilde U^\mu \tilde f_\mu^H = 0
    \,. \label{rel_cont_cov}
\end{align}
As a side note, notice that promoting derivatives
into covariant derivatives, the following continuity equation
is valid for any metric tensor \cite{Rezzolla:2013dea},
\begin{equation}
    -U_\nu T^\munu\covderT{\mu} = U^\mu \partial_\mu \rho + (p + \rho)\, \theta = 0\,,
\end{equation}
where we have defined
the relativistic fluid expansion scalar $\theta = U^\mu\covderV{\mu}$.

Going back to the FLRW metric tensor,
taking $\tilde p = \cs^2 \tilde \rho$ with constant $\cs^2$,
one finds an alternative version of the energy conservation
equation
\begin{equation}
    D_\tau \ln \tilde \rho + (1 + \cs^2) D_\tau \ln \gamma + (1 + \cs^2) \nab \cdot \uu = 
    \frac{1}{\tilde \rho} (\tilde f_H^0 - \uu \cdot \tff_H)\,.
    \label{energy_alternative}
\end{equation}
In order to get rid of the $D_\tau \ln \gamma^2 = 2 D_\tau \ln \gamma$ term,
we can directly combine \Eqs{cont0}{energy_alternative}, since both equations need to be
satisfied.
Actually, we can observe that taking $D_\tau \ln \gamma \to 0$ in the
subrelativistic limit leads to a different result in \Eqs{cont0}{energy_alternative},
already proving that this assumption cannot be correct.
Some cumbersome but direct algebra brings us the relativistic conservation equation,
\begin{equation}
    \partial_\tau \ln \tilde \rho + \frac{1 + \cs^2}{1 - \cs^2 u^2} \nab \cdot \uu
    + \frac{1 - \cs^2}{1 - \cs^2 u^2} (\uu \cdot \nab) \ln \tilde \rho = {\cal F}_H^0\,,
    \label{cont_relativistic0}
\end{equation}
where the Hubble friction in the energy equation is
\begin{equation}
    {\cal F}_H^0 = \frac{1}{1 - \cs^2 u^2}
    \frac{1}{\tilde \rho} \bigl[\tilde f_H^0 (1 + u^2) - 2 \uu \cdot \tff_H\bigr] =
    \Bigl[(\beta - 4) + \frac{1 + u^2}{1 - \cs^2 u^2} (1 - 3 \cs^2) \Bigr] \, \HH\,. \label{FFH0}
\end{equation}
This Hubble term depends on $\beta$ and only for $\beta = 4$ (which yields $\tff_H = 0$)
and $\cs^2 = \third$ it
vanishes in the fully
relativistic case.
However, we
keep $\beta$ in this term to allow for a general choice of the scaling of $\tilde T^0_{\ \, 0}$ using
\Eq{Hubble_force}.
Note that the first equality in this expression can be used to add any external
forces to the relativistic energy equation, e.g., imperfect
 and electromagnetic forces; see \Secs{viscosity}{four_lorentz},
respectively.

Taking the $u^2 \ll 1$ limit in \Eq{cont_relativistic0}
shows the correction in the subrelativistic energy equation
[cf. \Eq{continuity_new}] that has been
omitted in previous work,
\begin{equation}
    \lim_{u^2\ll 1} \partial_\tau \ln \tilde \rho = -(1 + \cs^2) \nab \cdot \uu -
    \red{(1 - \cs^2)} (\uu \cdot \nab)
    \ln \tilde \rho + \bigl[\beta - 3 \, (1 + \cs^2)\bigr]\,\HH \,.
\end{equation}
Furthermore, as discussed in \Sec{generic_scaling}, contrary to the fully relativistic case,
it is now possible to choose $\beta = 3 \, (1 + \cs^2)$ to get rid of
$\tilde f_H^0$.

\subsubsection*{3.5.2 \ Relativistic momentum equation}

Let us now proceed to compute the momentum equation, given by the
spatial components of \Eq{mom_aux}.
After dividing the equation with $(\tilde p + \tilde \rho) \gamma^2$,
one finds
\begin{equation}
    D_{\tau} \uu + \uu D_{\tau} \ln \bigl[(\tilde p + \tilde \rho) \gamma^2 \bigr] + \uu\, (\nab \cdot \uu) = -
    \frac{\nab \tilde p - \tff_H}{(\tilde p + \tilde \rho) \gamma^2} \,.
    \label{mom_eq0}
\end{equation}
For an equation of state with a constant
speed of sound, $\tilde p = \cs^2 \tilde \rho $,
\begin{equation}
    D_{\tau} \uu  + \uu \bigl(D_{\tau}
    \ln \tilde \rho \,
    + D_{\tau} \ln \gamma^2 +
    \nab \cdot \uu \bigr) = -  \frac{\cs^2}{1 + \cs^2} \frac{\nab \ln \tilde \rho}{\gamma^2}  + \frac{1}{1 + \cs^2}\frac{\tff_H}{\tilde \rho \gamma^2} \,.
    \label{mom_1}
\end{equation}
Using the energy conservation equation 
(\ref{cont0}),
Euler equation becomes
\begin{equation}
    D_{\tau} \uu + \frac{\uu}{(1 + \cs^2)\gamma^2} \biggl( \cs^2  \, \partial_\tau \ln \tilde \rho
     + \frac{\tilde f_H^0}{\tilde \rho} \biggr) = -  \frac{\cs^2}{1 + \cs^2} \frac{\nab \ln \tilde \rho}{\gamma^2}  + \frac{1}{1 + \cs^2}\frac{\tff_H}{\tilde \rho \gamma^2} \,. \label{mom_2}
\end{equation}
Again, we note that previous work has considered the
subrelativistic limit of this equation setting $\gamma^2 = 1$,
and used $\beta = 4$,
such that $\tilde f_H^i = 0$.
Similarly as with the energy equation, the subrelativistic term
contained in $\partial_\tau \gamma^2$ that modifies the energy
equation also enters the momentum equation.

The relativistic Euler equation is then obtained using the energy equation to substitute
$\partial_\tau \ln \tilde \rho$ [cf.~\Eq{cont_relativistic0}] into \Eq{mom_2},
\begin{align}
    D_\tau \uu =&\, \frac{\uu \, \cs^2}{(1 - \cs^2 u^2) \gamma^2} \biggl[ \nab \cdot \uu + 
    \frac{1 - \cs^2}{1 + \cs^2} (\uu \cdot \nab) \ln \tilde \rho   \biggr]
    - \frac{\cs^2}{1 + \cs^2} \frac{\nab \ln \tilde \rho}{\gamma^2}
    +\pmb{\cal F}_H\,,
    \label{mom_relativistic0}
\end{align}
where the Hubble friction in the momentum equation is
\begin{equation}
    \pmb{\cal F}_H = - \frac{\uu}{1 - \cs^2 u^2} \frac{1}{\tilde \rho \gamma^2}
    \biggl(\tilde f_H^0 - \frac{2 \cs^2}{1 + \cs^2} \uu \cdot \tff_H\biggr) + \frac{1}{1 + \cs^2}
    \frac{\tff_H}{\tilde \rho \gamma^2} = \frac{3 \cs^2 - 1}{1 - \cs^2 u^2} \frac{\uu \, \HH}{\gamma^2}\,.
    \label{Hubble_mom}
\end{equation}
We note that the Hubble friction in the momentum equation becomes
independent of the value of $\beta$ and vanishes when $\cs^2 = \third$.
The subrelativistic limit of the momentum equation is found taking $u^2 \ll 1$ in \Eqs{mom_relativistic0}{Hubble_mom},
\begin{equation}
    \lim_{u^2 \ll 1} D_\tau \uu = \uu \, \cs^2 \biggl[
    \nab \cdot \uu + \red{\frac{1 - \cs^2}{1 + \cs^2}}
    (\uu \cdot \nab) \ln \tilde \rho  \biggr] - \frac{\cs^2}{1 + \cs^2}
    \nab \ln \tilde \rho
    + (3 \cs^2 - 1) \, \uu \, \HH \,,
\end{equation}
where the correction with respect to previous work is indicated in red.

The system of \Eqs{cont_relativistic0}{mom_relativistic0} is found to be remarkably simplified after the
procedure described in this section, as one finds that their modification
with respect to their subrelativistic counterpart is restricted to the
prefactors $(1 - \cs^2 u^2)^{-1}$ and $(1 - u^2)(1 - \cs^2 u^2)^{-1}$ in the
energy and momentum equations, respectively, as well as an additional $\gamma^{-2} = 1 - u^2$
in the forces of the momentum equation (e.g., pressure
gradient and Hubble friction).
Taking the subrelativistic limit of the energy and momentum equations, they
are found to be 
those given in \Eqq{eqs_subrelativistic_1},
computed in the previous section directly from taking the
subrelativistic limit of the conservation form of the fluid
equations.
This again confirms the necessity to keep terms of order $r^2$ in the conservation
formalism to recover the correct subrelativistic limit.

Alternatively, we can find an equivalent momentum equation contracting the conservation
laws with the projection tensor $\tilde h_{\nu i} = \tilde U_\nu \tilde U_i + \eta_{\nu i}$.
In the first place, a direct evaluation shows
\begin{equation}
    \tilde h_{\nu}^{\ \,i} \bigl(\partial_\mu \tilde T^\munu - \tilde f_H^\nu \bigr) =
    \tilde U^i \tilde U_\nu \bigl(\partial_\mu \tilde T^\munu - \tilde f_H^\nu \bigr) +
    \bigl(\partial_\mu \tilde T^{\mu i} - \tilde f_H^i \bigr) = 0\,.
\end{equation}
This is equivalent to subtracting the energy equation [cf.~\Eq{energy_alternative}]
multiplied by $\uu$ [note the negative sign in \Eq{rel_cont_cov}]
and add the momentum equation [cf.~\Eq{mom_1}] multiplied by $(1 + \cs^2)\gamma$.
This procedure provides a simplified momentum equation.
To directly show this, we can use the properties of the projection tensor, $\tilde U^\mu \tilde h_{\munu} = 0$
and $\tilde h^\munu \tilde h_{\nu \lambda} = \tilde h^\mu_{\ \lambda}$, and
express $\tilde T^\munu = \tilde \rho \tilde U^\mu \tilde U^\nu + \tilde p \tilde h^\munu$.
Hence, applying these properties and taking into account that $\tilde U_\mu \partial_\nu \tilde U^\mu = 0$,
we find
\begin{equation}
    \tilde h_{\nu i} (\partial_\mu \tilde T^\munu - \tilde f^\nu_H)
    = (\tilde p + \tilde \rho) \tilde U^\mu \partial_\mu \tilde U_i + \tilde h^\mu_{\ \, i}
    \partial_\mu \tilde p - \tilde U^\mu \tilde U_i \tilde f^H_\mu - \tilde f^H_i = 0\,.
\end{equation}
As done above, this equation can be generalized for any metric tensor \cite{Romatschke:2009im,Rezzolla:2013dea}
\begin{equation}
    (p + \rho) a^\mu + \nabla^\mu p = 0\,,
\end{equation}
where $a^\mu = U^\nu U^\mu\covderV{\nu}$ is the acceleration four-vector and $\nabla^\mu = h^{\mu \nu} \partial_\nu$.

\subsubsection*{3.5.3 \ Evolution of the Lorentz factor}

We proceed to explicitly compute an evolution equation for $u^2$ or, equivalently,
for the Lorentz factor, as this equation is in general enlightening for studying the kinetic energy,
$\tilde \rho_\kin = (\tilde p + \tilde \rho) \gamma^2 u^2$,
and to explicitly show its correction in the subrelativistic limit.
The relativistic evolution equation of the Lorentz
factor can be computed from the momentum equation [cf.~\Eq{mom_relativistic0}],
by contracting it with the velocity field $\uu$,
as done in the subrelativistic limit in \Eq{dt_u2},
\begin{equation}
    D_\tau \ln \gamma^2 = 2 \gamma^2 \, \uu \cdot (D_{\tau}
    \uu) =  \frac{2 \cs^2}{1 - \cs^2 u^2} \biggl[u^2 \nab \cdot \uu - \frac{1}{(1 + \cs^2)\gamma^2}
    (\uu \cdot \nab) \ln \tilde \rho \biggr]
    + 2 \gamma^2 \uu \cdot \pmb {\cal F}_H\,,
    \label{lorentz_evol}
\end{equation}
where the term due to the Hubble friction is
\begin{equation}
    2 \gamma^2 \uu \cdot \pmb {\cal F}_H = - \frac{2 u^2}{1 - \cs^2 u^2}
    \frac{\tilde f_H^0}{\tilde \rho} +
    \frac{2}{1 + \cs^2} \frac{1 + \cs^2 u^2}{1 - \cs^2 u^2}
    \frac{\uu \cdot \tff_H}{\tilde \rho}
    = 2 \,
    \frac{3 \cs^2 - 1}{1 - \cs^2 u^2} \, u^2  \HH\,.
\end{equation}
Taking the subrelativistic limit of \Eq{lorentz_evol} we find
\begin{equation}
    \lim_{u^2 \ll 1} D_\tau \ln \gamma^2 = - \frac{2 \cs^2}{1 + \cs^2} (\uu \cdot \nab) \ln \tilde \rho 
    +  \frac{2}{1 + \cs^2}\frac{\uu \cdot \tff_H}{\tilde \rho}\,,
    \label{deta_gamma2_subrel}
\end{equation}
explicitly showing that this term cannot be neglected in the subrelativistic limit.

As already discussed, comparing
\Eqq{eqs_subrelativistic_1}
to the equations used in previous work for a radiation-dominated
fluid with $\cs^2 = \third$ and $\beta = 4$, such that $\tilde f^\mu_H = 0$
[cf.~\Eqq{eqs_old}], we find an additional $\half$ factor multiplying one of the terms in both the energy and the momentum
equations, as indicated explicitly in red in \Eqq{eqs_new}.
This is due to the fact that even if $\gamma^2 \to 1$, $D_\tau \ln \gamma^2$ contains subrelativistic
corrections, as indicated in \Eq{deta_gamma2_subrel}.
Only when $\cs^2 \ll 1$ (i.e., matter-dominated fluid) and $\beta = 4$ ($\tilde f_H^i = 0$), this term can be
neglected in the subrelativistic limit and taking $\gamma^2 \to 1$
from the beginning is justified.
This term leads to the following correction in \Eqq{eqs_old} with respect to the equations found
when $\gamma^2 \to 1$ and $\partial_\tau \gamma^2 \to 0$ in previous work,
\begin{equation}
   (\uu \cdot \nab) \ln \tilde \rho \to 
    \frac{1 - \cs^2}{1 + \cs^2} \, (\uu \cdot \nab) \ln \tilde \rho\,,
\end{equation}
in both the energy and momentum equations.

\subsection{Conservation laws for specific values of $\cs^2$}
\label{values_cs2_hydro}

Let us now consider two special cases: fluids dominated by radiation
($\cs^2 = \third$) and matter ($\cs^2 \ll 1$).

\subsubsection*{3.6.1 \ Equations for radiation-dominated fluids}

For $\cs^2 = \third$, the relativistic energy and 
momentum equations are
\begin{subequations}
\begin{align}
    \partial_{\tau} \ln \tilde \rho = &  - \frac{4}{3 - u^2} \bigl[  \nab \cdot \uu + \half  \bigl(\uu \cdot \nab \bigr) \ln \tilde \rho \bigr] + (\beta - 4) \, \HH \,, \\
    D_{\tau} \uu = &\, \frac{1 - u^2}{3- u^2} \uu 
    \bigl[\nab \cdot \uu + \half \bigl(\uu \cdot \nab \bigr)\ln \tilde \rho \bigr] - \frac{1 - u^2}{4}
    \nab \ln \tilde \rho\,,
\end{align}
\end{subequations}
where the momentum equation is conformally flat for any choice of $\beta$, while
the Hubble term in the energy equation vanishes when $\beta = 4$.
These conservation laws reduce to those in \Eqq{eqs_new}
in the subrelativistic limit.
In the ultrarelativistic limit $u^2 \to 1$, the equations become
\begin{equation}
    \lim_{u^2 \to 1} D_\tau \ln \tilde \rho = -2 \, \nab \cdot \uu \,, \qquad
    \lim_{u^2 \to 1} D_\tau \uu = 0\,.
\end{equation}

\subsubsection*{3.6.2 \ Equations for matter-dominated fluids}

For $\cs^2 \ll 1$, the fluid equations become those for usual fluid dynamics with an additional Hubble friction term that takes into account the Universe
expansion,
\begin{equation}
    D_{\tau} 
    \ln \tilde \rho +  \nab \cdot \uu =  (\beta  -3 + u^2) \, 
    \HH \,, \qquad
    D_{\tau} \uu = - (1 - u^2) \biggl(\uu \, \HH  + \frac{\nab \tilde p}{\tilde \rho} \biggr) \,,
\end{equation}
where the Hubble friction that appears in
the momentum equation is independent of the choice of
$\beta$.
In the subrelativistic limit,
the equations reduce to
\begin{equation}
    \lim_{u^2 \ll 1} D_\tau \ln \tilde \rho = -  \nab \cdot \uu +
    (\beta - 3) \, \HH \,,
    \qquad  \lim_{u^2 \ll 1} D_\tau \uu  = - \uu \,
    \HH - \frac{\nab \tilde p}
    {\tilde \rho}\,.
\end{equation}
Then, for subrelativistic flows with $\cs^2 \ll 1$, it is clear that it is more
convenient to choose $\beta = 3$ in the scaling, as it allows to get rid of the
Hubble term in the energy equation.
As discussed in \Sec{cons_perf_fluid}, the choice of super-comoving
variables $\tilde \uu = a \uu$
($\delta = 1$), $\alpha = 2$, and $\tilde p = a^5 p$ ($\chi = 5$), together
with $\beta = 3$, allows us to find conformally-flat equations,
\begin{equation}
    \lim_{u^2 \ll 1} D_{\tau_\alpha} \ln \tilde \rho = - \nab \cdot \uu\,,
    \qquad \lim_{u^2 \ll 1} D_{\tau_\alpha} \tilde \uu = - \frac{\nab \tilde p}{\tilde \rho}\,.
\end{equation}

\subsection{Sound waves}
\label{sound_waves}

In this section, linear perturbations of the perfect
fluid equations of motion are studied.
For this purpose, we consider that the background energy density
is homogeneous and
evolves with the scale factor as $\bar \rho \sim a^{-3 \,(1 + w)}$
(see \Sec{Friedmann}) for a background equation of state $\bar p = w \bar \rho$.
Let us consider perturbations $\rho_1$ and $\uu$ over the background
at rest.
The energy and momentum equations up to first order in perturbations are [cf.~\Eqq{eqs_subrelativistic_1}],
\begin{subequations}
\begin{align}
    \partial_\tau \tilde \rho_1 = &
    - (1 + \cs^2) \, \tilde \rho_0 \, \nab \cdot \uu
    +  \bigl[\beta - 3 \, (1 + \cs^2)\bigr] \, \tilde \rho_0\, \HH\,, \\ \partial_\tau \uu = &
    - \frac{\cs^2}{1 + \cs^2} \frac{\nab \tilde \rho_1}{\tilde \rho_0} +
    ( 3\cs^2 - 1) \, \uu \, \HH\,,
\end{align}
\end{subequations}
where $\tilde \rho_0 = a^{3\,(1 + w)} \, \bar \rho$ corresponds
to the comoving energy density of the background, constant in time.
In first place, we note that the correction described in previous
sections, which applies to the term $\uu \cdot \nab \ln \tilde \rho$,
does not affect the linearized equations but it becomes relevant
in the non-linear regime.
Secondly, it is convenient to choose
$\beta = 3\, (1 + \cs^2)$ to get rid of the Hubble
friction in the energy equation,
such that $\tilde \rho_1 = a^\beta \rho_1$.
We note that $\cs^2$ and $w$ are not, in general, required to be equal,
allowing the total energy of the Universe to contain
additional contributions besides the one from the fluid.
In terms of the normalized energy density perturbations $\lambda$, defined such
that
\begin{equation}
    \lambda = \frac{1}{1 + \cs^2}\frac{\tilde \rho_1}{\tilde \rho_0} = \frac{1}{1 + \cs^2}
    \frac{\rho_1}{\bar \rho} \, a^{3 \, (\cs^2 - w)} \,,
\end{equation}
 where $(1 + \cs^2)\, \tilde \rho_0$ is the comoving background
enthalpy, the energy and momentum equations are
\begin{equation}
    \partial_\tau \lambda = - \nab \cdot \uu\,, \qquad
    \label{soundwaves_pert}
    \partial_\tau \uu =  (3 \cs^2 - 1) \, \uu \, \HH - \cs^2 \, \nab \lambda \,.
\end{equation}
We note that for any other choice of $\beta$, the energy
equation would present an additional source term
\begin{equation}
    \partial_\tau \lambda = - \nab \cdot \uu + \frac{\beta - 3\, (1 + \cs^2)}{1 + \cs^2}\, \HH\,.
\end{equation}

One directly finds from the momentum equation
that fluid perturbations are aligned with the
gradients of the energy density perturbations (corresponding to sound waves).
Equivalently, in Fourier space,
$\uu = u_\parallel \hat \kk$.
On the other hand,
perpendicular fluid perturbations follow an exponential
decay (for $\cs^2 < \third$) or growth (for a stiff equation of state $\cs^2 > \third$)  due to the Hubble friction,
\begin{equation}
    u_\perp = u_\perp (\tau_0) \exp \Bigl[(3 \cs^2 - 1) \int_{\tau_0}^\tau \HH(\tau') \dd \tau'\Bigr]\,.
\end{equation}
Combining both energy and momentum
equations, we find damped wave equations for $\lambda$
and $u_\parallel$ (expressed in momentum space $\kk$),
\begin{subequations}
\begin{align}
    \partial^2_\tau \lambda \, -  &\, (3 \cs^2 - 1) \, \HH \,
    \partial_\tau \lambda + \cs^2 k^2 \lambda =
    0 \,, \\
    \partial_\tau^2 u_\parallel \, -  &\, (3 \cs^2 - 1) \, \HH \,
    \partial_\tau u_\parallel  + \bigl[\cs^2 k^2 - \HH'(3 \cs^2 - 1)\bigr] u_\parallel = 0\,.
\end{align}
\end{subequations}
When $\cs^2 = \third$, the system reduces to free-propagating sound
waves with no Hubble friction, whose dispersion relation is $\omega^2 = \cs^2 k^2$.
When $\cs^2 \neq \third$,
the sound-wave perturbations present a
time-dependent Hubble friction $(3 \cs^2 - 1) \, \HH \, \partial_\tau {\cal U}^a$ with
${\cal U}^a = \{\lambda, u_\parallel\}$,
and the parallel velocity field has an additional modification to the
angular frequency $\omega^2 \equiv \cs^2 k^2 - \HH'\, (3\cs^2 - 1)$.
In general, for a background equation of state $\bar p = w \bar \rho$, when
$w > -\third$ we can express the Hubble rate and its
derivative in the following way
\begin{equation}
    \HH = \frac{a'}{a} = \frac{2}{1 + 3 w} \frac{1}{\tau} \,,
    \qquad \HH' = - \frac{2}{1 + 3w} \frac{1}{\tau^{2}}\,.
\end{equation}
Therefore, the homogeneous wave equations
become
\begin{subequations}
\begin{align}
    \lambda'' \, -  2 \sigma \lambda'/\tau + \omega^2 \lambda = 0\,, \qquad
    u_\parallel'' \, - 2 \sigma u_\parallel'/\tau + (\omega^2 + 2 \sigma/\tau^2) u_\parallel = 0\,,
\end{align}
\end{subequations}
where $\sigma = (3 \cs^2 - 1)/(1 + 3w)$.
Their analytical solution is
\begin{figure}[t]
    \centering
    \includegraphics[width=0.325\linewidth]{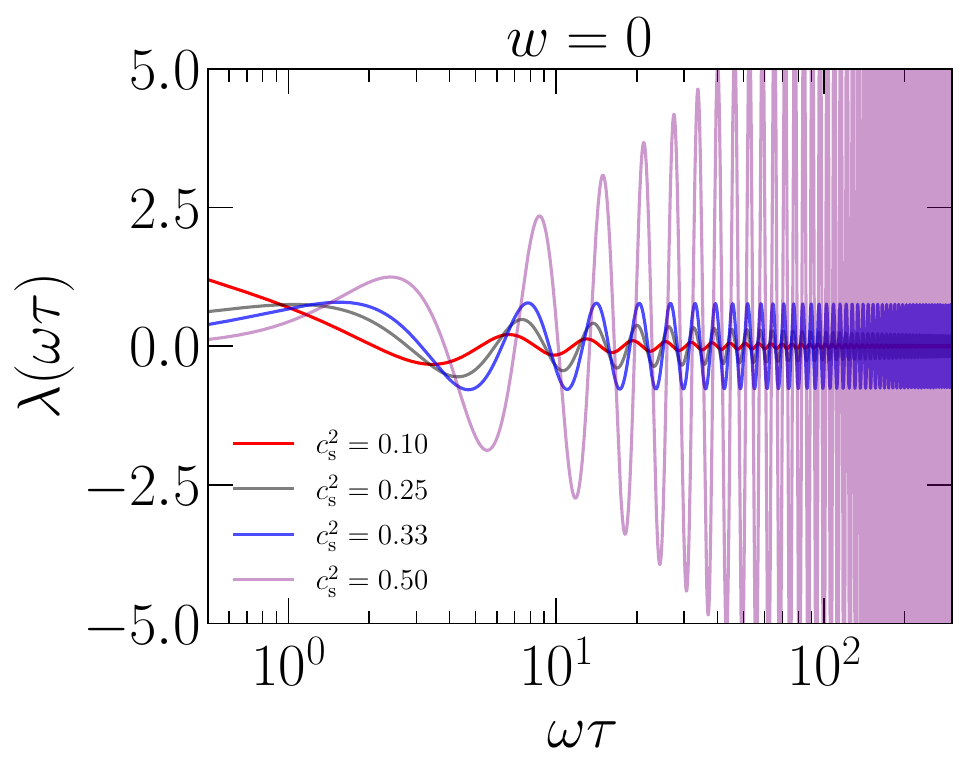}
    \includegraphics[width=0.325\linewidth]{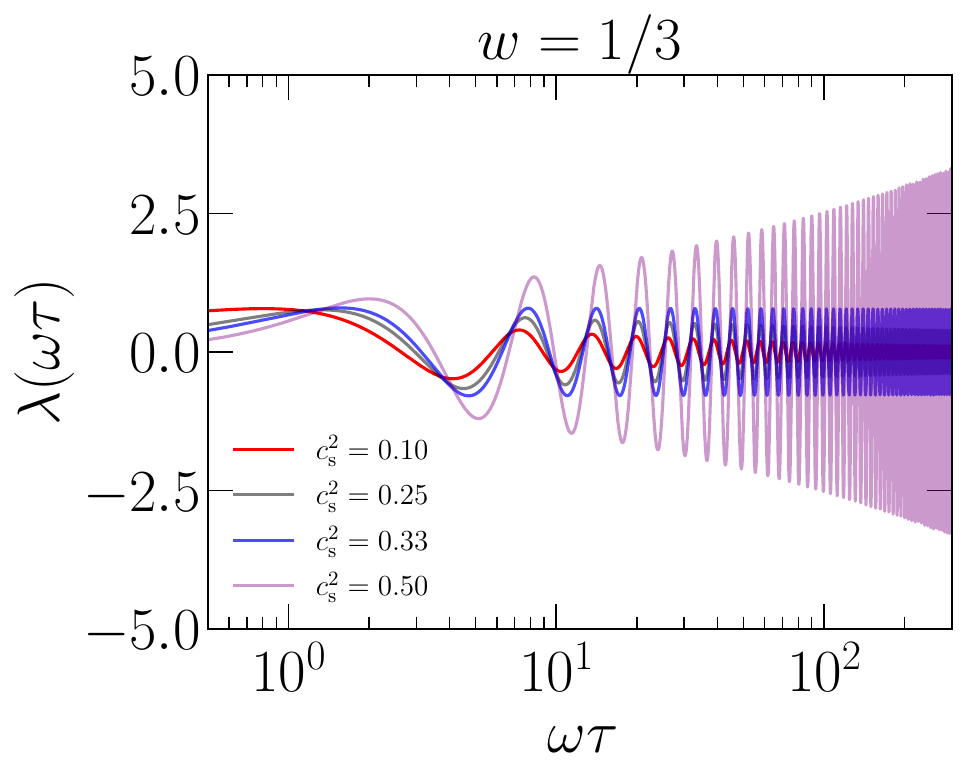}
    \includegraphics[width=0.325\linewidth]{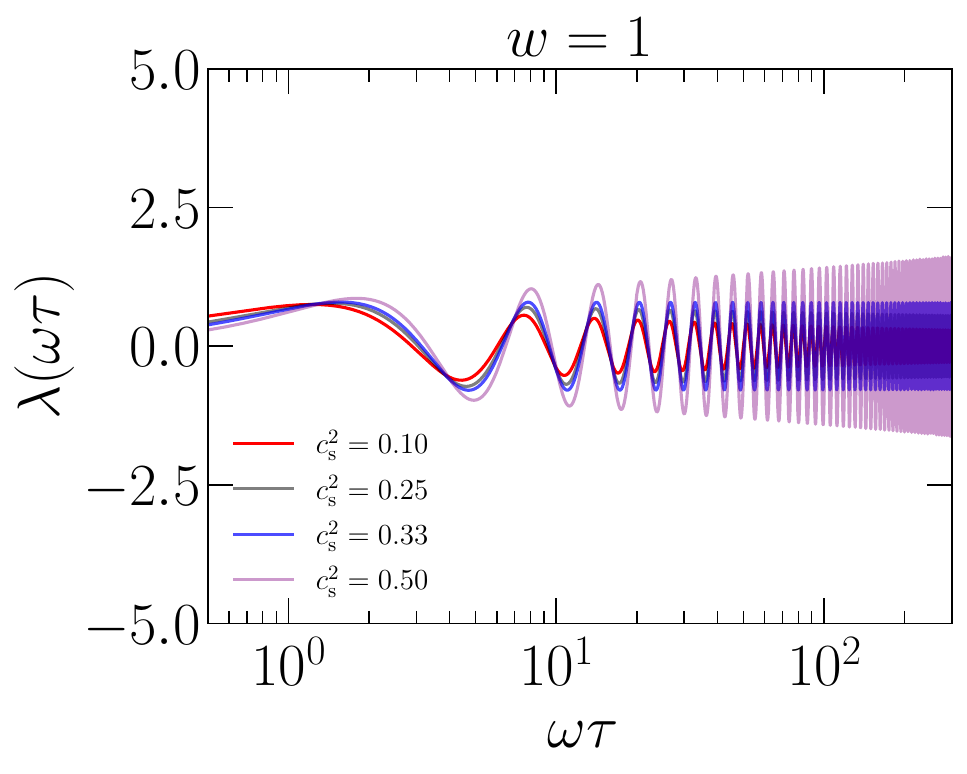}
    \caption{Evolution of the sound-wave energy density
    perturbations $\lambda$ for an arbitrary choice
    of the initial conditions, $c_2 = 0$.
    Free-propagating waves are found when $\cs^2 = \third$.}
    \label{sws_evol}
\end{figure}
\begin{subequations}
\begin{align}
    \lambda (\omega \tau) = & \, c_1 \,
    (\omega \tau)^{\sigma + 1} j_\sigma (\omega \tau)
    + c_2 \,
    (\omega \tau)^{\sigma + 1} y_\sigma (\omega \tau)
    \,, \label{sol_sws} \\ 
    u_\parallel (\omega \tau) = &\, c_3 \,
    (\omega \tau)^{\sigma + 1} \, j_{\sigma - 1} (\omega \tau)
    + c_4 \,
    (\omega \tau)^{\sigma + 1} \, y_{\sigma - 1} (\omega \tau)\,,
\end{align}
\end{subequations}
where
$j_a$ and $y_a$ are the spherical
Bessel functions of the first and
the second kind, respectively.
These solutions have been recently considered in \cite{Giombi:2024kju,Giombi:2025tkv} to study the production of gravitational
waves from sound waves in an expanding background dominated
by the fluid with
$w = \cs^2 \neq \third$.
\FFig{sws_evol} shows the evolution of $\lambda$ for arbitrary initial conditions such that $c_2 = 0$
for different values of $\cs^2$ and $w$.

\section{Relativistic imperfect fluid dynamics}
\label{viscosity}

The zero-th order approximation in fluid dynamics has been
used in \Sec{perfect_fluid_hydro} to characterize perfect fluids.
This approximation is valid under the
assumption that the distribution function $f$ of the
fluid particles is the one found in local thermal equilibrium (LTE), 
i.e., the Maxwell-Boltzmann distribution in the
subrelativistic limit.
In the relativistic regime, one finds that the LTE
distribution can be Maxwell-J\"uttner, Bose-Einstein,
or Fermi-Dirac; see  \cite{Cercignani2002RelativisticBoltzmann,Rezzolla:2013dea}
for details.

The perfect fluid description holds when the collisions in the
system are frequent, such that they drive the fluid parcels
to LTE
in a time scale $l_{\rm mfp}/v$ that is shorter than the time scales
characterizing the fluid, being $v$ the typical particle velocity and $l_{\rm mfp}$ the mean free path.
This corresponds to the limit of small Knudsen
number, Kn $= l_{\rm mfp}/L \ll 1$, i.e., small mean free
path of the fluid particles compared to a characteristic length
scale $L$ of the fluid fields.

In \Sec{NavierStokes},
the distribution function is expanded using the Knudsen number
as the perturbative parameter, following Chapman-Enskog theory
\cite{1970mtnu.book.....C}.
This allows to describe subrelativistic imperfect fluids in the first-order
fluid dynamics approximation, leading to Navier-Stokes
shear stress tensor
describing viscosity, and to heat fluxes, described
by Fourier's law of conductivity.
In \Sec{transport_coeffs}, we review the estimates of the viscous shear and bulk coefficients
and of the thermal conductivity in the early Universe.
In \Sec{conformal_NS}, Navier-Stokes viscosity
and Fourier's heat fluxes
are presented in a covariant relativistic formulation,
following the classical
irreversible thermodynamics (CIT) approach
\cite{PhysRev.58.919,Landau1987Fluid,Rezzolla:2013dea}.
A known problem of the CIT approach
is that it leads to
fluid perturbations that are allowed to propagate at unbounded
speeds, violating
the postulates of special relativity.
This is a consequence of the parabolic nature
of the diffusion operators that
describe Navier-Stokes
viscosity in the momentum equation, $\nu \,\nab^2 \uu$,
and Fourier's heat flux in the energy equation,
$\kappa \,\nab^2 T$,
being $\nu$ and $\kappa$ the coefficients of shear viscosity
and heat-flux conductivity, respectively.
As we show in \Sec{NavierStokes}, these terms appear due to
the inclusion of
the following stress-energy fluxes from
deviations with respect to a pefect fluid in LTE:
$\partial_i \tilde T^{ij} \supset - \tilde \nu \, \partial_i \partial^i u^j$
and $\partial_i \tilde T^{0i} \supset \partial_i \tilde q^i = - \tilde \kappa \, \partial_i \partial^i \tilde T$,
respectively, with the comoving
variables $\tilde \nu$, $\tilde \qq$, $\tilde \kappa$, and $\tilde T$ defined in \Sec{NavierStokes}.

A possible solution is to introduce relaxation times $\tau_r$
to the equations describing the deviations with respect to LTE, motivated by the results found in kinetic theory,
as proposed in \cite{Cattaneo:1948} for the heat flux (expressed in Minkowski space-time),
\begin{equation}
    \tau_r \, \partial_t \qq + \qq = - \kappa \nab T\,,
\end{equation}
where the inclusion of the $\tau_r \, \partial_t \qq$ term
makes the equation hyperbolic and, hence, it represents a causal
description, introducing a response time
to changes in the temperature.
Following the pioneer work of \cite{Israel:1976tn,Israel:1976efz}, Israel-Stewart theory
is described by dynamical equations for the transport coefficients (e.g., shear and bulk viscosity, and thermal conductivity), introduced together with
relaxation times, also known as the Maxwell-Cattaneo form of the relativistic viscous fluid dynamic equations \cite{Rezzolla:2013dea,Romatschke:2009im}.
For example, the deviatoric viscous tensor that is subtracted to
the stress-energy tensor of a perfect fluid, $T^{ij} \supset
T^{ij}_{\rm pf} - \Pi^{ij}_2$, is described using the
following dynamical equation (expressed in Minkowski space-time)
\begin{equation}
    \tau_\Pi \, \partial_t \Pi^{ij}_2 + \Pi^{ij}_2 = \Pi^{ij}_1\,,
\end{equation}
where $\Pi^{ij}_1$ corresponds to the deviatoric viscous
tensor found under the CIT approach (see \Secs{NavierStokes}{conformal_NS}).
These models are known as extended irreversible thermodynamics,
or second-order fluid dynamics.
They are an active topic
of research (see the review \cite{Romatschke:2009im}) and go beyond
the scope of this work.

For simplicity and to allow to express the resulting MHD equations (see \Sec{conclusions}) in a form that is suitable to directly apply time integrators
like Runge-Kutta methods, i.e., writing one equation for the time
derivative of each variable,
we will restrict ourselves to the CIT
approach,
leading to the Navier-Stokes
description of the viscous shear and bulk stresses
and the Fourier's heat flux, in the subrelativistic limit.
For simulations of MHD in the early Universe,
from the onset of radiation
domination down to neutrino decoupling, as we show in \Sec{transport_coeffs}, the actual values of the
transport coefficients are very small when the primordial
plasma is near equilibrium, since they are proportional to the mean free path of the plasma
particles \cite{Weinberg:1971mx,Jedamzik:1996wp,Baym:1997gq,Ahonen:1998iz,Arnold:2000dr,Arnold:2003zc,Arnold:2006fz,Durrer:2013pga,Uchida:2024ude}, affecting the dynamics
at scales several orders of magnitude smaller than the Hubble horizon \cite{Caprini:2009yp,RoperPol:2018sap}.
Therefore, realistically viable simulations usually take viscous coefficients that are small enough
to not affect the large-scale dynamics, but still orders of magnitude larger than their actual
values (cf.~\cite{Brandenburg:1996fc,Christensson:2000sp,
Brandenburg:2017rnt,Brandenburg:2017neh,Brandenburg:2017rcb,Schober:2018wlo,RoperPol:2019wvy,RoperPol:2021xnd,RoperPol:2022iel}).
In these situations, as the simulations are not resolving the 
small scales, the relativistic description is not
expected to affect the macroscopic dynamics, and this approach is justified.

For further details on relativistic fluid dynamics of imperfect
fluids we recommend the reader \cite{Romatschke:2009im,Rezzolla:2013dea}
and references therein.

\subsection{First-order imperfect fluids}
\label{NavierStokes}

Viscous effects and heat fluxes
arise due to collisions of the fluid particles that
perturb the distribution function with respect to LTE,
such that $\dd f/\dd t \neq 0$.
The resulting distribution function can be expanded in terms of the
Knudsen number
\begin{equation}
    f (X^\mu, p^\mu) = f_0 (X^\mu, p^\mu) + \delta f (X^\mu, p^\mu)
    + {\cal O} ({\rm Kn}^2)\,,
\end{equation}
where $p^\mu$ is the four-momentum, $f_0$ is the LTE distribution (e.g., for fluids made of classical particles, Maxwell-Boltzmann, or
Maxwell-J\"uttner in the relativistic limit \cite{MaxwellJuttner}), and $\delta f$ is the
first-order perturbation in the distribution function.
This approach follows Chapman-Enskog theory \cite{1970mtnu.book.....C},
and provides a fluid dynamics description
for imperfect fluids
using the Knudsen number as a perturbative parameter. 
In particular, the Knudsen numbers
defined using the gradients of the fluid velocity
and the temperature to estimate the characteristic fluid
length scale $L$
yield Navier-Stokes viscosity and Fourier's law of conductivity,
respectively.

In the subrelativistic limit, the additional contribution to the stress tensor
due to first-order imperfect fluids is described by the deviatoric viscous stress tensor $\Pi^{ij}$,
such that
$T^{ij} = T^{ij}_{\rm pf} - \Pi^{ij}_\visc$
\cite{Landau1987Fluid,Arnold:2000dr,2008Kundu,Rezzolla:2013dea,Durrer:2003ja},
\begin{equation}
    \frac{\Pi^{ij}_\visc}{p + \rho} = 2 \, \nu \,
    \sigma^{ij} + \xi   \, \theta
    \, g^{ij} \,, \label{Piij_NS} \qquad \text{with \ }
    \sigma^{ij} = S^{ij} - \onethird \theta \, g^{ij}\,,
\end{equation}
where $\nu = \eta_\visc/(p + \rho)$ and $\xi = \zeta_\visc/(p + \rho)$ are the 
kinematic shear and bulk viscosities ($\eta_\visc$ and $\zeta_\visc$ being the dynamic
viscous coefficients).
The traceless rate-of-strain tensor is $\sigma^{ij}$
with
\begin{equation}
    S^{ij} = \frac{1}{2}
    \bigl(U^{i;j} + U^{j;i}\bigr) = 
    \half a^{-3}
    (\partial^i u^j + \partial^j u^i + 2\, \HH\,  \delta^{ij})\,,
\end{equation}
and $\theta = S^i_{\ i} = a^{-1} (\nab \cdot \uu +
3 \HH)$
is the fluid expansion scalar
(already defined in \Sec{rel_hydrodynamics}).
The comoving rate-of-strain tensor and expansion scalar are
$\tilde S^{ij} = a^3 S^{ij}$
and $\tilde \theta = a \theta$.
A kinematic approach to describe the anisotropic
stress tensor, $T_{ij}$, as the most general rank-2 tensor to describe linear
constitutive relations (i.e., Newtonian fluids) also leads to \Eq{Piij_NS} and corresponds
to the original derivation of the Navier-Stokes equations \cite{Navier,Stokes}.
We note that the comoving deviatoric viscous 
stress tensor is
\begin{equation}
    \tilde \Pi^{ij}_\visc = a^6 \Pi^{ij}_\visc =  2 \, \tilde
    \nu \, (\tilde p + \tilde \rho) \, \tilde \sigma_{ij} +
    \tilde \xi \, (\tilde p + \tilde \rho) \, \tilde \theta \,
    \delta^{ij}
    \,,
\end{equation}
where the comoving kinematic and bulk viscosity coefficients
are $\tilde \nu = a^{-1} \nu$
and $\tilde \xi = a^{-1} \xi$.
The comoving viscous energy flux in the subrelativistic
limit is non-zero for an expanding Universe (see details
in \Sec{conformal_NS}),
\begin{equation} \label{equation_Pii0}
    \tilde \Pi^{i0}_\visc = a^6 \Pi^{i0}_\visc = 3 \, \tilde \xi \,
    \HH \, (\tilde p + \tilde \rho)\, u^i\,.
\end{equation}
As we show in \Sec{transport_coeffs}, the bulk viscosity is, in general,
several orders of magnitude smaller than the Hubble time,
so we will assume $\xi H = \tilde \xi \HH \ll 1$ and,
hence, $\tilde \Pi^{i0}_\visc \simeq 0$.

On the other hand, the contribution to the energy conservation equation
due to heat fluxes $q_i$
is described by an additional term in the momentum density,
$T^{i0} = T^{i0}_{\rm pf} - \Pi^{i0}_{\rm heat}$.
According to the Fourier's law
of conductivity, also found following first-order Chapman-Enskog theory,
the subrelativistic limit of the deviatoric momentum tensor is
\cite{Ahonen:1998iz,Romatschke:2009im,Rezzolla:2013dea}
\begin{equation}
    \Pi^{i0}_{\rm heat} = - q^i U^0 \simeq - a^{-1} q^i\,, 
    \qquad \text{where \ } q^i = - \kappa\, g^{ij} \partial_j T
    = - a^{-2} \, \kappa \, \partial_i T\,,
\end{equation}
being $\kappa$ the thermal conductivity.
The comoving deviatoric momentum density is

\begin{equation}
\tilde \Pi^{i0}_{\rm heat} = a^6 \Pi^{i0}_{\rm heat} = -\tilde q^i =
\tilde \kappa \, \partial^i \tilde T\,,
\end{equation}
where $\tilde q^i = a^5 q^i = a^3 q_i$
is the comoving heat flux, $\tilde \kappa = a^2 \kappa$ the comoving thermal conductivity, and $\tilde T = a T$ the comoving temperature.

\subsubsection*{4.1.1 \ Conservation form of Navier-Stokes equations with thermal conductivity}

For a fluid description of the primordial plasma in an expanding Universe, 
Navier-Stokes and Fourier descriptions for viscosity
and heat fluxes
lead to the
inclusion of \Eq{Piij_NS} in the momentum equation [cf.~\Eq{mom_aux}],
\begin{equation}
    \partial_\tau \tilde T^{0i}_{\rm pf} +
    \partial_j \tilde T^{ij}_{\rm pf} = \tilde f_H^i + \partial_j \tilde \Pi^{ij}_\visc
    = \tilde f_H^i + \tilde f^i_{\ipf}\,, \label{cont_aux_visc}
\end{equation}
where we have defined the force exerted in the imperfect (ipf) fluid due to out-of-equilibrium
effects\footnote{We have neglected in $\tilde f_{\rm ipf}^\mu$ the force arising from the heat
conductivity, $\tilde f^i_{\rm heat} = \partial_\tau \tilde{\Pi}^{i0}_{\rm heat} = \tilde \kappa  \,
\partial_\tau \partial^i \tilde T
\simeq 0$.
In general, from the heat equation (i.e., for a fluid at rest), $\partial_\tau \tilde T \propto \partial_i \partial^i \tilde T$, such that $\tilde f_{\rm heat}^i$
becomes proportional to the gradients of $\tilde T$ at cubic order, which is suppressed compared to the quadratic
order we consider for first-order hydrodynamics.} $\tilde f_\ipf^\mu \equiv \partial_i \tilde \Pi^{i\mu}$
and
we set $\alpha = 1$ (conformal time) for compactness.
Then, the viscous force $\tff_\ipf$ corresponds to the divergence of the deviatoric 
viscous tensor and takes the following form,
\begin{align}
    \frac{\tff_{\ipf}}{\tilde p + \tilde \rho} = 
    \frac{\nab \cdot
    \tilde {\pmb \Pi}_\visc}{\tilde p + \tilde \rho} = &\,
    \tilde \nu \nab^2 \uu + \bigl(\onethird \tilde \nu +
   \tilde \xi \bigr) \nab \tilde \theta + \bigl[\bigl(
    2 \, \tilde \nu \, \pmb{\tilde \sigma} + \tilde \xi \, \tilde \theta \, {\pmb I} \bigr)
    \cdot \nab\bigr] \ln \tilde \rho\,,
    \label{visc_forc}
\end{align}
where we already normalize with the enthalpy $\tilde p + \tilde \rho$ as this
term will appear in the momentum equation \Eq{mom_subrelativistic_1},
and assume homogeneous
$\tilde \nu$, $\tilde \xi$, and
$\cs^2$ in the last equality.

The energy equation also includes the effect of
energy dissipation, $\partial_i \tilde \Pi^{0i}_{\rm heat} =
\tilde f_\ipf^0 = - \nab \cdot \tilde \qq$,
which irreversibly converts mechanical to thermal energy,
\begin{equation}
    \partial_\tau \tilde T^{00}_{\rm pf} + \partial_i \tilde T^{0i}_{\rm pf} =
    \tilde f_H^0 + \tilde f_\ipf^0 \,,
    \label{mom_aux_visc}  \qquad \text{where \ }
    \tilde f_\ipf^0 = - \nab \cdot \tilde \qq = 
    \tilde \kappa \nab^2 \tilde T\,,
\end{equation}
and we have assumed homogeneous $\tilde \kappa$ in the last equality.
The temperature can be directly related to the energy density using \Eq{rho_rad}
for a radiation-dominated fluid.

To solve this system of equations in the conservation form, one can follow the procedure
described in \Sec{conservation_rel_perf} to relate $\tilde T^{ij}_{\rm pf}$ to $\tilde T^{0\mu}_{\rm pf}$, and then construct the viscous imperfect
four-force $\tilde f^\mu_\ipf$ from \Eqs{visc_forc}{mom_aux_visc},
where the velocity field can be obtained from $\tilde T^{0\mu}_{\rm pf}$ using \Eq{u_i_conservative}.
Then, the viscous force  $\tilde f^\mu_\ipf$ can be included in the force array ${\cal F}^a$
in \Eq{cons_form}, as well as any additional forces that might be
exerted on the fluid (e.g., the Lorentz force due to electromagnetic fields, as we will see
in \Sec{MHD_eom}).

\subsubsection*{4.1.2 \ Non-conservation form of Navier-Stokes equations with thermal conductivity}

In the following, we include the imperfect (viscous) force $\tilde f_\ipf^\mu$ in the
non-conservation form of the energy and momentum equations,
studied for perfect fluids in \Sec{rel_hydrodynamics}.
To find the effect of viscous forces,
we first note that any external forces lead to an additional term in the evolution
equation of the Lorentz factor [cf.~\Eq{lorentz_evol}] as they
produce work against the fluid.
The resulting dynamical equation for $\gamma^2$ is
\begin{equation}
    [D_\tau \ln \gamma^2]_\ipf = [D_\tau \ln \gamma^2]_{\rm pf} - \frac{2 \, u^2}{1 - \cs^2 u^2} \frac{\tilde f_\ipf^0}{\tilde \rho} + \frac{2}{1 + \cs^2} \, \frac{1 + \cs^2 u^2}{1 - \cs^2 u^2} \, \frac{\uu \cdot 
    \tff_\ipf}{\tilde \rho} \,. \label{lorentz_viscs}
\end{equation}
The energy dissipation due to the deviations with respect to LTE
is usually neglected in numerical studies of MHD evolution in the primordial Universe
(cf.~\cite{Brandenburg:1996fc,Brandenburg:2017neh}), although it has been
incorporated in theoretical studies \cite{Jedamzik:1996wp}.
However, since the shear viscosity and thermal conductivity are in general small,
imperfect fluid effects on the evolution of $u^2$ can be neglected at large scales.
The dissipation due to thermal conductivity corresponds to $\tilde f_\ipf^0 = \tilde \kappa \nab^2 \tilde T$ [cf.~\Eq{mom_aux_visc}].
The dissipation due to viscous forces can be computed
taking into account the following property of the viscous deviatoric
stress tensor under the CIT approach, $U_\nu \Pi_\visc^\munu\covderT{\mu}
\equiv U_\nu f_{\rm visc}^\nu = -
2 \, \eta_\visc \, \sigma^\munu S_\munu - \zeta_\visc \, \theta^2$ \cite{Romatschke:2009im,Rezzolla:2013dea}, where $\sigma^\munu = S^\munu - \third \, \theta \, h^\munu$
is the covariant generalization of the Navier-Stokes viscosity given
in \Eq{Piij_NS} (see \Sec{conformal_NS}).
Omitting $\tilde f_\visc^0 = \partial_i \tilde \Pi^{0i}_\visc$ for negligible
bulk viscosity [cf.~\Eq{equation_Pii0}],
the subrelativistic version of this identity allows us to compute
the work done by the viscous forces
\begin{equation}
    \label{work_visc}
    \frac{\uu \cdot \tff_\visc}{\tilde p + \tilde \rho}
    \simeq - 2 \, \tilde\nu \, \tilde \sigma^{ij}
    \tilde S_{ij} - \tilde \xi \, \tilde \theta^2  = -
    2 \, \tilde\nu \, \tilde
    S^{ij} \tilde S_{ij} +
    \bigl(\twothird \tilde \nu - \tilde \xi\bigr)
    \,
    \tilde \theta^2\,,
\end{equation}
which corresponds to the classical energy dissipation that appears when one takes
the second moment of the subrelativistic Boltzmann equation \cite{2008Kundu,Rezzolla:2013dea,Cercignani2002RelativisticBoltzmann}.
It
represents the irreversible conversion of mechanical to thermal
energy through the action of viscous stresses.
As these correspond to out-of-equilibrium effects, both $\tilde f_\ipf^0 = - \nab \cdot \tilde \qq$
and $\uu \cdot \tff_\ipf = - 2 \,\tilde \eta_\visc\, \tilde \sigma^{ij}\, \tilde S_{ij}
- \tilde \zeta_\visc \, \tilde \theta^2$ are irreversible processes leading to
an increase of entropy.

The resulting relativistic energy and momentum equations are found including the
divergence of the imperfect fluid stresses and momentum fluxes
to \Eqs{cont_relativistic0}{mom_relativistic0},
\begin{subequations}
\label{eqs_relativistic_visc}
\begin{align}
    \partial_\tau \ln \tilde \rho \, = &\, - 
    \frac{1 + \cs^2}{1 - \cs^2 u^2} \nab \cdot \uu
    - \frac{1 - \cs^2}{1 - \cs^2 u^2} (\uu \cdot \nab) \ln \tilde \rho \nonumber \\ &\, 
    + \frac{1}{1 - \cs^2 u^2} \frac{1}{\tilde \rho} \bigl[
    \tilde f_\ipf^0 (1 + u^2) - 2 \, \uu \cdot \tff_\ipf \bigr] + 
    {\cal F}_H^0 
    \,, \label{cont_relativistic_visc} \\
    D_\tau \uu = &\, \frac{\uu}{(1 - \cs^2 u^2) \gamma^2} \biggl[ \cs^2 \nab \cdot \uu + \cs^2 
    \frac{1 - \cs^2}{1 + \cs^2} (\uu \cdot \nab) \ln \tilde \rho - \frac{1}{\tilde \rho}
    \biggl(\tilde f_\ipf^0
    - \frac{2 \, \cs^2}{1 + \cs^2} \uu \cdot \tff_\ipf \biggr) \biggr] \nonumber \\
    &\, - \frac{\cs^2}{1 + \cs^2} \frac{\nab \ln \tilde \rho}{\gamma^2} + \frac{1}{1 + \cs^2}
    \frac{\tff_\ipf}{\tilde \rho \gamma^2} + \pmb{\cal F}_H \,, \label{mom_relativistic_visc}
\end{align}
\end{subequations}
where the Hubble friction ${\cal F}_H^0$ appearing in the energy equation
is given for a generic choice of $\beta$ in \Eq{FFH0}, and
for $\beta = 4$, it can be simplified to \Eq{FFH0}.
Note that the system of \Eqq{eqs_relativistic_visc} can be generalized to any set
of external forces $\tilde f^\mu_{\rm tot}$ by setting $\tilde f_\ipf^\mu \to \tilde f_\tot^\mu$.
In \Sec{MHD_eom}, we will include the
contribution from the Lorentz force.

In the subrelativistic limit, we find
\begin{subequations}
\label{eqs_subrelativistic_visc}
\begin{align}
    \lim_{u^2 \ll 1}
    \partial_{\tau} \ln \tilde \rho \, = - &\,
    (1 + \cs^2)\, \nab \cdot \uu -
    (1 - \cs^2) (\uu \cdot \nab ) \ln \tilde \rho \nonumber \\ + &\, \frac{1}
    {\tilde \rho} (\tilde f_\ipf^0 - 2 \, \uu \cdot \tff_\ipf)
    + \bigl[\beta - 3\, (1 + \cs^2) \bigr] \, \HH
    \,,
    \label{cont_subrelativistic_visc}
    \\ 
    \lim_{u^2 \ll 1} D_{\tau} \uu  = &\,
    \uu \, \cs^2 \biggl[
    \nab \cdot \uu + 
    \frac{1 - \cs^2}{1 + \cs^2} \, (\uu \cdot \nab) \ln \tilde \rho\biggr]
    - \frac{\uu}{\tilde \rho} \biggl(\tilde f_\ipf^0 - \frac{2 \, \cs^2}{1 + \cs^2}
    \uu \cdot \tff_\ipf \biggr) \nonumber \\
    & - \frac{\cs^2}{1 + \cs^2} \nab \ln \tilde \rho +
    \frac{1}{1 + \cs^2}
    \frac{\tff_\ipf}{\tilde \rho} + (3 \cs^2 - 1) \, \uu \, \HH\,,
    \label{mom_subrelativistic_visc}
\end{align}
\end{subequations}
where the term in the momentum equation proportional to $\tff_\ipf$ is the term that had been included in previous work studying Navier-Stokes in a radiation-dominated expanding Universe.
Considering Navier-Stokes description for the viscosity
and Fourier's law for the thermal conductivity,
the imperfect (viscous) force can be described
using \Eq{visc_forc}, and the heat dissipation $\tilde f_\ipf^0$ using \Eq{mom_aux_visc}.
The work done by the viscous forces $\uu \cdot \tff_\ipf$ can be computed
using \Eq{work_visc}.
Under the assumption that the
shear viscosity is very small in the early Universe \cite{Arnold:2000dr},
previous work neglected the imperfect-fluid dissipation terms in
the energy and momentum equations.

\subsection{Transport coefficients in the primordial plasma}
\label{transport_coeffs}

The viscous transport coefficients $\nu$ and $\xi$,
and the thermal conductivity $\kappa$,
are proportional to the mean free path $l_{\rm mfp}$ of a
plasma composed by radiation particles
\cite{Weinberg:1971mx,Jedamzik:1996wp,Ahonen:1998iz},
\begin{equation}
\label{viscosities}
    \nu = \frac{4}{15}
    \frac{l_{\rm mfp}}{1 + \cs^2}\,, \qquad
    \xi = 4 \,
    \bigl(\onethird - \cs^2 \bigr)
    \frac{l_{\rm mfp}}{1 + \cs^2}\,, \qquad
    \kappa = \frac{4}{3} \frac{\rho}{T} l_{\rm mfp}\,.
\end{equation}

\begin{figure}[t]
    \centering
     \includegraphics[width=0.9\textwidth]{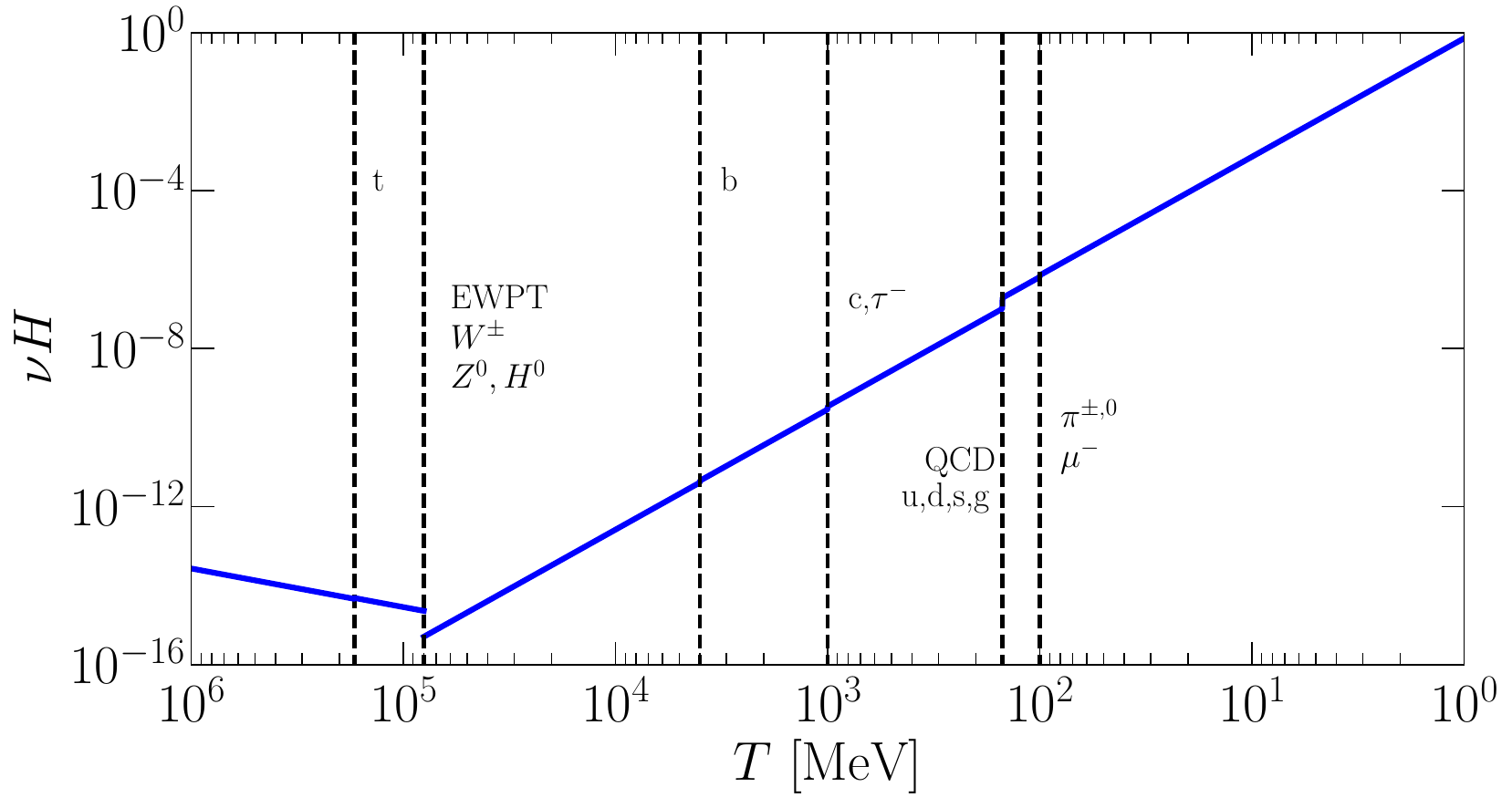}
     \caption{ 
     Ratio of the shear viscosity $\nu$ [cf.~\Eq{viscosities}] to the Hubble time $H^{-1}$ [cf.~\Eq{rho_rad}] at temperature scales above neutrino decoupling $T > T_\nu \sim 1$ MeV.
     The temperatures at which the top (t), bottom (b), and charm (c) quarks, and tauons ($\tau^-$)
     and muons ($\mu^-$) become
     non-relativistic ($T < m$) are indicated with vertical lines, as well as the temperature
     scales of the electroweak (EW) and QCD phase transitions (PTs). During the EWPT, around 100 GeV,
     the
     electroweak gauge ($W^\pm$, $Z^0$) and Higgs ($H^0$) bosons become massive, and during the QCD transition ($T \simeq$ 150 MeV),
     the up (u), down (d), and strange (s) quarks and gluons (g) confine, forming baryons
     and mesons, from which only pions $\pi^{\pm, 0}$ remain relativistic.
     For temperatures below the W-boson mass, $T < 80$ GeV, the shear viscosity is
     determined by the neutrino mean free path, given in \Eq{lmfp_nu}.
     For larger $T$, we use $\eta_\visc \simeq C_\visc \, g^{-4}_{\rm hyper} \, T^3/\ln g^{-1}_{\rm hyper}$
     \cite{Arnold:2000dr}
     with $g_{\rm hyper} \simeq 0.36$
     being the hypercharge coupling at the EW scale, resulting in $\nu \simeq 21/T$ (see \Eq{nu_T_hyper} and \cite{Durrer:2013pga}).
     }
     \label{shear_viscosity}
\end{figure}

In the early Universe, the viscosity and thermal conductivity
are expected to be determined by the
neutrinos mean free path for temperatures below the weak W-boson mass, $T \lesssim 80$
GeV \cite{Arnold:2000dr}, down to the neutrino decoupling at $T_\nu \simeq$ 1 MeV.
The mean free path of neutrinos can be expressed as \cite{Heckler:1993nc,Jedamzik:1996wp,Durrer:2013pga}
\begin{equation} \label{lmfp_nu}
    l_{\rm mfp} (80 \, {\rm GeV} > T > 1 \, {\rm MeV})
    \approx l_{\nu, {\rm mfp}} \approx n_f^{-1}
    G_F^{-2} T^{-2} \simeq  11 \, {\rm cm} \,   \biggl(\frac{14}{g_f}\biggr)   \,
    \biggl(\frac{100 \, {\rm MeV}}{T} \biggr)^5\,,
\end{equation}
where $G_F \simeq 1.166 \times 10^{-5} \, {\rm GeV}^{-2}$ is the Fermi constant
that determines the strength of the weak interactions at $T \lesssim 80$ GeV,
$\zeta$ is the Riemann zeta function, and the number density of fermions is $n_f = 3\, \zeta(3)/(4\pi^2)
\, g_f \, T^3$, with $g_f = 14$ accounting for
muons, neutrinos, and electrons.
Therefore, the
shear viscosity is in general tiny
at high temperatures when compared to the Hubble scale
\cite{Caprini:2009yp,Durrer:2013pga,Brandenburg:2017neh,RoperPol:2019wvy},
\begin{equation}
    \nu H (80 \, {\rm GeV} > T > 1 \, {\rm MeV})
    \simeq 6.5 \times 10^{-7} \biggl(\frac{14}{g_f}\biggr)
    \biggl(\frac{g_\ast}{17.25}\biggr)^{1/2} \biggl(\frac{100 \, {\rm MeV}}{T}\biggr)^3\,,
      \label{nu_times_h}
\end{equation}
where $g_\ast = g_b + \tfrac{7}{8} g_f$ is the number of relativistic degrees of freedom,
with $g_b = 5$ accounting for photons and pions $\pi^{\pm, 0}$ ($g = 3$), which are the lightest mesons
and remain
relativistic at $T \simeq 100$ MeV \cite{Kolb:1990vq}.
The Hubble rate $H$
can be computed using \Eq{eq:Friedmann} taking into account that the
energy density is dominated by radiation,
\begin{equation}
    H = \sqrt{\frac{\rho}{3 \Mpl^2}}\,, \qquad \text{with \ }
    \rho = \frac{\pi^2}{30} g_\ast T^4\,. \label{rho_rad}
\end{equation}
Indeed, we find that $\nu H \sim 10^{-7}$ around the QCD scale at 100 MeV and
it decreases proportional to $T^{-3}$ at higher temperatures, with
some moderate dependence on the degrees of freedom $g_\ast$ and $g_f$,
becoming $\nu H \sim
10^{-16}$ around the electroweak scale at 100 GeV when $g_\ast = 96.25$ and $g_f = 78$, counting
for all quarks, leptons, and bosons of the Standard Model, excluding the top quark.
We show the value of $\nu H$ for a temperature range of $T_\nu \simeq 1 {\rm \, MeV} < T < 1$ TeV in \Fig{shear_viscosity}.
Therefore, the primordial plasma can be closely treated as a perfect fluid.
However, dissipative viscous and resistive
effects need to be included for a realistic turbulent
description of the evolution of the
fluid perturbations, especially in the context of MHD
(see \Sec{cov_ohms})
and decay of primordial
magnetic fields \cite{Durrer:2013pga}.

At temperatures higher than 80 GeV, neutrino interactions are no longer suppressed and their
mean free path stops being the longest one in the
primordial Standard Model plasma.
At these temperatures, the shear dynamic viscosity is estimated to be
$\eta_\visc \simeq C_\visc \, g^{-4}_{\rm hyper} \, T^3/\ln g^{-1}_{\rm hyper}$ \cite{Arnold:2000dr,Arnold:2003zc},
being $g_{\rm hyper} \simeq 0.36$
the hypercharge coupling at the EW scale
and $C_\visc \simeq 16$ a constant
that depends on the plasma particle content
(see Eq.~(108) of App.~A1 in \cite{Durrer:2013pga}).
Since this estimate omits the collisions
of neutrinos, it is no longer valid to describe the
shear and bulk viscosity
at $T \lesssim 80$ GeV, as discussed above.
In this regime, the shear viscosity becomes \cite{Durrer:2013pga}
\begin{equation}
    \nu (T > 80 \, {\rm GeV}) \simeq \frac{20.6}{T} \frac{100}{g_\ast}\,, 
    \label{nu_T_hyper}
\end{equation}
which is shown in \Fig{shear_viscosity}, where $\nu H \sim T$.

%%%%%%%%%%%%%%%%%%%%%%%%%%%%%%%

On the other hand, for a perfectly conformal theory, $\cs^2 = \third$
and the bulk viscosity
vanishes, $\zeta_\visc = 0$.
However, contributions from particle number violation yield
a leading-log order contribution at large temperatures,
$\zeta_\visc \sim g^2 T^3/\ln g^{-1}$ \cite{Arnold:2006fz}, with $g$ being the
appropriate coupling constant at large $T$.
This is in any case generally suppressed compared to the shear viscosity
$\eta_\visc/\zeta_\visc = \nu/\xi \sim g^{-6} \gg 1$.
When the bulk viscosity is zero (case classically known in fluid dynamics
as the Stokes assumption),
the thermodynamic and
mechanical pressures are equivalent \cite{2008Kundu}
and the trace of the deviatoric stress tensor vanishes,
$\Pi^i_{\ i} = 3 \, \zeta_\visc \, \theta = 0$.
Although this is true at small and high temperatures,
there is no particular reason to assume that the
Stokes assumption is generically satisfied in astrophysical systems  \cite{Weinberg:1971mx}.
Larger values of $\zeta_\visc$ can also arise during periods of
conformal violation, for example during phase transitions
in the early Universe \cite{Arnold:2006fz}.
Therefore, although $\zeta_\visc$ is negligible in most scenarios, we keep it
in our equations for generality.

\subsection{Covariant formulation of relativistic
imperfect fluids}
\label{conformal_NS}

In the following, the covariant
generalization of the Navier-Stokes equations with Fourier heat
flux
is given, following the pioneer work of \cite{PhysRev.58.919}
under the CIT
approach \cite{Misner:1973prb,Landau1987Fluid,Jedamzik:1996wp,Font:2008fka,Subramanian:2009fu,Romatschke:2009im,Rezzolla:2013dea}.
For imperfect relativistic fluids, the definition of the four-velocity is
not unique.
Two common choices correspond to the
Eckart $(U^N_\mu,$ where N stands for number density) \cite{PhysRev.58.919} or the Landau
($U^E_\mu$, where E stands for energy) \cite{Landau1987Fluid} frames
(see discussion in \cite{Romatschke:2009im,Rezzolla:2013dea}).
These frames can be understood as those where, respectively, the
charge number density $N^\mu$ or the stress-energy tensor $T^\munu$
are at rest, i.e.,
$U_\mu^N N^\mu = -n$ in the Eckart frame,
and the four-velocity is an eigenvector of $T^\munu$ ($U_\mu^E T^\munu = -\rho U^\nu$)
in the Landau frame.
This implies that either
$U_\mu^N \delta N^\mu = 0$
or $U_\mu^E \Pi^\munu = 0$, where
$\delta N^\mu$
corresponds to the modifications of $N^\mu$ due to deviations with respect to LTE.
For a perfect fluid, both choices coincide $U_\mu^N = U_\mu^E$ and there is
no ambiguity on the four-velocity definition, but in general, the two velocities
are related to each other by the heat flux four-vector $q^\mu$ \cite{Israel:1976efz,Rezzolla:2013dea}
\begin{equation} \label{relation_velocities}
    U^\mu_E = U^\mu_N + \frac{q^\mu}{p + \rho} + {\cal O} (u^2)\,.
\end{equation}

In a system without conserved charges (zero chemical potential),
as approximately realized in the primordial quark-gluon plasma,
the Eckart frame is not well defined and the natural choice
is the Landau frame \cite{Romatschke:2009im}.
In this frame, the heat current $q^\mu$ does not appear in the
deviatoric stress-energy tensor $\Pi^\munu$ but instead it modifies
the charge number density $N^\mu = N^\mu_{\rm pf} + n q^\mu/(p + \rho)$.
Then, both the charge density and the stress-energy tensor are modified,
$N^\mu = N^\mu_{\rm pf} + \delta N^\mu$ and 
$T^\munu = T^\munu_{\rm pf} - \Pi^\munu$,
with $U_\mu \Pi^\munu = 0$.
In the following,
we will consider the Eckart
frame, which naturally incorporates the heat flux.
In this frame, the charge current is that of a perfect fluid
and the deviatoric stress-energy tensor,
$\Pi^\munu$, includes the heat current
\cite{PhysRev.58.919,Ahonen:1998iz,Landau1987Fluid,Weinberg:1972kfs,Jedamzik:1996wp,Font:2008fka,Romatschke:2009im,Rezzolla:2013dea},
\begin{align}
    \Pi^\munu = 2 \, \eta_\visc \, \sigma^\munu + &\, \zeta_\visc \, \theta\, h^\munu
    - 2\, q^{(\mu} U^{\nu)}
    \,, \nonumber \\ &\,
    \text{with \ } \sigma^\munu = S^\munu - \frac{1}{3} \theta \, h^\munu\,,
    \quad \text{and \ } q^\mu = - \kappa \bigl(\nabla^\mu T + T a^\mu\bigr) \,,
    \label{visc_NS}
\end{align}
where $\theta = U^\mu\covderV{\mu}$ is the relativistic fluid expansion scalar,
the
parenthesis indicates symmetrization (i.e., $2 q^{(\mu} U^{\nu)} = q^\mu U^\nu + q^\nu U^\mu$),
$\nabla^\mu T = h^\munu \partial_\nu T$, $h^\munu = g^\munu + U^\mu U^\nu$ is the
velocity projection tensor, and $a^\mu = U^\nu U^\mu\covderV{\nu}$ is the four-vector
acceleration.
$S^\munu$ is the relativistic rate-of-strain tensor,
\begin{equation}
    S^\munu =
    \nabla^{(\nu} U^{\mu)}\,, \qquad \text{where \ }
    \nabla^\nu U^\mu = h^{\nu\lambda} U^\mu\covderV{\lambda}\,.
    \label{rate_strain}
\end{equation}
We can rearrange \Eq{visc_NS} in the following way
\begin{equation}
    \Pi^\munu = 2 \, \eta_\visc \, S^\munu + \lambda_\visc \, \theta \, h^\munu
    + 2 \, \kappa \, U^{(\mu} \nabla^{\nu)} T + 2 \, \kappa \, T
    \, U^{(\mu} a^{\nu)}\,,
\end{equation}
where $\lambda_\visc = \zeta_\visc - \twothird \nu_\visc$.
The terms $\sigma^\munu$ and $q^\mu U^\nu$ are traceless by construction,
\begin{equation}
    \sigma^\mu_{\ \, \mu} = S^{\mu}_{\ \, \mu} - \third \theta h^\mu_{\ \mu} = 0 \,,
    \qquad q^\mu U_\mu = \kappa \, U^\mu h_{\mu\nu} \partial^\nu T + \kappa \, U^\mu U^\nu
    U_{\mu\, ; \nu} = 0\,,
\end{equation}
while the bulk viscosity yields the following trace of the
deviatoric tensor,
\begin{equation}
    \Pi^\mu_{\ \mu} = 3\, \zeta_\visc \, \theta \ .
\end{equation}
Therefore, only when the bulk viscosity is zero, the equations of motion of a radiation-dominated imperfect fluid in an expanding background are conformally flat \cite{Subramanian:1997gi}.
Subtracting the deviatoric tensor
$\tilde \Pi^\munu$ to the stress-energy tensor of 
a perfect fluid
$\tilde T^\munu = \tilde T^\munu_{\rm pf} - \tilde \Pi^\munu$, we can express the relativistic energy and
momentum equations from \Eq{cons_FLRW},
\begin{equation}
    \partial_\tau \tilde T^{0\mu}_{\rm pf} + \partial_i \tilde T^{i\mu}_{\rm pf} =
    \tilde f_H^\mu + \partial_\nu \tilde \Pi^{\nu \mu} = \tilde f_H^\mu + \tilde f_\ipf^\mu\,.
\end{equation}
The solution to this system of equations becomes more cumbersome
than its subrelativistic counterpart, as it involves time
derivatives of the primitive fluid variables in the viscous forces and heat
fluxes \cite{Rezzolla:2013dea},
\begin{equation}
    \tilde f^\mu_\ipf =
    2 \, \partial_\nu (\tilde \eta_\visc \, \tilde S^\munu)
    + \tilde \lambda_\visc \, \tilde \theta
    \, \partial_\nu (\tilde U^\mu \tilde U^\nu)
    + \tilde h^\munu \partial_\nu (\tilde \lambda_\visc \, 
    \tilde \theta) - \partial_\nu (\tilde q^\mu \tilde U^\nu 
    + \tilde q^\nu \tilde U^\mu)
    \,,
\end{equation}
where the comoving variables are $\tilde \theta = a \theta$,
$\tilde S^\munu = a^3 S^\munu$,
$\tilde h^\munu = a^2 h^\munu$, $\tilde q^\mu = a^5 q^\mu$,
$\tilde \eta_\visc = a^3 \eta_\visc$, and
$\tilde \lambda_\visc = a^3 \lambda_\visc$.
The time derivatives over $\tilde \Pi^{00}$ and
$\tilde \Pi^{0i}$ lead in fact to second-order derivatives of the
fluid variables $\gamma$ and $\uu$ in the energy and momentum
equations.
This would require to readapt the fluid equations if we want to describe
them in a suitable way for direct numerical time integration, such that only one time derivative
appears per equation.
However, since this
is still not a suitable relativistic theory due
to violating causality, as discussed above (see also \cite{Romatschke:2009im} for a review),
in the following we will always consider the subrelativistic limit
of the covariant Navier-Stokes viscosity and Fourier's thermal conductivity
for simplicity.

\section{Standard Electromagnetism: Maxwell equations}
\label{Maxwell_sec}

In order to describe the dynamics of charged fluids, we first
need to introduce Maxwell equations in an expanding Universe
that will later be coupled to the equations of motion of the
fluid, following an MHD description in \Sec{MHD_eom}.
In the following, we set $m = 1$, i.e.,
the space-like signature $(- + + +)$, $X^0 = \tau$, and $\alpha = 1$, for simplicity.

\subsection{Faraday tensor and covariant electromagnetic fields}
\label{faraday_tensor}

We will only consider the $U(1)_{\rm EM}$ sector of gauge fields as the
classical electromagnetic fields,
such that
we describe electromagnetism at temperature scales below the electroweak symmetry breaking.
At larger temperatures, the electroweak force is described by the group of symmetries
$SU(2)_{\rm L} \times U(1)_{\rm Y}$.
A $U(1)$ Abelian gauge field $A^\mu = (\chi, A^i)$ is described
by the Faraday tensor,
\begin{equation}
    F_\munu = \partial_\mu A_\nu - \partial_\nu A_\mu\,,
\end{equation}
which is, by construction, conformal invariant, i.e.,
$F_\munu = A_{\nu\,;\mu} - A_{\mu\,;\nu}$.
The covariant electromagnetic fields can then be defined in terms of the Faraday tensor in the following way
\cite{Font:2008fka,Subramanian:2009fu}
\begin{equation}
    E_\mu = F_\munu \, U^\nu\,, \qquad B^\mu = {\cal F}^\munu \, U_\nu\,, \label{EB_rel}
\end{equation}
where ${\cal F}^\munu = \half {\cal E}^{\munu \alpha \beta} F_{\alpha \beta}$
is the Hodge dual of the
Faraday tensor, $U^\mu$ represents the four-velocity
of the observer that measures the electromagnetic fields,
and ${\cal E}^{\munu \alpha \beta}$ is the Levi-Civita
tensor,
\begin{equation}
    {\cal E}^{\munu \alpbet} = - \frac{1}{\sqrt{-g}} \varepsilon^{\munu \alpbet} = - a^{-4} \varepsilon^{\munu \alpbet}\,,
\end{equation}
being
$\varepsilon^{\munu \alpbet} = \varepsilon_{\munu \alpbet}$
the usual Levi-Civita symbol, taking the value $+1 \, (-1)$ for even (odd) permutations of $\munu \alpbet$ with respect to 
$1230$ and vanishing when two indices are repeated.

This allows for a tensor definition of the electromagnetic fields,
whose vector description is otherwise not invariant under
Lorentz boosts.
In fact, an electric field can be transformed into an electric
and a magnetic field component, and viceversa,
in a boosted reference frame.
We note that the covariant electric and magnetic fields
are orthogonal to $U^\mu$,
\begin{equation}
    E_\mu U^\mu = B^\mu U_\mu = 0\,.
\end{equation}
Inverting \Eq{EB_rel}, one finds
\begin{equation}
    F_\munu = {\cal E}_{\mu\nu\alpha\beta} B^\alpha
    U^\beta + U_\mu E_\nu - U_\nu E_\mu\,, \qquad 
    {\cal F}^\munu = {\cal E}^{\mu \nu \alpha \beta} U_\alpha E_\beta +
    U^\mu B^\nu - U^\nu B^\mu\,.
\end{equation}

For the comoving Hubble observer with
$U^\mu = (1, {\pmb 0})/a$, we find the usual definitions of the electromagnetic fields,
\begin{equation}
    a\, E_i = F_{i0} = \nabla_i \chi - \partial_\tau A_i\,, \qquad 
    a^3 B^i = \half  \varepsilon^{ijk} F_{jk} =
    \nab \times \AAA\,,
    \label{EB_subrel}
\end{equation}
and zero temporal components, $E_0 = B_0 = 0$.
We denote as $\chi$ the scalar potential and $\AAA$ the vector potential.
As a consequence of the $U(1)$ symmetry of the gauge field $A_\mu$,
any transformation of the type $\chi \to \chi + \partial_\tau \psi$
and $\AAA \to \AAA + \nab \psi$ maintains the physical degrees of freedom (the
electric and magnetic fields) the same.
Raising indices in \Eq{EB_subrel} with the FLRW metric tensor
$g_\munu$ of the space components of the four-electromagnetic fields,
$E^i = a^{-2} E_i$ and $B^i = a^{-2} B_i$.
From \Eq{EB_subrel}, we also note that
\begin{equation}
    F_{ij} = a^3 \varepsilon_{ijk} B^k\,.
\end{equation}

The electromagnetic four-vectors measured in the reference frame
of a fluid with peculiar velocity $\uu = \dd \xx (\tau)/\dd \tau$
with respect to the Hubble observer, such that
$U^\mu = \gamma (1, u^i)/a$, have the following components
\begin{equation}
    \tilde E^\mu \equiv a^3 E^\mu =
    \gamma  \, (\tEE \cdot \uu, \tEE + \uu \times \tBB)\,,
    \qquad  \tilde B^\mu \equiv a^3 B^\mu = 
    \gamma \, (\tBB \cdot \uu, \tBB - \uu \times \tEE)\,, \label{comv_em}
\end{equation}
where $\tilde E_i = \tilde E^i = a E_i = a^3 E^i$
and $\tilde B_i = \tilde B^i = a B_i = a^3 B^i$,
given in \Eq{EB_subrel},
are the comoving electric and magnetic
fields, as we will justify in the following
section.

We note that the spatial components of the covariant electromagnetic
fields $E^\mu$ and $B^\mu$ are not equivalent to the fields boosted
to the fluid reference frame \cite{Jackson:1998nia},
\begin{subequations}
\begin{align}
    \EE' = &\, \gamma \, (\EE + \uu \times \BB) - (\gamma - 1)
    (\EE \cdot \hat \uu) \, \hat \uu\,, \\
    \BB' = &\, \gamma \, (\BB - \uu \times \EE) - ( \gamma - 1)
    (\BB \cdot \hat \uu) \, \hat \uu\,,
\end{align}
\end{subequations}
and they only coincide in the limit of subrelativistic bulk motion,
such that $\gamma - 1 \sim {\cal O} (u^2) \ll 1$.

\subsection{Maxwell equations and comoving electromagnetic fields}
\label{comoving_EM}

Let us now proceed to describe Maxwell equations,
which are found from minimization of the electromagnetism
action,
\begin{equation}
    S = \int \dd^4 \xx \, \sqrt{-g} \, \LL (A^\mu, \partial_\mu A^\nu)\,,
    \label{action_EM}
\end{equation}
where the Lagrangian in electromagnetism is \cite{Landau1987Fluid,Jackson:1998nia}
\begin{equation}
    \LL = - \fourth \, F^\munu F_\munu  +
    J^\mu A_\mu\,, \label{Lag_EM}
\end{equation}
and an effective current $J^\mu$
is introduced to model the current
density produced, for example, by the charged particles in a plasma.
Notice that the term $J^{\mu}A_{\mu}$ does not
lead to a gauge invariant definition of the Lagrangian.
However, the current density $J^{\mu}$ is assumed to describe a sector whose interaction with the gauge fields is written in the minimal coupling scheme, such that the term $J^{\mu}A_{\mu}$ is included in the gauge invariant kinetic term of this additional sector (with $A_{\mu}$ coming from the gauge covariant derivative) \cite{Weinberg:1995mt,CosmoLattice_MHD}.

Minimizing the action of \Eq{action_EM} in a curved space-time
yields the Euler-Lagrange equations,
\begin{equation}
    \biggl[\frac{\partial \LL}{\partial\bigl(\partial_\nu A_\mu
    \bigr)} \biggr]\covderT{\nu} - \frac{\partial \LL}{\partial A_\mu} = 0\,.
\end{equation}
From this equation, the covariant formulation of
Maxwell equations is found,
\begin{equation}
    F^\munu\covderT{\nu}
    = \frac{1}{\sqrt{-g}} \partial_\nu \bigl( \sqrt{-g} F^\munu \bigr)
    = J^\mu\,, \label{Max}
\end{equation}
where the covariant derivative is described using \Eq{Fmunu}.

As presented in \Sec{cons_perf_fluid}, when two metric tensors are related by
a Weyl transformation $g^\munu = \Omega^2 (x^\mu)\, \tilde g^\munu$,
then the equations of motion, given by the conservation of a {\em symmetric}
tensor $T^\munu\covderT{\mu} = 0$,
are invariant under the conformal transformation
$\tilde T^\munu = \Omega^{-6} \, T^\munu$ if the trace of $T^\munu$ is zero.
On the other hand, the covariant derivative of an {\em antisymmetric} tensor
can always be conformally transformed $\tilde F^\munu = \Omega^{-4} F^\munu = a^4 F^\munu$, as can be directly seen in \Eq{Max}.
For an expanding background, this allows us to map
the results in the FLRW $g^\munu$ and the Minkowski
$\eta^\munu = \tilde g^\munu = a^2 g^\munu$
metric tensors using $\Omega = a^{-1} (\tau)$.

Therefore, Maxwell equations in an expanding background
can be described as in Minkowski space-time after a conformal transformation \cite{Parker:1968mv},
\begin{equation}
    \partial_\nu \tilde F^{\munu} = \tilde J^{\mu}\,, \label{Fmunu_comoving}
\end{equation}
where $\tilde J^{\mu} = a^4 J^\mu$ is  the comoving current density. 
Furthermore, we will show in \Sec{MHD_eom} that the stress-energy tensor contributions from electromagnetism are traceless.
Therefore, as long as the trace of the stress-energy tensor associated to
the fluid vanishes, which occurs for radiation-dominated fluids with
vanishing bulk viscosity, then
the MHD equations are conformally flat \cite{Brandenburg:1996fc}.

From the conformal transformation $\tilde F^\munu = a^4 F^\munu$
and taking into account $\tilde U^\mu = a U^\mu = \gamma \, (1, \uu)$,
a consistent
comoving transformation of the covariant electromagnetic fields
description becomes $\tilde E^\mu = a^3 E^\mu$ and
$\tilde B^\mu = a^3 B^\mu$,
as already introduced in \Eq{comv_em}.

From these expressions, for steady comoving magnetic fields,
the physical components decay with the expansion of the Universe,
$B_i \sim a^{-1}$ and $B^i \sim a^{-3}$, such that
the energy density decays like radiation,
$\rho_B \sim B_i B^i \sim a^{-4}$ (see \Sec{Friedmann}).
As expected, the decay of the electromagnetic energy density
is absorbed in the comoving
magnetic fields, $\tilde B_i \tilde B^i = \tilde B^2 \sim a^0$.
These results are identical for electric fields.
We also note that the comoving electric and magnetic fields
can be expressed from the covariant
terms of the Faraday tensor $F_\munu$
and hence, the covariant gauge field
components $A_\mu$ are already comoving with the Universe expansion.
However, the comoving electromagnetic fields do not transform like tensors.

For a proper definition of the physical magnetic fields
in the rest frame,
the orthonormal basis of tetrads $e_{(i)}^\mu$ is introduced,
with the following properties
\begin{equation}
    g_\munu e^\mu_{(i)} e^\nu_{(j)} = \eta_{ij}\,, \qquad
    \eta^{ij} e^\mu_{(i)} e^\nu_{(j)} = g^\munu\,,
\end{equation}
such that the magnetic field can be expressed as $B^\mu = \bar B_i\, e^\mu_{(i)}$
\cite{Subramanian:2009fu,Durrer:2013pga}.
Then, one finds that $\bar B_0 = \bar B^0 = 0$ and the spatial components become
\begin{equation}
    \bar B_i = \bar B^i = a B^i \sim a^{-2}\,.
\end{equation}
The projection of $B^\mu$ along tetrads allows to properly define
the components of the comoving magnetic field
as $\tilde B_i = a^2 \bar B_i = a B_i$.
Then, the comoving magnetic energy density is
$\tilde \rho_B = a^4 \rho_B = \half\, a^4 \bar B^2 = \half \tilde B^2 \sim a^{0}$.

\subsection{Covariant generalized Ohm's law}
\label{cov_ohms}

\begin{figure}[t]
   \centering
    \includegraphics[width=0.9\textwidth]{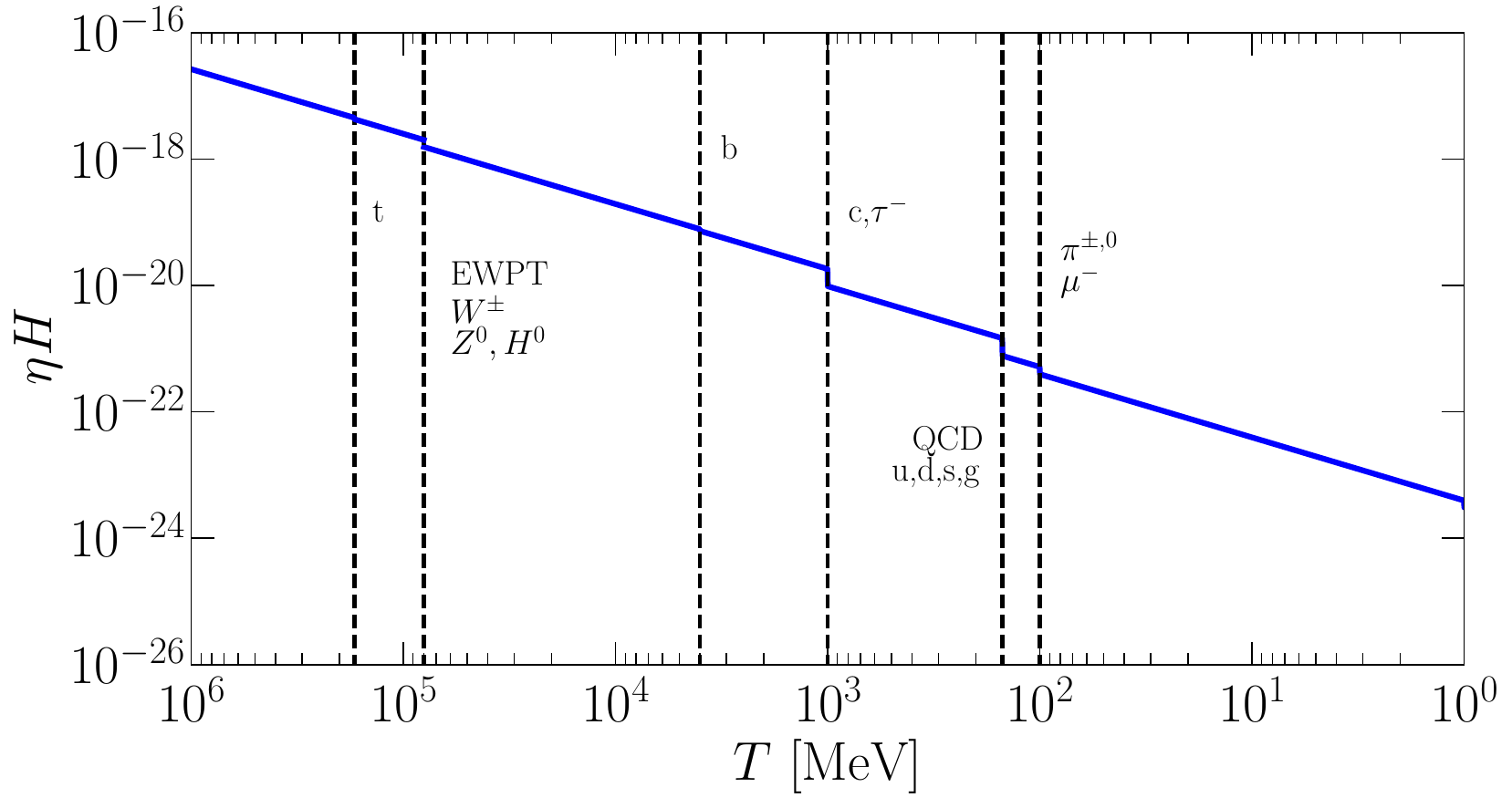}
    \caption{
    Ratio of the magnetic diffusivity $\eta \equiv 1/\sigma$ [cf.~\Eqs{electric_conductivity}{hypercharge_conductivity}] to the Hubble time $H^{-1}$ [cf.~\Eq{rho_rad}] at temperature scales above neutrino decoupling $T > T_\nu \sim 1$ MeV.
    The vertical lines are explained in the caption of \Fig{shear_viscosity}.
    }
    \label{conductivity}
\end{figure}

The current density induced by the charged particles of a fluid is described by the covariant
formulation of the generalized Ohm's law \cite{Landau1987Fluid,Subramanian:2009fu},
\begin{equation}
    J^\mu = \rho_e \, U^\mu + \sigma \, E^\mu\,, \label{Ohms_covariant}
\end{equation}
where $\rho_e$ and $\sigma$ are respectively the charge density and the conductivity as measured in the fluid reference frame.
The conductivity is indicated with $\sigma$, not to be confused with
the traceless rate-of-strain tensor
${\pmb \sigma}$ defined in \Sec{viscosity}.
In the radiation-dominated era, at high temperatures above the electron mass
$m_e \simeq 511$ keV, the leading-log form of the electrical conductivity is approximately given by \cite{Arnold:2000dr,Uchida:2024ude}
\begin{align}
    \sigma (100 \, {\rm GeV} > T >
    m_e) = C_{\rm cond} \, \frac{T}{{\rm e}^2 \ln {\rm e}^{-1}}\,, \qquad 
    \text{with \ }
    C_{\rm cond} \simeq \frac{12^4 \zeta(3)^2 \pi^{-3} N_{\rm leptons}}{3\pi^2 + 32 \, N_{\rm species}} \,,
    \label{electric_conductivity}
\end{align}
where e $ \simeq 0.3$ is the electromagnetic coupling constant,
$N_{\rm leptons}$ is the number of leptonic charge carriers and
$N_{\rm species}$ is the number of Dirac fermions weighted by the
square of their electric charge (see Table~4 in \cite{Uchida:2024ude}).
The value of $C_{\rm cond} \simeq 12$ at $T \simeq 100$ MeV and 7
at $T\simeq 100$ GeV.
The ratio of the electrical conductivity to the inverse
Hubble time is then extremely large in the radiation-dominated era,
\begin{equation}
    \frac{\sigma}{H} \simeq 5 \times 10^{17} \biggl( \frac{100 \, {\rm GeV}}{T} \biggr)
    \biggl(\frac{100}{g_\ast}\biggr)^{1/2}\,.
\end{equation}
We show in
\Fig{conductivity} the magnetic diffusivity, $\eta = 1/\sigma$, normalized by
the Hubble time $H^{-1}$
for the range $1 \, {\rm MeV} \leq T \leq 1 \, {\rm TeV}$.
Given the large values of the conductivity (small values of the diffusivity),
it is in general
justified considering the ideal MHD limit to study the interplay of fluid and magnetic perturbations at cosmological scales (see \Sec{induction_sec}).
Even at later times, i.e., at temperatures below the electron mass,
the conductivity of the Universe is still in general very large \cite{Ghosh:2025bqp}.
\EEq{electric_conductivity} is valid at temperatures below the EWPT, above which we need to consider the hypercharge conductivity \cite{Arnold:2000dr},
\begin{align}
    \sigma_{\rm hyper} = C_{\rm hyper} \, \frac{T}{g_{\rm hyper}^2 \ln g_{\rm hyper}^{-1}}\,, \qquad \text{with \ }
    C_{\rm hyper} = 6^4 \zeta(3)^2 \pi^{-3} \biggl( \frac{\pi^2}{8}+\frac{20}{3}+\frac{2}{3}n_s \biggr)^{-1} \,,
    \label{hypercharge_conductivity}
\end{align}
where $n_s$ is the number of Higgs doublets (i.e., $1$ in the Standard Model).

We note that the charge density and the electrical
conductivity are both Lorentz scalars obtained by projecting $J^\mu$
in the parallel and transverse directions with respect to $U^\mu$,
\begin{equation}
    \rho_e = - J_\mu U^\mu\,, \qquad \sigma \, E^\mu = J_\nu h^\munu\,.
\end{equation}
Notice that $U^\mu$ in the generalized Ohm's law corresponds to the fluid four-velocity.
Let us first consider a fluid at rest $U^\mu = (1, {\pmb 0})/a$, such that
the components of the comoving four-current are
\begin{equation}
    \tilde J^\mu = a^4 J^\mu = a^3 (\rho_e, \sigma E^i)
    = (\tilde \rho_e, \tilde \sigma \tEE)\,,
\end{equation}
where we have defined the comoving charge density and conductivity,
$\tilde \rho_e = a^3 \rho_e$ and $\tilde \sigma = a\, \sigma$, respectively.

%%%%%%%%%%%%%%%%%%%%%%%%%%%%%%%%
%%%%%%%%%%%%%%%%%%%%%%%%%%%%%%%%

For a fluid with a peculiar velocity $\uu$ with respect to the Hubble
observer,
$U^\mu = \gamma(1, u_i)/a$,
the components of the comoving four-current density are
\begin{equation}
    \tilde J^0 = \gamma (\tilde \rho_e + \tilde \sigma\, \uu \cdot \tEE)\,, \qquad \tilde J^i =
    \gamma (\tilde \rho_e \uu + \tilde \sigma[\tEE + \uu \times \tBB ])\,.  \label{Ohms_law}
\end{equation}
The temporal component can alternatively be expressed in terms of the current density,
\begin{equation}
    \tilde J^0 = \tilde \rho_e/\gamma +  \uu \cdot \tJJ\,.
\end{equation}
Note that in the subrelativistic limit $\uu \cdot \tEE \sim \uu \cdot \tJJ  \sim u^2 \ll 1$
and we recover $\tilde J^0 \to \tilde \rho_e$.
The current density components
will enter Maxwell equations in \Eq{Fmunu_comoving} and the
fluid equations via the Lorentz force
when we consider charged fluids with
peculiar velocities $\uu$ with respect to the Hubble observer
in \Sec{MHD_eom}.
The electric field can be expressed from the current density
inverting \Eq{Ohms_law},
\begin{equation}
    \tEE = \frac{\tilde \eta}{\gamma} \, \tJJ - \tilde \eta \,
    \tilde \rho_e \,
    \uu - \uu \times \tBB\,, \label{E_Ohms}
\end{equation}
where $\eta = \sigma^{-1}$ is the magnetic diffusivity
(see \Fig{conductivity}) and $\tilde \eta = a^{-1}
\eta$ is the comoving diffusivity.
This second form of the generalized Ohm's law will become relevant
when studying MHD in the limit of large conductivity (see \Sec{induction_sec}) as this assumption will allow us to close
the system of MHD equations after expressing the electric field
as a function of the current density.

\subsection{Maxwell equations in an expanding Universe}
\label{maxwell_eqs}

We now present the Maxwell equations, with the electric
and magnetic fields computed in the reference
frame of the Hubble observer\footnote{For a generic observer, Maxwell equations in a curved
space-time are presented, e.g., in \cite{Subramanian:2009fu} 
or App.~B of \cite{Durrer:2013pga}.} with $U^\mu = (1, {\pmb 0})/a$
in \Eq{comv_em}.
Let us start with Gauss' law, found when $\mu = 0$ in \Eq{Fmunu_comoving},
\begin{equation}
    \partial_i \tilde F^{0i} = \nab \cdot \tEE = \tilde J^0 \,,
\end{equation}
where $\tilde E_i = F_{i0}$ and $\tilde J^0$ corresponds to the comoving charge density
of a fluid with peculiar velocity $\uu$
measured by a Hubble observer, given in \Eq{Ohms_law}.

Gauss' law provides a constraint on the electric field and, when expressed in terms
of the gauge field $A_\mu = (\chi, A_i)$, it yields an evolution
equation for $\Gamma = \nab \cdot \AAA$,
\begin{equation}
    \partial_\tau \Gamma = \nabla^2 \chi - \tilde J^0\,.
\end{equation}
Amp\`ere's law is found from \Eq{Fmunu_comoving} taking $\mu = i$,
\begin{equation}
    \partial_\tau \tilde F^{i0} + \partial_j \tilde F^{ij} = \tilde J^i
    \Rightarrow \partial_\tau \tEE = \nab \times \tBB - \tJJ \,,
    \label{Faraday1}
\end{equation}
where $\tilde B^i = \varepsilon^{ijk} F_{jk}$.
Taking the divergence of Amp\`ere's law,
we find the charge conservation equation,
\begin{equation}
    \partial_\mu \tilde J^\mu = \partial_\tau \tilde J^0 +
    \partial_i \tilde J^i = 0\,,
\end{equation}
which is a direct consequence of Maxwell equations and the antisymmetry of the Faraday
tensor: $\partial_\mu \tilde J^\mu = \partial_{\mu}\partial_{\nu} \tilde F^{\munu} = 0$.
In terms of the gauge field components $A_\mu$, we find a second-order differential
equation in time for $A_i$,
\begin{equation}
    \partial_\tau^2 \AAA - \partial_\tau \nab \chi + \nab
    \Gamma - \nab^2 
    \AAA = \tJJ\,,
\end{equation}
which can be solved for a particular gauge choice $\chi = \chi (A_i)$.
We note that this equation can be obtained directly from \Eq{Faraday1}
from the Faraday tensor components,
\begin{equation}
    -\partial_\tau F_{i0} + \partial_j F_{ij} = 
    \partial_\tau^2 A_i - \partial_\tau \partial_i A_0 + 
    \partial_j \partial_i A_j - \partial_j \partial_j A_i\,,
\end{equation}
or
taking into account that $\tilde B_i = \nab \times \AAA$
and using the following vectorial identity in \Eq{Faraday1},
\begin{equation}
    \nab \times (\nab \times {\pmb A}) = - \nabla^2 \AAA + \nab \Gamma\,.
    \label{double_vector}
\end{equation}

At this stage, although not necessary,
it is useful to introduce the remaining Maxwell equations using
Bianchi identities.
For this purpose, we note that any Faraday tensor described in terms of a gauge
field $A_\mu$ satisfies the Bianchi identities,
\begin{equation}
    F_{\munu\,; \sigma} +
    F_{\sigmu\,; \nu} + F_{\nusig\,; \mu} = 0\,.
\end{equation}
Contraction of these identities with $\half {\cal E}^{\mu \alpha \nu \sigma}$ yields a conservation law
for the dual Faraday tensor,
\begin{equation}
    {\cal F}^\munu\covderT{\mu} = 0\,.
\end{equation}
Noting that ${\cal F}^\munu$ is equivalent to $F^\munu$ under the transformation
$E^\mu \to -B^\mu$ and $B^\mu \to E^\mu$,
then the Gauss' law for magnetic fields and
the Faraday's law can be obtained under this transformation,
setting the magnetic sources to zero,
\begin{equation}
    \nab \cdot \tBB = 0\,, \qquad \partial_\tau \tBB = - \nab \times \tEE\,.
    \label{Faraday}
\end{equation}
We note that these laws are obtained due to the invariance of the Faraday tensor over
coordinate transformations and they do not add new dynamics.
Indeed, we find that these equations are trivially satisfied
when we introduce the gauge field $A_\mu$ and its relation
to the electromagnetic fields \Eq{EB_subrel},
\begin{equation}
    \nab \cdot \tBB = \nab \cdot (\nab \times \AAA) = 0\,, \qquad
    \partial_\tau \tBB = \nab \times \nab \chi - \nab \times \tEE = - \nab \times \tEE \,.
\end{equation}

In summary, one can describe the dynamical evolution of Maxwell equations evolving the
electric and the gauge fields, together with a Gauss constraint,
\begin{equation}
    \partial_\tau \tEE = - \nab^2 \AAA + \nab \Gamma - \tJJ\,, \qquad
    \partial_\tau \AAA = \nab \chi - \tEE\,, \qquad \nab \cdot \tEE = \tilde J^0\,,
\end{equation}
where the magnetic Gauss constraint is automatically satisfied.
Otherwise, an equivalent system can be solved in terms of the electric
and magnetic fields,
\begin{equation}
    \partial_\tau \tEE = \nab \times \tBB - \tJJ \,, \qquad
    \partial_\tau \tBB = - \nab \times \tEE\,, \qquad \nab \cdot \tEE = \tilde J^0\,, \qquad 
    \nab \cdot \tBB = 0\,.
\end{equation}
This system of equations is then closed using \Eq{Ohms_law} for the components of the
four-current $\tilde J^\mu$, as long as we can describe the evolution
of the peculiar velocity.

\subsection{Displacement current and magnetic induction equation}
\label{induction_sec}

The conductivity of the primordial plasma became
very large after the period of inflation, during the radiation-dominated epoch
of the early Universe \cite{Arnold:2000dr,Durrer:2013pga}
(see \Fig{conductivity}).
In this limit, it is possible to neglect the displacement current $\partial_\tau \tEE$ and the charge density $\tilde \rho_e$
with respect to the conductivity, which is a common assumption
made in MHD \cite{Brandenburg:1996fc,Biskamp2003,Brandenburg:2004jv,Durrer:2013pga,Shukurov_Subramanian_2021}.
This can be understood introducing Ohm's law in Amp\`ere's law,
\begin{equation}
    (\partial_\tau + \gamma \tilde
    \sigma) \tEE + \gamma \tilde \rho_e \uu = \nab \times \tBB -
    \gamma \tilde \sigma \uu \times \tBB\,.
\end{equation}
Comparing the two terms in brackets, we find that the displacement
current can be neglected when the time scale of the electric field
oscillations is much larger than the Faraday time
$\tau_{\rm Far} = \tilde \eta/\gamma$ \cite{Brandenburg:2004jv}.
Under this assumption, the current density can be directly
obtained from the magnetic field using Amp\`ere's law,
\begin{equation}
    \tJJ = \nab \times \tBB \,. \label{current}
\end{equation}
We note that this is valid in the large conductivity limit unless the frequency
of the electric field oscillations becomes very large.
In this limit, Amp\`ere's law [cf.~\Eq{current}] becomes a constraint equation
and the
remaining dynamical equation can be obtained introducing  Ohm's law [cf.~\Eq{E_Ohms}]
in Faraday's law [cf.~\Eq{Faraday}],
\begin{equation}
    \partial_\tau \tBB = - \nab \times \tEE = \nab \times \bigl(\uu \times \tBB - \tilde \eta \tJJ/\gamma + \tilde \eta \, \tilde \rho_e \uu
    \bigr)\,.
\end{equation}
This equation is commonly known as the induction equation in MHD and using $\tJJ = \nab \times \tBB$
gives an evolution equation for $\tBB$ that only depends on $\tBB$ and the peculiar velocity $\uu$,
\begin{equation}
    \partial_\tau \tBB = \nab \times (\uu \times \tBB) + \frac{\tilde \eta}
    {\gamma} \, \nab^2 \tBB - \nab \times [\tBB \times \nab(\tilde \eta
    / \gamma)] + \nab \times (\tilde \eta \, \tilde \rho_e  \uu)\,.
    \label{rel_induction}
\end{equation}
In terms of the vector potential, the induction equation is
\begin{equation}
    D_\tau \AAA = \uu \cdot (\nab \AAA) + \frac{\tilde \eta}{\gamma}
    \, (\nab^2 \AAA - \nab \Gamma) + \tilde \eta \, \tilde \rho_e
    \uu \,.
\end{equation}

In the subrelativistic limit ($\gamma \to 1$),
for neutral plasmas ($\tilde \rho_e = 0$) and homogeneous $\tilde \eta$,
the induction equation for $\tBB$ simplifies to
\begin{equation}
    \lim_{\gamma \to 1} \partial_\tau \tBB = \nab \times (\uu \times \tBB) +
    \tilde \eta \, \nab^2 \tBB\,. \label{induc1}
\end{equation}
Expanding the curl in \Eq{induc1}, the induction equation can be expressed as
\begin{equation}
   \lim_{\gamma \to 1} D_\tau \tBB = (\tBB \cdot \nab) \uu - \tBB (\nab \cdot\uu) + \tilde \eta \nab^2 \tBB\,.
\end{equation}

\section{Relativistic MHD in an expanding Universe}
\label{MHD_eom}

In \Sec{perfect_fluid_hydro}, the equations of motion of a
fluid that does not interact with gauge fields have been presented.
For a single-fluid description of charged species, the
equations of motion become coupled to Maxwell equations, studied in an expanding background in
\Sec{Maxwell_sec}, leading to the MHD equations.

In this section, the MHD equations are described in an expanding homogeneous and isotropic FLRW background, and the
results are extended with respect to previous work
to relativistic fluid peculiar
velocities, presenting the set of equations for 
the stress-energy tensor components $\tilde T^{0\mu}$ (conservation form)
and for the fluid primitive variables
$\tilde \rho$ and $\uu$ (non-conservation form), together with the electromagnetic fields $\tEE$
and $\tBB$.
As in previous sections, we consider $X^0 = \tau$, $m = 1$
(space-like signature), and $\alpha = 1$ in this section,
for compactness.

In previous numerical work using the {\sc Pencil Code} (cf.~\cite{Brandenburg:1996fc,Brandenburg:2017neh,RoperPol:2019wvy}), the MHD equations were considered for a
radiation-dominated fluid with $\cs^2 = \third$,
such that the system of equations becomes
conformally flat (see discussion in \Sec{cons_perf_fluid}).
The equations were considered
in the subrelativistic limit of fluid
perturbations in the following form,
\begin{subequations}
\label{old_eqs_MHD}
\begin{align}  
    \partial_\tau \ln \tilde \rho = -  \frac{4}{3} \bigl[&\nab \cdot \uu
    + (\uu \cdot \nab) \ln \tilde \rho\bigr] + \frac{1}{\tilde \rho}
    \bigl[\tilde \eta \tJJ^2 + \uu \cdot (\tJJ \times \tBB)
    \bigr]\,, \label{old_cont_MHD} \\ 
    \label{old_mom_MHD}
    \partial_\tau \uu + (\uu \cdot \nab) \, \uu =
    \frac{\uu}{3}
    \bigl[& \nab \cdot \uu + (\uu \cdot \nab) \ln \tilde \rho \bigr]
    - \frac{\uu}{\tilde \rho} \bigl[\tilde \eta \tJJ^2  + 
    \uu \cdot (\tJJ \times \tBB) \bigr] - \frac{1}{4} \nab \ln \tilde \rho  \nonumber \\   + &\,
    \frac{3}{4 \tilde \rho} \tJJ \times \tBB + \frac{2}{\tilde \rho} \,
    \nab \cdot (\tilde \rho \, \tilde \nu \, \pmb{\tilde \sigma} )\,,
\end{align}
\end{subequations}
where $\tff_\ipf = \fourthird \,
\nab \cdot (2\, \tilde \rho \, \tilde \nu \,
\pmb{\tilde \sigma})$ corresponds to the subrelativistic Navier-Stokes description
of the viscosity with zero bulk viscosity,
as discussed in \Sec{viscosity} [cf. \Eq{mom_subrelativistic_visc}].

It has been already shown in \Secs{conservation_rel_perf}{rel_hydrodynamics} that this system of equations is missing
a few corrections in the purely fluid limit (i.e., in the
absence of electromagnetic fields) due to neglecting $\partial_\tau \gamma^2$, and the system of equations has been
extended to the fully relativistic regime in \Eqs{cont_relativistic0}{mom_relativistic0}.
Taking into account these corrections,
we find in this section the following subrelativistic
MHD equations (for $\cs^2 = \third$) [cf.~\Eqq{eqs_cont2}],
where the corrections with respect to \Eqq{old_eqs_MHD} are indicated in red
\begin{subequations}
\label{corr_eqs_MHD}
\begin{align}
    \partial_\tau \ln \tilde \rho =  -
     \frac{4}{3}
    \bigl[&\nab \cdot \uu + \red{\bf \half} (\uu \cdot \nab) \ln \tilde \rho\bigr] + \frac{1}{\tilde \rho} \bigl[
    \tilde \eta \tJJ^2 \red{\bf - } \, \uu \cdot (\tJJ \times
    \tBB) \bigr]\,, \label{corr_cont_MHD} \\
    \partial_\tau \uu + (\uu \cdot \nab) \, \uu =
    \frac{\uu}{3} \bigl[& \nab \cdot \uu + \red{\bf \half} (\uu \cdot \nab) \ln \tilde
    \rho \bigr] - \frac{\uu}{\tilde \rho} \bigl[\tilde \eta \tJJ^2  + \red{\bf \half}\,
    \uu \cdot (\tJJ \times \tBB) \bigr] - \frac{1}{4} \nab \ln \tilde \rho \nonumber \\
     + &\,
    \frac{3}{4 \tilde \rho} \tJJ \times \tBB + \frac{2}{\tilde \rho} \,
    \nab \cdot (\tilde \rho \, \tilde \nu \, \pmb{\tilde \sigma}) \,.  \label{corr_mom_MHD}
\end{align}
\end{subequations}
The work and energy dissipation due to viscous stresses, $\uu \cdot \tff_\ipf$,
and thermal conductivity, $\tilde f_\ipf^0$ 
[cf.~\Eqq{eqs_subrelativistic_visc}],
have been neglected
under the assumption of small shear and bulk viscosities,
and small thermal conductivity (see \Sec{transport_coeffs}).
However, we include these contributions in the form of $\tilde f^\mu_\ipf$,
discussed in \Sec{viscosity}, in this section for generality.
Furthermore, these equations are extended to include Hubble friction
terms for deviations of 
a constant $\cs^2$ with respect to $\third$ in this work.
In the following, we also extend the MHD equations including electromagnetic
stresses in the equations of motion in the fully relativistic regime
as in \Secs{conservation_rel_perf}{rel_hydrodynamics}.

The Maxwell equations discussed in \Sec{Maxwell_sec}
become coupled to the fluid equations
of motion via the Lorentz force $\tilde f^\mu_\Lor$ as it is discussed
in \Sec{four_lorentz}.
When the displacement current can be neglected in the limit
of large conductivity (see \Sec{induction_sec}), the current density is $\tJJ = \nab \times \tBB$, and Maxwell equations can be combined
with Ohm's law (see \Sec{cov_ohms})
into the induction equation [cf.~\Eq{rel_induction}],
\begin{equation}
    D_\tau \tBB = (\tBB \cdot \nab) \uu - \tBB (\nab \cdot \uu) + \frac{\tilde \eta}{\gamma} \nab^2 
    \tBB - \nab \times \bigl[\tBB \times \nab(\tilde \eta/\gamma) + \tilde \eta \tilde \rho_e \uu\bigr]\,. \label{induc2}
\end{equation}
On the other hand, when the displacement current cannot be neglected,
e.g., due to fast electromagnetic oscillations,
the fluid equations need to be solved together with Maxwell equations,
\begin{equation}
    \partial_\tau \tBB = - \nab \times \tEE\,, \qquad \partial_\tau \tEE = 
    \nab \times \tBB - \tJJ\,, \qquad \nab \cdot \tEE = \tilde J^0\,, \qquad \nab \cdot \tBB = 0\,,
\end{equation}
and the generalized Ohm's law [cf.~\Eq{Ohms_law}], which describes the
space-time components of $\tilde J^\mu$.

\subsection{Electromagnetic stress-energy tensor}
\label{sec:Tmunu}

The stress-energy tensor from electromagnetism can be directly computed from the Lagrangian taking
into account that Einstein equation is obtained from the Einstein-Hilbert action,
\begin{equation}
    T_\munu = g_\munu \LL - 2 \frac{\delta \LL}{\delta g^\munu}\,,
\end{equation}
where the Lagrangian of electromagnetism is given in \Eq{Lag_EM},
leading to the following stress-energy tensor,
\begin{equation}
    T_\munu^{\rm EM} = F_\mu^{\  \sigma} F_{\nu \sigma} -
    \fourth \, g_\munu F^{\lamsig} F_{\lamsig}
    \,.
\end{equation}
The electromagnetic stress-energy tensor
is traceless by construction,
\begin{equation}
    T^{\rm EM} = {T^\mu_{\ \mu}}^{\rm EM} = F^{\musig} F_\musig - \fourth
    g^\mu_{\ \mu} F^\lamsig F_\lamsig = 0\,.
\end{equation}
Therefore, the equations of motion in an expanding FLRW Universe are conformally flat
when the fluid is radiation-dominated with $\cs^2 = \third$ and the bulk
viscosity $\zeta_\visc$ is zero,
making the trace of the total stress-energy tensor to vanish,
as shown in \Eq{cons_expand}.
The comoving electromagnetic stress-energy tensor is
\begin{equation}
    \tilde T^\munu_{\rm EM} = a^6 T^\munu_{\rm EM} 
    = \tilde F^{\mu \sigma} \tilde F^\nu_{\ \, \sigma} - \fourth \, \eta^\munu
    \tilde F^\lamsig F_\lamsig\,,
\end{equation}
where $\tilde F^{\munu} = a^4 F^\munu$ and $\tilde F^\mu_{\ \, \nu} = a^2 F^\mu_{\ \, \nu}$
are the comoving components of the Faraday tensor.
The contraction of the Faraday tensor with itself is
\begin{equation}
    \tilde F^{\lamsig} F_\lamsig = 2 \tilde F^{0i} F_{0i} +
    \tilde F^{ij} F_{ij} = -2 \, \tEE^2 + \varepsilon_{ijl}
    \varepsilon_{ijk} \tilde B_k \tilde B_l = - 2 \, 
    (\tEE^2 - \tBB^2)\,.
\end{equation}
Note that we omit a tilde for the $F_{\mu\nu}$
components since they already correspond to the comoving fields
without rescaling.
The comoving components of the electromagnetic stress-energy
tensor are
\begin{subequations}
\label{components_Tmunu_EM}
\begin{align}
    \tilde T^{00}_{\rm EM} = &\, \tilde F^{0i} \tilde F^{0}_{\ \, i} + \fourth
    \tilde F^\lamsig F_\lamsig
    = \half (\tEE^2 + \tBB^2)\,, \\  
    \tilde T^{0i}_{\rm EM} = & - \tilde F^{0 j} \tilde F^{i}_{\ \, j}
    = \tEE \times \tBB \,, \\  
    \tilde T^{ij}_{\rm EM} = &\, \tilde F^{i0} \tilde F^{j}_{\ \, 0} + 
    \tilde F^{il} \tilde F^{j}_{\ \, l} - \fourth \tilde F^\lamsig F_\lamsig
    \, \delta^{ij} = - \tilde E^i \tilde E^j - \tilde B^i \tilde B^j
    + \tilde T^{00}_{\rm EM}  \delta^{ij} \,,
\end{align}
\end{subequations}
corresponding $\tilde T^{00}_{\rm EM}$ to the electromagnetic energy density,
$\tilde T^{0i}_{\rm EM}$ to the Poynting vector, and
$\tilde T^{ij}_{\rm EM}$ to the electromagnetic
stresses.

The equations of motion are then equivalent to those in \Eqs{cont_aux_visc}{mom_aux_visc}, or \Eq{mom_aux} for perfect
fluids, after including the electromagnetic stress-energy tensor
components,
\begin{equation}
    \partial_\mu \tilde T^\munu = \partial_\tau \tilde T^{0\nu} + \partial_j \tilde T^{j\nu} = \tilde f_H^\nu
    + \tilde f_\ipf^\nu \,,
    \label{mom_aux_MHD}
\end{equation}
where $\tilde T^\munu = \tilde T^\munu_{\rm pf} + 
\tilde T^\munu_{\rm EM}$, being $\tilde T^\munu_{\rm pf} = \tilde \rho \tilde U^\mu \tilde U^\nu
+ \tilde p \tilde h^\munu$ the stress-energy
tensor of a perfect fluid.
We note that, as before, the equations of motion are conformally flat with
$\tilde f_H^\mu = 0$ when
$\beta = 4$, $\alpha = 1$,
$\tilde \rho = 3 \tilde p$, and $\tilde \zeta_\visc = 0$.

\subsection{Lorentz force in covariant formulation}
\label{four_lorentz}

The electromagnetic contribution to the conservation laws in an expanding
background can be recast in the form of a four-vector, corresponding to the
Lorentz force:
\begin{equation}
    \partial_\mu \tilde T^{\munu}_{\rm EM} = - \tilde f^\nu_{\rm Lor} = - \tilde J_\mu \tilde F^\numu\,, \label{Lor_force}
\end{equation}
where we have defined the comoving
Lorentz force $\tilde f^\mu_{\rm Lor} = a^6 f^\mu_{\rm Lor} = \tilde J_\nu
\tilde F^\munu$.
To prove that the Lorentz force is equivalent to
$- \partial^\mu \tilde T_\munu^{\rm EM}$, the following relation
needs to be shown,
\begin{equation}
    - \partial_\mu \tilde T^\munu_{\rm EM} = - \partial_\mu (
    \tilde F^\musig \tilde F^{\nu}_{\ \, \sigma}) - \fourth \,
    \eta^\munu \partial_\mu (\tilde F^\lamsig F_\lamsig ) = -
    \tilde F^\nu_{\ \, \sigma} \,
    \partial_\mu \tilde F^\musig = \tilde F^\nu_{\ \, \sigma}
    \tilde J^\sigma\,, \label{aux5}
\end{equation}
where we have introduced Maxwell equations $\partial_\mu \tilde F^\musig = - \tilde J^\sigma$.
The following relation implies the validity of \Eq{aux5},
\begin{equation}
    \tilde F^\musig \partial_\mu F_\nusig =
    \fourth \partial_\nu (\tilde F^\lamsig F_\lamsig)
    = \half \tilde F^\musig \partial_\nu F_\musig\,. \label{aux4}
\end{equation}
To prove this relation, and hence \Eq{aux5}, we first exploit the antisymmetry of the
Faraday tensor to write
\begin{equation}
    \tilde F^\musig (\partial_\mu F_\nusig - \half \partial_\nu F_\musig
    ) = \half \tilde F^\musig (\partial_\mu F_\nusig - \partial_\mu
    F_\signu + \partial_\nu F_\sigmu)\,.
\end{equation}
Then, we can apply Bianchi identities $\partial_\mu F_\nusig + \partial_\nu F_\sigmu + \partial_\sigma F_\munu = 0$ to
prove \Eq{aux4},
\begin{equation}
    \tilde F^\musig \bigl(\partial_\mu F_\nusig + \half \partial_\nu F_\musig
    \bigr) = - \half \tilde F^\musig \bigl(\partial_\mu F_\signu + \partial_\sigma F_\munu\bigr) = 0\,,
\end{equation}
which is zero as it corresponds to the product of a symmetric and
an antisymmetric tensor.
Therefore, \Eq{Lor_force} is proven. $\square$

The components of the Lorentz force are
\begin{equation}
\tilde f^0_\Lor = \tilde J_i \tilde F^{0i} = \tEE \cdot \tJJ\,,
\qquad
\tilde f^i_\Lor = \tilde J_0 \tilde F^{i0} + \tilde J_j \tilde F^{ij} 
= \tilde J^0 \tEE + \tJJ \times \tBB\,. \label{lorentz_comps}
\end{equation}
Using the covariant generalized Ohm's law given in \Eq{Ohms_covariant}, the Lorentz
force can also be expressed as
\begin{equation}
    \tilde f^\mu_\Lor = \tilde J_\nu \tilde F^{\munu} =
    \tilde \rho_e
    \tilde E^\mu + \tilde \sigma \tilde E_\nu \tilde F^\munu = \tilde E_\nu \,
    (\tilde \rho_e \eta^\munu + \tilde \sigma \tilde F^\munu) \,.
\end{equation}
In particular, the temporal component of the Lorentz force becomes
\begin{equation}
    \tilde f_\Lor^0 = \frac{\tilde \eta}{\gamma} \, \tJJ^2 - \tilde \eta
    \,
    \tilde \rho_e \, \uu \cdot \tJJ + \uu \cdot (\tJJ \times \tBB)\,,
    \label{f0_MHD_rel}
\end{equation}
when expressed as a function of the current density, and
\begin{equation}
    \tilde f_\Lor^0 = \gamma \tilde \rho_e \uu \cdot \tEE + \gamma \tilde \sigma \tEE^2 - \gamma \tilde \sigma u_i \tilde {T}^{0i}_{\rm EM} \,,
    \label{f0_MHD_rel2}
\end{equation}
when expressed as a function of the electric field and the Poynting vector $\tilde T^{0i}_{\rm EM} = (\tEE \times \tBB)^i$.
In the subrelativistic limit, and assuming a neutral plasma ($\tilde \rho_e = 0$), it reduces to
\begin{equation}
    \lim_{u^2 \ll 1} \tilde f_\Lor^0 = \tilde \eta \JJ^2 + \uu \cdot (\tJJ \times \tBB) = \tilde \sigma \tEE^2 - \tilde \sigma
   u_i \tilde T^{0i}_{\rm EM}\,.
    \label{f0_MHD}
\end{equation}
On the other hand, the space components of the Lorentz force,
applying Ohm's law, become
\begin{equation} \label{fi_MHD_rel}
    \tff_\Lor = \biggl(\frac{\tilde \rho_e}{\gamma} + \uu \cdot 
    \tJJ \biggr) \biggl(\frac{\tilde \eta}{\gamma} \tJJ - \tilde \eta 
    \tilde \rho_e \uu - \uu \times \tBB \biggr) + \tJJ \times \tBB\,,
\end{equation}
expressed as a function of the current density and
\begin{equation}
    \tff_\Lor = \gamma \tilde \rho_e (\tEE + \uu \times \tBB)
    + \gamma \tilde \sigma \bigl[(\uu \cdot 
    \tEE) \tEE + \tEE \times \tBB - \tBB^2 \uu + (\uu \cdot \tBB) \tBB\bigr]\,,
\end{equation}
as a function of the electric field.
Again, in the subrelativistic limit, and assuming a neutral plasma, we
find
\begin{equation}
    \lim_{u^2 \ll 1} \tff_\Lor = \tJJ \times \tBB = \tilde \sigma
    \bigl[\tEE \times \tBB - \tBB^2 \uu + (\uu \cdot \tBB)
    \bigr]\tBB\,,
\end{equation}
where we have assumed that $|\tJJ| \sim |\tEE| \sim {\cal O} (u)$ and
omitted terms of order ${\cal O} (u^3)$ and higher.
In this limit, $\tilde f_\Lor^0 = \tilde \eta \tJJ^2 + \uu \cdot \tff_\Lor$.
As in the case of imperfect fluids with viscous forces [cf.~\Eq{mom_aux_visc}], the temporal
component of the Lorentz force $\tilde f_\Lor^0$ corresponds to irreversible energy
dissipation, Joule heating, and includes irreversible work done by the Lorentz force
converting electromagnetic into thermal and kinetic
energy respectively [cf.~\Eq{corr_cont_MHD}].
This is due to the fact that the electromagnetic stress-energy tensor cannot be
recast in the form of a perfect fluid, as can be seen from the off-diagonal non-zero
components in \Eqq{components_Tmunu_EM}.
Hence, the stress-energy tensor adds a contribution to the deviatoric
stress tensor described in \Sec{viscosity} that leads to the production of entropy.
At the same time, this implies that net electromagnetic fields at the largest
scales of the Universe are required to vanish, to avoid breaking down the assumptions
of homogeneity and isotropy described in \Sec{FLRW}.
Note that this does not imply that their squared fluctuations, which can be
correlated at large scales, are necessarily vanishing.

Note that we can alternatively show the relation $\partial_\mu \tilde T^{\mu \nu}_{\rm EM} = - \tilde f_\Lor^\nu$
for each component
using Maxwell equations, $\partial_\tau \tEE - \nab \times \tBB = - \tJJ$ and $\partial_\tau \tBB + \nab \times \tEE = \pmb{0}$.
These relations will be useful in the following.
Let us start with the temporal component,
\begin{align}
    \partial_\tau \tilde T^{00}_{\rm EM} + \partial_i \tilde T^{0i}_{\rm EM} = &\, \half \partial_\tau \bigl(\tEE^2 + \tBB^2 \bigr)
    + \varepsilon_{ijl}\, \partial^i (\tilde E^j \tilde B^l)
    \nonumber \\ = &\, \tEE \cdot \bigl(\partial_\tau \tEE
    - \nab \times \tBB\bigr)
    + \tBB \cdot \bigl(\partial_\tau \tBB + \nab \times \tEE \bigr)
    =
     - \tEE \cdot \tJJ\,. \hspace{23mm} \blacksquare
\end{align}
This expression corresponds to Poynting's theorem, describing the evolution
of electromagnetic energy density.
For the spatial components, let us first compute the time derivative
of the Poynting flux
\begin{align}
    \partial_\tau \tilde T^{0i}_{\rm EM} = &\, \partial_\tau (\tEE \times \tBB)^i
    = \varepsilon^{ijl} \, \tilde B_l \, \partial_\tau \tilde E_j
    + \varepsilon^{ijl} \, \tilde E_j \, \partial_\tau \tilde B_l \nonumber \\
    = &\, \tilde B^j \, \partial_j \tilde B^i -
    \tilde B^j \, \partial^i \tilde B_j - \tilde E^j \, \partial^i \tilde E_j + \tilde E^j \, \partial_j \tilde E^i - (\tJJ \times \tBB)^i\,.
\end{align}
Then, $\partial_\mu \tilde T^{\mu i}_{\rm EM}$ becomes
\begin{align}
    \partial_\tau \tilde T^{0i}_{\rm EM} + \partial_j \tilde T^{ij}_{\rm EM} = &\, \partial_\tau (\tEE \times \tBB)^i -
    \partial_j (\tilde E^i \tilde E^j) - \partial_j (\tilde B^i \tilde B^j) + \half \delta^{ij}  \, \partial_i (\tEE^2 + \tBB^2) \nonumber \\ = &
    - \tilde{E}^i \, (\nab   \cdot \tEE)  - \tilde{B}^i \, (\nab \cdot \tBB) -
    (\tJJ \times \tBB)^i =
    - \tilde f^{i}_\Lor \,,  \hspace{30mm}  \blacksquare \label{fi_dmu}
\end{align}
where we have used Gauss laws, $\nab \cdot \tEE = \tilde J^0$ and
$\nab \cdot \tBB = 0$.

\subsection{Conservation form of relativistic magnetohydrodynamics}
\label{conservation_MHD}

The MHD equations of motion given in \Eq{mom_aux_MHD} already constitute
a solvable system expressed in the conservation form.
The contribution to the stress-energy tensor from electromagnetic fields
can either be included in the dynamical variables $\tilde T^{0\mu}$ solved in
the conservation form or be added as an external force $\tilde f^\mu_\Lor$,
as done in \Sec{conservation_rel_perf} for perfect fluids
and in \Sec{viscosity} for imperfect fluids including Navier-Stokes viscosity
and Fourier's thermal conductivity.
In the latter case, the procedure is indentical to that described in
\Sec{conservation_rel_perf}, and the Lorentz force is computed using
\Eq{lorentz_comps} after solving Maxwell equations.

Alternatively, if we solve for $\tilde T^{0\mu} = \tilde T^{0\mu}_{\rm pf} + \tilde T^{0\mu}_{\rm EM}$ as the dynamic variables, we need to express
$\tilde T^{ij}$ as a function
of $\tilde T^{0\mu}$, as done in \Sec{conservation_rel_perf}
for the perfect fluid system.
As we know the relation between the ratio $r^2$ and the Lorentz factor
in terms of $\tilde T^{0 \mu}$ of the perfect fluid,
given in \Eq{r2_vs_gamma},
one finds
\begin{equation}
    r^2 = \frac{\tilde T^{0i}_{\rm pf}\, \tilde T^{0i}_{\rm pf}}{\bigl(\tilde T^{00}_{\rm pf}\bigr)^2}
    = \frac{(\tilde T^{0i} - \tEE \times \tBB) \,
    (\tilde T^{0i} - \tEE \times \tBB)}{(\tilde T^{00} - \tilde T^{00}_{\rm EM})^2}  = \frac{\gamma^2 (\gamma^2 - 1)}{\Bigl(\gamma^2 - \frac{\cs^2}{1 + \cs^2}\Bigr)^2}\,,
\end{equation}
where the electromagnetic energy density is
$\tilde T^{00}_{\rm EM} = \half (\tilde E^2 + \tilde B^2)$,
and we have used an equation of state such that  $\tilde p
= \cs^2 \tilde \rho$.
Hence, once the electromagnetic fields are known, $r^2$ can be computed,
and then one can
solve for the Lorentz factor using \Eq{gamma2_r2}.
Finally, $\tilde T^{ij}$ can be computed in the following way
\begin{equation}
    \tilde T^{ij} = \tilde T^{ij}_{\rm pf} + \tilde T^{ij}_{\rm EM}
    =
    \bigl[(1 + \cs^2) \gamma^2 u^i u^j + \cs^2 \delta^{ij}\bigr] \tilde \rho
    + \tilde T^{ij}_{\rm EM}\,,
\end{equation}
with
\begin{equation}
    \tilde \rho = 
    \frac{\tilde T^{00} - \tilde T^{00}_{\rm EM}}{(1 + \cs^2)\gamma^2 - \cs^2}\,, \qquad 
    u^i =
    \frac{\tilde T^{0i} - (\tEE \times \tBB)^i}{(1 + \cs^2) \tilde \rho \gamma^2}\,.
\end{equation}
This procedure allows us to compute the fully relativistic system of MHD
equations in their conservation form in an alternative way (compared to the
one in which electromagnetic contributions are included via the Lorentz
force).
Note that this procedure is generic for any contribution to $\tilde T^{ij}$
in addition to that of a perfect fluid that does not depend on the
fluid variables.

The equations of motion of the stress-energy tensor components
are coupled to Maxwell equations
and the generalized Ohm's law, which have been described in
\Sec{Maxwell_sec}.
We then need to solve the MHD equations coupled to the
Maxwell equations, as given in
\Sec{maxwell_eqs},
to evolve the electric and magnetic fields,
together with the equations of motion of $\tilde T^{\mu 0}$ or $\tilde T^{\mu 0}_{\rm pf}$.
One also needs to use
Ohm's law to decribe the components of the current
density, given in \Eq{Ohms_law}, and to compute the components
of the Lorentz force, given in \Eq{lorentz_comps},
when $\tilde T^{\mu 0}_{\rm pf}$ are evolved.

On the other hand, when the displacement current can be neglected, $\tJJ = \nab \times \tBB$,
and it is enough to evolve the magnetic field using the induction
equation [cf.~\Eq{rel_induction}], where Ohm's law has already been applied
to describe the electric field in Maxwell equations.
The electric field can then be computed using \Eq{E_Ohms}.
In this case, the electromagnetic components of the stress-energy
tensor or the components of the Lorentz force need to be expressed
in terms of the current density applying Ohm's law.

\subsection{Non-conservation form of relativistic magnetohydrodynamics}
\label{nonconservation_MHD}

\subsubsection*{6.4.1 \ Relativistic equations}

The computation of the non-conservation form of the fluid equations follows an
analogous calculation to those of \Secs{rel_hydrodynamics}{viscosity}, where we now
include the electromagnetic Lorentz forces, setting $\tilde f_\ipf^\mu \to \tilde f_{\rm tot}^\mu = \tilde f_\ipf^\mu + \tilde f^\mu_{\rm Lor}$ in \Eqq{eqs_relativistic_visc}.
This leads to the relativistic energy and momentum MHD equations, presented
in the introduction in \Eqq{summary_rel},
\begin{subequations}
\label{eqs_relativistic_mhd}
\begin{align}
    \partial_{\tau} \ln \tilde \rho \, = &\, -
    \frac{1 + \cs^2}{1 - \cs^2 u^2}
    \nab \cdot \uu
    -
    \frac{1 - \cs^2}{1 - \cs^2 u^2} \, (\uu \cdot \nab) \ln \tilde \rho
    \nonumber \\
    &\, + 
    \frac{1}{1 - \cs^2 u^2} \frac{1}{\tilde \rho} \bigl[(\tilde f_\ipf^0 + \tilde f_\Lor^0)
    (1 + u^2) - 2 \, \uu \cdot (\tff_\ipf + \tff_\Lor) \bigr] + 
    {\cal F}_H^0
    \,, \label{cont_relativistic_mhd} \\
    D_{\tau} \uu  = &\,
    \frac{\uu}{(1 - \cs^2 u^2)\gamma^2} \biggl[
    \cs^2 \, \nab \cdot \uu + 
    \cs^2 \frac{1 - \cs^2}{1 + \cs^2} (\uu \cdot \nab) \ln \tilde \rho -
    \frac{1}{\tilde \rho} \biggl(\tilde f_\ipf^0 + \tilde f_\Lor^0 -
    \frac{2 \cs^2}{1 + \cs^2} \uu \cdot [\tff_\ipf + \tff_\Lor] \biggr)
    \biggr]  \nonumber \\ 
    &\,  -  \frac{\cs^2}{1 + \cs^2}  \frac{\nab \ln \tilde \rho}{\gamma^2}
     + \frac{1}{1 + \cs^2} \,
    \frac{\tff_\ipf + \tff_\Lor}{\tilde \rho \gamma^2} +  \pmb{\cal F}_H\,,   \label{mom_relativistic_mhd}
\end{align}
\end{subequations}
where the Hubble friction ${\cal F}_H^\mu$
appearing in the energy and momentum equations
is given for a generic choice of $\beta$ in \Eqs{FFH0}{Hubble_mom}.
The imperfect (viscous) contributions to the energy and momentum equations
are described by the four-force $\tilde f^\mu_\ipf = \partial_\nu \tilde \Pi^\munu$, whose
components
for the subrelativistic Navier-Stokes viscosity and Fourier's conductivity description are given in \Eqs{visc_forc}{mom_aux_visc}.
The work done by the viscous forces, $\uu \cdot \tff_\ipf$, is given in \Eq{work_visc}.
The contributions from the Lorentz force
to the energy conservation are given in \Eq{lorentz_comps}.
Using Ohm's law, $\tilde f^0_\Lor$ is given by \Eq{f0_MHD_rel}, and the
work done by the Lorentz force, $\uu \cdot \tff_\Lor$, becomes
\begin{equation}
    \uu \cdot \tff_\Lor = \frac{\tilde \eta}{\gamma} (\uu \cdot \tJJ)^2 +
    (1 - 2u^2) \, \tilde \eta \, \tilde \rho_e \,
    \uu \cdot \tJJ
    - \frac{\tilde \eta}{\gamma} \, \tilde \rho_e^2 \, u^2 + \uu \cdot (\tJJ \times \tBB)\,.
    \label{dissi_Lor}
\end{equation}

The energy conservation equation contains irreversible Joule heating
$\tilde \eta \tJJ^2 \supset \tilde f_\Lor^0$, already introduced
in previous work with {\sc Pencil Code}, while it also contains the work
exerted by the Lorentz force $\uu \cdot (\tJJ \times \tBB) \supset \uu \cdot \tff_\Lor$, which converts electromagnetic energy into kinetic energy
when $\uu \cdot \tff_\Lor > 0$ and kinetic into electromagnetic
when $\uu \cdot \tff_\Lor < 0$.
In this work, we also include the viscous contributions, as well as the
general $\tilde f^0_\Lor$ and $\uu \cdot \tff_\Lor$ for relativistic
bulk motion and $\tilde \rho_e \neq 0$.

\subsubsection*{6.4.2 \ Evolution of the Lorentz factor}

Similarly as for perfect and imperfect fluids, the time derivative of the
squared Lorentz factor contains subrelativistic terms.
The generic time derivative of the Lorentz factor, in the full relativistic
regime, is obtained by adding $\tilde f^\mu_\Lor$
to the Hubble and viscous forces in
\Eqs{lorentz_evol}{lorentz_viscs}, respectively
\begin{equation}
    D_\tau \ln \gamma^2 \to [D_\tau \ln \gamma^2]_{\rm ipf}
    - \frac{2\, u^2}{1 - \cs^2 u^2} \frac{\tilde f_\Lor^0}{\tilde \rho}
    + \frac{2}{1 + \cs^2} \, \frac{1 + \cs^2 u^2}{1 - \cs^2 u^2} \, \frac{\uu \cdot 
    \tff_\Lor}{\tilde \rho}  \,.
\end{equation}
In the subrelativistic limit, the contributions to $\partial_\tau \ln \gamma^2$ due to electromagnetic forces (and any other potential
forces in the system) are then non-negligible,
\begin{equation}
    \lim_{u^2 \ll 1} D_\tau \ln \gamma^2 = - \frac{2\cs^2}{1 + \cs^2}
    \bigl( \uu \cdot \nab \bigr)
    \ln \tilde \rho + \frac{2}{1 + \cs^2} \frac{\uu \cdot (\tff_\ipf + \tff_\Lor)}
    {\tilde \rho} \,,
\end{equation}
leading to a correction in the $\uu \cdot (\tJJ \times \tBB)$ terms
in \Eqq{old_eqs_MHD}, as shown in red in \Eqq{corr_eqs_MHD},
that was overlooked in previous work.

\subsubsection*{6.4.3 \ Subrelativistic limit}

The subrelativistic limit of \Eqq{eqs_relativistic_mhd} is
\begin{subequations}
\label{subrel_eqs}
\begin{align}
   \lim_{u^2\ll 1} \partial_{\tau} \ln \tilde \rho \, =  &\,-
   (1 + \cs^2)\, \nab \cdot \uu -
    (1 - \cs^2) (\uu \cdot \nab ) \ln \tilde \rho \nonumber \\
    & \, + 
    \frac{1}{\tilde \rho} \bigl[\tilde f_\ipf^0 + \tilde f_\Lor^0 - 2 \uu \cdot(\tff_\ipf + \tff_\Lor)\bigr]
    + \bigl[\beta - 3\, (1 + \cs^2) \bigr] \HH\,,
    \label{subrel_cont} \\
    \lim_{u^2\ll1} D_{\tau} \uu = & \,  \uu \, \cs^2 \biggl[
    \nab \cdot \uu + 
    \frac{1 - \cs^2}{1 + \cs^2} (\uu \cdot \nab) \ln \tilde \rho \biggr] -
    \frac{\uu}{\tilde \rho} \biggl[\tilde f_\ipf^0 + \tilde f_\Lor^0 -\frac{2 \, \cs^2}{1 + \cs^2}
    \uu \cdot (\tff_\ipf + \tff_\Lor)\biggr]
    \nonumber \\
    & \, 
    - \frac{\cs^2}{1 + \cs^2} \nab \ln \tilde \rho + \frac{1}{1 + \cs^2}
    \frac{\tff_\ipf + \tff_\Lor}{\tilde \rho} +  (3 \cs^2 - 1) \, \uu \, \HH \,.
    \label{subrel_mom}
\end{align}
\end{subequations}

The Lorentz-force components $\tilde f^0_\Lor$ [cf.~\Eq{f0_MHD_rel}]
and $\uu \cdot \tff_\Lor$ [cf.~\Eq{dissi_Lor}]
can be expressed up to leading-order in $u^2$ as
\begin{equation}
    \tilde f_\Lor^0 = \tilde \eta \, \tJJ^2 + \uu \cdot (\tJJ \times \tBB)
    - \tilde \eta \, \tilde \rho_e \, \uu \cdot \tJJ\,,
    \qquad \uu \cdot \tff_\Lor = \uu \cdot (\tJJ \times \tBB) +
    \tilde \eta \, \tilde \rho_e \,
    (\uu \cdot
    \tJJ - \tilde \rho_e u^2) + {\cal O} (u^3)\,,
\end{equation}
where we keep terms up to order ${\cal O} (u^2)$ under
the assumption $|J| \sim {\cal O} (u)$.
Then, the terms appearing in the subrelativistic energy and momentum equations are
\begin{subequations}
\begin{align} 
    \tilde f_\Lor^0 - 2 \uu \cdot \tff_\Lor = &\,
    \tilde \eta \tJJ^2 - \uu \cdot (\tJJ \times \tBB)
    - 3 \tilde \eta \tilde \rho_e \uu \cdot \tJJ + 2 \tilde \eta \tilde \rho_e^2 u^2
    + {\cal O} (u^3)
    \,, \\
    \tilde f_\Lor^0 - \frac{2 \, \cs^2}{1 + \cs^2} \uu \cdot \tff_\Lor
    =  &\, \tilde \eta \tJJ^2 
    + \frac{1 - \cs^2}{1 + \cs^2} \, \uu \cdot (\tJJ \times \tBB)
    \nonumber \\ &\, - \frac{\tilde \eta \tilde \rho_e
    }{1 + \cs^2} \bigl[(1 + 3 \cs^2) \, \uu \cdot \tJJ  +
    2 \cs^2 \tilde \rho_e u^2 \bigr]
    + {\cal O} (u^3)\,.
\end{align}
\end{subequations}

In particular, when the plasma is quasi-neutral ($\tilde \rho_e \simeq
0$), the subrelativistic MHD equations in an expanding background
become
\begin{subequations}
\label{eqs_cont2}
\begin{align}
    \lim_{u^2 \ll 1} \partial_{\tau} \ln \tilde \rho \, =  &\, - (1 + \cs^2) \,
    \nab \cdot \uu -
    \red{(1 - \cs^2)} \, (\uu \cdot \nab) \ln \tilde \rho
    + \frac{1}
    {\tilde \rho} (\tilde f_\ipf^0 - 2 \uu \cdot \tff_\ipf)  \nonumber \\
    &\, 
    + \frac{1}{\tilde \rho} \bigl[\tilde \eta \, \tJJ^2
    \red{-} \uu \cdot (\tJJ \times \tBB ) \bigr] + \bigl[\beta - 3 (1 + \cs^2)\bigr] \, \HH \,,
    \label{subrel_cont2}
    \\
    \lim_{u^2 \ll 1} D_{\tau} \uu  =  &\,
    \uu \, \cs^2 \biggl[
    \nab \cdot \uu + 
    {\red{\frac{1 - \cs^2}{1 + \cs^2}}} (\uu \cdot \nab) \ln \tilde \rho \biggr] - 
    \frac{\uu}{\tilde \rho} \biggl[\tilde f_\ipf^0 - 
    \frac{2 \, \cs^2}{1 + \cs^2}  \uu \cdot \tff_\ipf + 
    \tilde \eta \tJJ^2 + \red{\frac{1 - \cs^2}{1 + \cs^2}} \, \uu \cdot 
    \bigl( \tJJ \times \tBB \bigr) \biggr]
    \nonumber \\  &\,
     - \frac{\cs^2}{1 + \cs^2} 
    \nab \ln \tilde \rho + \frac{1}{1 + \cs^2} \frac{\tff_\ipf + \tJJ \times \tBB}
    {\tilde \rho} + (3 \cs^2 - 1) \, \uu\, \HH \,,
    \label{subrel_mom2}
\end{align}
\end{subequations}
where the correction with respect to previous work is indicated in red.
These equations

reduce to \Eqq{corr_eqs_MHD} for a viscous
fluid with $\cs^2 = \third$, omitting viscous effects in the
energy conservation equation.

\subsection{Transverse Alfv\'en waves}
\label{alfven_waves}

In this section, we study one of the most relevant solutions
in an MHD system, transverse
Alfv\'en waves, with the objective to review the
dependence of the Alfv\'en speed on the
value of $\cs^2$ for a constant equation of state $\tilde p = \cs^2 \tilde \rho$
\cite{Jedamzik:1996wp,Subramanian:1997gi} and show how the Alfv\'en speed can become
superluminal when one neglects the displacement current, even if
the bulk velocity is subrelativistic.
This issue can be overcome using the Boris correction \cite{jay_p__boris_1970}, which
will be discussed in \Sec{boris_correction} and extended to the
relativistic MHD equations in the early Universe.

For simplicity, we will ignore dissipative effects like viscosity and
magnetic diffusivity and consider a neutral plasma with $\tilde \rho_e = 0$.
We also neglect the Hubble friction, which vanishes for a radiation-dominated fluid and, otherwise, complicates the computation of the
perturbations due to the time-dependence of the Hubble rate $\HH$.
However, the procedure to compute the perturbations including the expansion
of the Universe follows the one presented in \Sec{sound_waves}.
For this reason, in this and \Sec{magnetosonic},
we drop the tildes in the comoving variables,
as the Universe expansion is neglected.
Then, the subrelativistic ideal
MHD equations are
[cf.~\Eqq{eqs_cont2}]
\begin{subequations}
\begin{align}
    \partial_{\tau} \ln \rho \, + &\,
    (1 + \cs^2) \, \nab \cdot \uu +
    (1 - \cs^2) \, \bigl(\uu \cdot \nab\bigr) \ln \rho 
       = -  \frac{1}{\rho} \, \uu \cdot \bigl(\JJ \times \BB \bigr) \,, \\ 
    D_{\tau} \uu = &\,
    \uu \, \cs^2 \biggl[
    \nab \cdot \uu + 
    \frac{1 - \cs^2}{1 + \cs^2} (\uu \cdot \nab) \ln \rho 
     \biggr] - \frac{\uu}{\rho} \, \frac{1 - \cs^2}{1 + \cs^2}
     \, \uu \cdot \bigl(\JJ \times \BB \bigr) \nonumber \\ 
     &\, - \frac{\cs^2}{1 + \cs^2} \, \nab \ln \rho + \frac{1}{1 + \cs^2} \frac{\JJ \times \BB}{\rho} \,.
\end{align}
\end{subequations}

When the displacement current can be neglected, in the limit $\eta \to 0$,
the magnetic field evolution is governed by the ideal induction equation [cf.~\Eq{induc2}],
\begin{equation}
    D_\tau \BB = (\BB \cdot \nab) \uu - \BB(\nab \cdot \uu)\,.
\end{equation}

Let us consider linear perturbations above a background
homogeneous magnetic field $\BB_0 = B_0  \, \hat e_z$,
$\BB = \BB_0 + \BB_1 e^{i \kk \cdot \xx - i \omega \tau}$, $\uu = \uu_1 e^{i\kk \cdot \xx - i \omega \tau}$, $\rho = \rho_0 + \rho_1 e^{i\kk \cdot \xx - i \omega \tau}$, where the
perturbations occur in the direction of the homogeneous magnetic
field and in the plane
perpendicular to it, $\kk = k_\parallel \hat e_z + k_\perp \hat e_y$.
The induction equation yields,
\begin{equation}
    \omega \BB_1 = - (\BB_0 \cdot \kk) \uu_1 + \BB_0 (\kk \cdot \uu_1)
    = B_0\, \left(\begin{array}{c}
         - k_\parallel u_x \\ - k_\parallel u_y \\ \ \, k_\perp u_y
    \end{array} \right)\,.
\end{equation}
The current density is in this case
\begin{equation}
    \JJ = \nab \times \BB = i \, \left( \begin{array}{c}
        k_\perp B_z - k_\parallel B_y  \\ \ k_\parallel B_x \\ - k_\perp B_x
    \end{array} \right)\,,
\end{equation}
such that the Lorentz force components are (up to first-order), $f_\Lor^0 =  0$, and
\begin{equation}
    \ff_\Lor = \JJ \times \BB = i B_0 \, \left(\begin{array}{c}
         k_\parallel B_x \\ k_\parallel B_y - k_\perp B_z \\ 0
    \end{array}\right)\,.
\end{equation}
Note that $\uu \cdot \ff_\Lor$
only contains terms of second and higher order, so it can be neglected.

Transverse Alfv\'en waves correspond to fluid perturbations
in the direction perpendicular to both $\kk$ and $\BB_0$, i.e., $u_x$.
The momentum equation in this direction, neglecting all terms of
second or higher order, is
\begin{equation}
    \omega u_x 
    +  k_\parallel \frac{B_x B_0}{(1 + \cs^2)\rho_0}
    = 0
    \Rightarrow u_x \bigl(\omega^2 
    - k_\parallel^2 \vA^2
    \bigr) = 0\,,
     \label{dispersion}
\end{equation}
which corresponds to transverse Alfv\'en waves with group
speed
\begin{equation}
    \vA^2 = \frac{B_0^2}{(1 + \cs^2) \rho_0}\,. \label{Alfven}
\end{equation}
Note that in regions where the energy density is very small or the magnetic field is very large, the
Alfv\'en speed can grow unbounded, potentially leading to superluminal Alfv\'en speeds.
This is a consequence of neglecting the displacement current,
as indicated in \cite{jay_p__boris_1970}, and the Alfv\'en speed
can be corrected including the Boris correction in the momentum
equation, as we will show in the next section.
Magnetosonic waves, i.e., perturbations in the plane formed by $\kk$ and $\BB_0$,
are considered in \Sec{magnetosonic}.

\subsection{Boris correction for relativistic Alfv\'en speeds}
\label{boris_correction}

In MHD simulations, it is common to neglect the displacement current,
as the conductivity is usually larger than the oscillation frequency
of the electric field (see \Sec{induction_sec}).
This is particularly true in the early Universe, where $\eta$ is usually
several orders of magnitude smaller than the Hubble size (see \Fig{conductivity}).
However, we have shown that the resulting
Alfv\'en waves can propagate
faster than the speed of light, independently of the values
of the bulk velocity [cf.~\Eq{Alfven}].
To correct this effect in the limit of subrelativistic bulk motion,
$u^2 \ll 1$, the Boris correction was proposed in \cite{jay_p__boris_1970}.
In the following, we reproduce the
Boris correction adapted to the MHD
system in an expanding Universe with a relativistic equation of state,
such that $\tilde p = \cs^2 \tilde \rho$ with constant $\cs^2$.
Note that in this section, all the assumptions taken to study
Alfv\'en waves in the previous section are dropped, and only recovered
at the end of the section when the application of Boris correction
to Alfv\'en waves is taken into account.

In first place, note that the Lorentz force that appears in the
momentum equation can be recast in terms of the divergence of the
electromagnetic tensor, $\tilde f^i_\Lor = - \partial_\mu \tilde T^{i\mu}_{\rm EM}$ [cf.~\Eq{fi_dmu}],
\begin{equation}
    \tilde f^i_\Lor = - \partial_\tau (\tEE \times \tBB)_i - \partial_j \tilde T^{ij}_{\rm EM}\,. \label{fi_real}
\end{equation}
In the limit of ideal MHD ($\tilde \eta \to 0$), we can set $\tilde \EE = - \uu \times \tBB$ from Ohm's law,
such that quadratic terms in $\tilde E_i$ are
proportional to $u^2$ and, hence, of a higher order than
the time derivative
of the Poynting vector and the magnetic stresses.

Then, the
subrelativistic momentum equation in this limit can be expressed as
[cf.~\Eq{subrel_mom}],
\begin{align}
    \partial_\tau \biggl[\uu - \frac{(\uu \times \tBB) \times \tBB}{(1 + \cs^2) \tilde \rho}
    \biggr] = & - (\uu \cdot \nab) \, \uu  + \uu \, \cs^2 \biggl[
    \nab \cdot \uu + 
    \frac{1 - \cs^2}{1 + \cs^2} (\uu \cdot \nab) \ln \tilde \rho \biggr] \nonumber \\ &
    - \frac{\uu}{\tilde \rho} \biggl[ \tilde f_\ipf^0 - \frac{2\, \cs^2}{1 + \cs^2}
    \uu \cdot \tff_\ipf +
    \tilde \eta \tJJ^2 + \frac{1 - \cs^2}{1 + \cs^2} \, \uu \cdot
    ( \tJJ \times \tBB ) \biggr] \nonumber \\ &
    -
    \frac{1}{(1 + \cs^2) \tilde \rho} \Bigl(\nab \cdot \pmb{\tilde P}
    + \bigl[(\uu \times \tBB) \times \tBB \bigr] \partial_\tau \ln \tilde \rho \Bigr)
    + (3 \cs^2 - 1) \, \uu \, \HH 
    \,,
    \label{momentum_Boris}
\end{align}
where the total comoving
pressure tensor $\pmb{\tilde P}$ incorporates the isotropic fluid pressure, the
deviatoric viscous stresses, and the magnetic field pressure,
\begin{equation}
    \tilde P_{ij} = \bigl(\tilde p -
    \tilde \zeta_\visc \, \theta + \half \tBB^2\bigr) \, \delta_{ij} - 2 \, \tilde \eta_\visc \, \tilde \sigma_{ij}
    - \tilde B_i \tilde B_j\,,
\end{equation}
omitting the electric field pressure $-\tilde E_i \tilde E_j + {1 \over 2} \delta_{ij}
\tilde E^2 \sim {\cal O} (u^2)$.
The $\partial_\tau \ln \tilde \rho$ term in \Eq{momentum_Boris} is obtained from the
subrelativistic energy equation [cf.~\Eq{subrel_cont2}].

The term in the left-hand-side of \Eq{momentum_Boris} can be expressed as \cite{jay_p__boris_1970,2002JCoPh.177..176G,2018arXiv180608166C}
\begin{equation}
    \uu - \frac{(\uu \times \tBB) \times \tBB}{(1 + \cs^2)\tilde \rho} = 
    \uu (1 + \vA^2) - \vA^2 (\hat {\pmb b} \cdot \uu) \,
    \hat {\pmb b} = u_j \bigl[\delta_{ij}
     + \vA^2(\delta_{ij} - \hat b_i \hat b_j)\bigr]\,, \label{boriss}
\end{equation}
where we have defined the Alfv\'en speed $\vA^2 = \tBB^2/[(1 + \cs^2) \tilde \rho]$
and $\hat {\pmb b} = \tBB/|\tBB|$ is the unitary magnetic field.
In matrix form, this can be expressed as
\begin{equation}
    \partial_\tau \biggl[\uu - \frac{(\uu \times \tBB) \times \tBB}{(1 + \cs^2)\tilde \rho}\biggr] = 
    \partial_\tau \uu \,
    \bigl[{\pmb I} + \vA^2 ({\pmb I} - \hat {\pmb b} \hat {\pmb b})\bigr]\,,
    \label{matrix}
\end{equation}
where we have neglected the following term
\begin{equation}
    \uu \, \partial_\tau \bigl[\vA^2({\pmb I} - \hat {\pmb b} \hat{\pmb b})\bigr] \sim u_j B^i \partial_\tau B^j \sim  u_j B^i
    \, (\nab \times E)^j \sim {\cal O} (u^2)\,,
\end{equation}
which is of second order in $u^2$ in the ideal MHD limit.
The matrix in \Eq{matrix}
has the following inverse
\begin{equation}
    \bigl[{\pmb I} + \vA^2 ({\pmb I} - \hat {\pmb b} \hat {\pmb b})\bigr]^{-1}
    = {\pmb I} - \vA^2 \gamma_A( {\pmb I}
    -
    \hat \bb \hat \bb) \,,
    \label{matrix_inverse}
\end{equation}
where
\begin{equation}
    \gamma_A = \frac{1}{1 + \vA^2}\,,
\end{equation}
is the relativistic Alfv\'en correction.
Then, the momentum equation can be expressed in the following way,
\begin{equation}
    \partial_\tau u^i =
    {\rm RHS}^i ({\uu}) - \vA^2 \gamma_A (\delta^{ij} - \hat b^i \hat b^j) \, {\rm RHS}_j
    ({\uu}) =
    {\rm RHS}^i (\uu) + F_{\rm corr}^i \,,
\end{equation}
where RHS$(\uu)$ is the right-hand side of \Eq{momentum_Boris}
and ${\pmb F}_{\rm corr}$ is the Boris correction for relativistic Alfv\'en speeds.

Let us now focus on Alfv\'en transverse waves, studied in \Sec{alfven_waves}, under this correction.
When we include the subrelativistic term of the displacement current,
i.e., the time derivative of the Poynting vector, in the momentum
equation, the additional term in \Eq{momentum_Boris} leads to the modification of the left-hand-side $\partial_\tau \uu \to \gamma_A^{-1}
\, \partial_\tau \uu$ (cf.~\Eq{boriss} applied to direction perpendicular
to the homogeneous magnetic field).
Then, the dispersion relation in \Eq{dispersion} becomes
\begin{equation}
    u_x\bigl(\omega^2 
    - k_\parallel^2 \vA^2 \gamma_A
    \bigr) = 0\,,
\end{equation}
and the transverse Alfv\'en speed gets corrected: $\vA^2 \to \vA^2 \gamma_{\rm A} = \vA^2/(1+\vA^2) \leq 1$ \cite{jay_p__boris_1970}, ensuring subluminal
propagation of the MHD perturbations.

\subsection{Magnetosonic waves}
\label{magnetosonic}

In this section, we focus on the perturbations introduced in \Sec{alfven_waves} when the velocity field is
in the plane perpendicular to the transverse Alfv\'en waves, i.e., the plane defined by the direction of the perturbations
and the direction of the homogeneous background magnetic field.
As done for sound waves in \Sec{sound_waves},
we define the energy density fluctuations,
$\lambda = \rho_1/[(1 + \cs^2) \rho_0]$.
In the linearized theory, the energy equation does not contain contributions
from the electromagnetic fields, as they are of higher-order and one
finds [cf.~\Eq{soundwaves_pert}],
\begin{align}
    \lambda = \frac{\kk \cdot \uu}
    {\omega}\,.
    \label{cont_fluctuations}
\end{align}
The momentum equation in the $y$-direction, parallel to the
perturbations, $\hat \kk_\perp$, is
\begin{align}
    - i \omega u_y \gamma_A^{-1} = &
    - i\, k_\perp \, \cs^2 \lambda 
     + i \, \frac{k_\parallel B_y - k_\perp B_z}
    {(1 + \cs^2) \rho_0} B_0 \nonumber \\
    &\, \Rightarrow u_y  \bigl( \omega^2 
    -  k^2 \cms^2\gamma_A + k_\parallel^2 \cs^2 \gamma_A \bigr) - u_z k_\perp k_\parallel \cs^2 \gamma_A =
    0\,,
\end{align}
where we have defined the magnetosonic speed $\cms^2 = \cs^2 + \vA^2$,
and already introduced the Boris correction.
The momentum equation in the parallel direction to $\BB_0$ is
\begin{align}
    -i \omega u_z \gamma_A^{-1} =- i\, k_\parallel \cs^2
    \lambda\, \Rightarrow u_z \bigl(\omega^2
    - k_\parallel^2 \cs^2 \gamma_A \bigr) - u_y k_\perp k_\parallel \cs^2
    \gamma_A=
    0 \,.
\end{align}
Hence, in matrix form, the perturbations can be expressed as
\begin{equation}
    \left(\begin{array}{cc}
       \omega^2
       - k^2 \cms^2 \gamma_A + k_\parallel^2 \cs^2 \gamma_A
       &  -k_\perp k_\parallel \cs^2 \gamma_A \\ - k_\perp k_\parallel \cs^2
       \gamma_A
         & \omega^2
         - k_\parallel^2 \cs^2 \gamma_A
    \end{array} \right) \left(\begin{array}{c}
         u_y  \\ u_z
    \end{array} \right) =
    \left( \begin{array}{c}
         0  \\ 0 
    \end{array}\right) \,.
\end{equation}
The eigenvalues of the equation yielding the dispersion relation are obtained
when the determinant vanishes
\begin{equation}
    \omega^4 - \omega^2 \gamma_A 
    k^2 \cms^2 
    + k^2 k_\parallel^2 \vA^2 \cs^2 
    \gamma_A^2 = 0\,.
\end{equation}
This expression
corresponds to magnetosonic waves.
The solution to the dispersion relation is
\begin{equation}
    \omega^2_\pm = \half \gamma_A k^2 \cms^2  
    \pm \gamma_A k
    \sqrt{\fourth k^2 \, \cms^4
    -
    k_\parallel^2 \vA^2 \cs^2}\,.
\end{equation}
The two solution branches of the dispersion relation correspond to the fast ($+$)
and slow ($-$) magnetosonic waves in the plasma, for the positive and negative
sign, respectively.

In particular, the perturbations propagating at an angle $\theta$
with respect to the homogeneous magnetic field
can be described setting $k_\parallel^2/k^2 = \cos^2 \theta$,
\begin{equation}
    \omega^2_\pm = \half \gamma_A k^2 \cms^2
    \pm \gamma_A k^2 \sqrt{\fourth \cms^4
    -
    \cos^2 \theta \vA^2 \cs^2}\,.
\end{equation}
In the perpendicular direction, $\theta = \pi/2$,
the slow magnetosonic wave does not propagate ($\omega_- = 0$)
and the angular frequency of the fast magnetosonic wave is
\begin{equation}
    \omega^2_+ = \gamma_A k^2 \cms^2\,.
\end{equation}

\section{Conclusions and summary}
\label{conclusions}

In the present work, the relativistic magnetohydrodynamic (MHD) equations for perfect and
imperfect (viscous) plasmas
in the early Universe are
studied in an expanding background described by the Friedmann-Lema\^itre-Robertson-Walker (FLRW) metric tensor under the assumptions of homogeneity and isotropy of the early
Universe.
The MHD equations in an expanding background presented in this work
are relevant to study the
dynamics of the primordial plasma during the epoch when the Universe
was dominated by radiation, allowing to include a matter component by considering small
deviations with respect to $\cs^2 = \third$, where the square of the speed of sound
is assumed to be a constant relating the pressure and total energy
density $p = \cs^2 \rho$ (see \Sec{eos} for details on the
equation of state).

After a brief review of the FLRW geometry in \Sec{FLRW},
the relativistic equations for perfect fluids are presented
both for the purely fluid dynamical
system (i.e., in the absence of electromagnetic fields or for non-conducting fluids)
and for the MHD system.
In terms of the conformal time $\tau$, the equations of motion of the fluid in an expanding
Universe
become equivalent to those in flat Minkowski space-time after a conformal mapping of the stress-energy tensor,
$\tilde T^\munu = a^6 T^\munu$, as long as the stress-energy tensor is traceless,
which is accomplished only when $p = \third \rho$,
i.e., $\cs^2 = \third$
\cite{Brandenburg:1996fc,Subramanian:1997gi}
and when the bulk viscosity 
$\zeta_\visc$ is zero (see Stokes assumption in \Sec{viscosity}).
The conformal transformation $\tilde T^{\munu} = a^6 T^{\munu}$ leads to the definition of the comoving energy density and pressure, $\tilde \rho = a^4 \rho$ and
$\tilde p = a^4 p$, and the comoving shear viscosity $\tilde \nu = a^{-1} \nu$.
In \Sec{generic_scaling}, we explore different scalings of the fluid variables that lead to equations
with a generic Hubble friction term.
In particular, when the bulk velocities are subrelativistic, $u^2 \ll 1$, it is found
that the scaling $\tilde p = a^\beta p$ and $\tilde \rho = a^\beta \rho$ with $ \beta = 3 (1 + \cs^2)$ is the most
convenient choice as it maximizes the number of
Hubble friction terms that vanish in the fluid
equations.
A special case corresponds to a fluid composed by massive particles (dust) with
$\cs^2 \ll 1$, for which it is convenient to separately scale the pressure and the energy
density, as they become decoupled in the fluid equations.
We show that
the equations become conformally flat choosing super-comoving
coordinates \cite{Martel:1997hk}, corresponding to
scaling $\tilde p = a^5 p$ and $\tilde u^i = a u^i$, together with
a scaled $\alpha$-time $\tau_\alpha$ with
$\alpha = 2$, defined such that $\dd \tau_\alpha = a^{\alpha - 1} \dd \tau$.
An alternative choice of super-comoving coordinates corresponds to scaling
$\tilde \rho = a^3 \rho$
and $\tilde p = a^4 p$ \cite{Banerjee:2004df},
together with a scaled $\alpha$-time, $\dd \tau_\alpha = a^{\alpha - 1} \dd \tau$ with $\alpha = \threehalf$.
This choice allows us to keep a constant value
of $\HH_\alpha = (\partial_{\tau_\alpha} a)/a$ and
minimizes the appearances of Hubble friction terms in the
equations if one further scales the velocity field
as $\tilde u^i = a^\delta u^i$ with $\delta = \half$.
We present a generalized version of the latter choice of super-comoving coordinates for
other values of the background equation of state $w \neq 0$
determining the evolution of the Universe (see \Sec{FLRW}).
These choices are explained in \Sec{cons_perf_fluid} and summarized in \Tab{tab:my_label}.
The choice $\beta = 3 \, (1 + \cs^2)$ is especially useful when studying sound
waves in an expanding Universe, considered in \Sec{sound_waves}.

In \Sec{conservation_rel_perf}, we compute the equations of motion considering
the $\tilde T^{0\mu}$ components of the perfect
fluid as the dynamical variables (conservation form).
We extend the conservation form of the equations of
motion to include out-of-equilibrium effects in \Sec{viscosity}.
We present the deviatoric stress-energy tensor arising in first-order
fluid dynamics
in \Sec{NavierStokes}, which corresponds to Navier-Stokes viscosity and Fourier's heat
conductivity, and review in \Sec{transport_coeffs}
the estimates for the transport coefficients describing out-of-equilibrium
effects in the fluid: kinematic shear viscosity ($\nu$),
shown in \Fig{shear_viscosity}, kinematic bulk viscosity ($\xi$), and
thermal conductivity ($\kappa$).
We include
the Lorentz force due to coupling of the fluid
with gauge fields in
\Sec{conservation_MHD}.
We summarize these equations in \Sec{MHD_summary_conservation}.
Then, in \Sec{rel_hydrodynamics}, we present the non-conservation form of the
equations of motion of a perfect fluid
in terms of the primitive fluid
variables: the comoving
energy density $\tilde \rho$ and the peculiar velocity $u^i$.
We again include imperfect (viscous) forces in \Sec{viscosity} and
extend to the non-conservation form of the MHD equations in \Sec{nonconservation_MHD}.
We summarize the non-conservation form of the MHD
equations in \Sec{MHD_summary_nonconservation}.
For both approaches, we correctly incorporate the leading-order corrections
in the subrelativistic limit, which had been overlooked in previous work
as a consequence of setting $D_\tau \gamma^2 \to 0$.
We explicitly show that $D_\tau \gamma^2$ contains terms that are of leading order
in the subrelativistic limit, and present the corresponding corrections to the
equations of conservation of energy and momentum.

When we include the Lorentz force (see \Sec{four_lorentz}) in the fluid
equations, the equations of motion become coupled to
Maxwell equations via the current density, which is described
using the generalized Ohm's law, reviewed in \Sec{cov_ohms}.
Since the trace of the electromagnetic stress-energy tensor
is zero (see \Sec{sec:Tmunu}), the MHD equations are still conformally flat when 
$\cs^2 = \third$ and $\zeta_\visc = 0$, and the
transformation $\tilde T^\munu = a^6 T^\munu$ leads to the
definition of comoving electric and magnetic fields,
$\tilde B^i = a^2 \bar B^i = \half \varepsilon^{ijk} F_{jk}$ and $\tilde E_i = a^2 \bar E_i = F_{i0}$
(see \Secs{faraday_tensor}{comoving_EM}).
We review in \Sec{maxwell_eqs}
Maxwell equations in an expanding Universe.
The conformal invariance of Maxwell equations implies a transformation
$\tilde J^\mu = a^4 J^\mu$ of the four-current,
which leads to the definition of a comoving charge density
$\tilde \rho_e = a^3 \rho_e$ and a comoving conductivity
$\tilde \sigma = a \sigma$.
In \Sec{induction_sec}, we discuss the large-conductivity limit,
commonly assumed in MHD,
in which we can neglect the displacement current and
reduce Maxwell equations to the magnetic induction equation.
We present an estimate of the conductivity in the early Universe (see \Fig{conductivity},
which shows the magnetic diffusivity $\eta = 1/\sigma$), showing
that this limit is, in general, valid during the radiation-dominated era.
We summarize Maxwell equations and the induction
equation in \Sec{maxwell_summary}.

We have reviewed the system of linearized waves in MHD.
In \Sec{sound_waves}, we describe sound waves in an expanding Universe,
and we describe transverse Alfv\'en waves and
magnetosonic waves in \Secs{alfven_waves}{magnetosonic} when
the equations of motion are conformally flat.
The resulting Alfv\'en speed $\vA$ is shown to potentially
become superluminal when $B_0/\rho_0 \gtrsim 1$ even
in the regime of linearized fluid perturbations.
In \Sec{boris_correction},
we provide a correction to the equation of momentum
conservation in the fluid that allows to find the correct
relativistic Alfv\'en speed, $\vA \to \vA \gamma_A$ with
$\gamma_A = (1 + \vA^2)^{-1}$, following
the original Boris correction \cite{jay_p__boris_1970}.

\subsection{Relativistic MHD equations in the conservation form}
\label{MHD_summary_conservation}

In the conservation form, the comoving $\tilde T^{0\mu}$ components of the
stress-energy tensor
can be dynamically evolved in conformal time
using the following conservation laws (see \Sec{MHD_eom}
for details),
\begin{equation}
    \partial_\tau \tilde T^{0\mu}_{\rm pf} + \partial_j \tilde T^{j\mu}_{\rm pf} = 
    \tilde f_H^\mu + \tilde f_\ipf^\mu + \tilde f_{\rm Lor}^\mu\,, \label{conservation_summary}
\end{equation}
where $\tilde T^{\munu}_{\rm pf} = a^4 T^\munu_{\rm pf}$ is the perfect
fluid stress-energy tensor (see \Sec{Tmunu_perf}),
\begin{equation}
    \tilde T^\munu_{\rm pf} = (\tilde p + \tilde \rho) \tilde U^\mu
    \tilde U^\nu + \tilde p \, \eta^\munu\,,
\end{equation}
with $\tilde U^\mu = \gamma(1, \uu)$ being the comoving four-velocity (see \Sec{four_velocity}).

The imperfect (viscous) forces have been considered following first-order fluid dynamics,
in which small deviations with respect to the perfect fluid local thermal equilibrium (LTE) are
included and modeled using the deviatoric stress tensor $\tilde \Pi^\munu$, such that
the imperfect fluid stress-energy tensor is 
$\tilde T_{\rm Fl}^\munu \equiv \tilde T_{\rm pf}^\munu - \tilde \Pi^\munu$, and incorporates Navier-Stokes viscosity and Fourier's heat conductivity (see \Sec{viscosity}).
Then, the imperfect four-force in \Eq{conservation_summary}
is $\tilde f^\mu_\ipf = \partial_\nu \tilde \Pi^\munu$.
In the subrelativistic limit, and for negligible
bulk viscosity, $\xi H \ll 1$,
the temporal component of the imperfect four-force corresponds to the divergence
of the heat flux and the spatial components to the Navier-Stokes viscous force (i.e., the divergence
of the shear stresses) [cf.~\Eqs{visc_forc}{mom_aux_visc}]
\begin{align}
    \label{visc_forces_sum}
    \tilde f_\ipf^0 =
    \tilde \kappa \, \nab^2 \tilde T\,, \qquad
    \frac{\tff_{\ipf}}{\tilde p + \tilde \rho} = &\,
    \tilde \nu\, \nab^2 \uu + \bigl(\third \tilde \nu + \tilde \xi \bigr)
    \nab \tilde \theta
    + \bigl[ (2 \, \tilde \nu \, \pmb{\tilde \sigma} + \tilde \xi \, \tilde \theta \, \pmb{I})
    \cdot \nab) \bigr] \ln \tilde \rho\,,
\end{align}
where $\tilde S^{ij} = \half (\partial^i u^j + \partial^j u^i
+ 2 \, \HH \, \delta^{ij})$
is the comoving
rate-of-strain tensor, $\tilde \sigma^{ij} = \tilde S^{ij}$ $ - \onethird \, \tilde \theta
\, \delta^{ij}$ its traceless counterpart, $\tilde \theta = \nab \cdot \uu + 3 \HH$ the fluid
expansion scalar, and homogeneous $\tilde \nu$, $\tilde \xi$, and $\tilde \kappa$ are assumed.
The comoving temperature $\tilde T = aT$ can be related to the (radiation) energy density via \Eq{rho_rad} to compute $\tilde f_\ipf^0$
when the thermal conductivity $\kappa$ is not vanishing.

The Hubble friction $\tilde f_H^\mu$ in \Eq{conservation_summary}
appears due to the expansion of the Universe [cf.~\Eq{fH_nu}]
\begin{equation}
    \tilde f_{H}^0 =
    \bigl[(\beta - 4) \, \tilde T^{00}_{\rm Fl}
    - \tilde T_{\rm Fl} \bigr] \HH\,,
     \qquad \tilde f_H^i =
    \, (\beta - 4)\, \tilde T^{0i}_{\rm Fl} \,  \HH\,,
\end{equation}
when the comoving energy density and pressure are rescaled as $\tilde p = a^\beta p$
and $\tilde \rho = a^\beta \rho$ (see \Sec{cons_perf_fluid}).
The choice $\beta = 4$
is an appropriate choice for relativistic fluids, especially when
the trace of the fluid stress-energy tensor $\tilde T_{\rm Fl} = 3 \, \tilde p - \tilde \rho - 3\, \tilde \zeta_\visc \, \tilde \theta$ vanishes, since then $\tilde f_H^\mu = 0$ and the equations
of motion become conformally flat [cf.~\Eq{cons_expand}].
This occurs for
an equation of state described by a constant $\cs^2 = \onethird$
(see \Sec{cons_perf_fluid}) and when the bulk viscosity is zero.
The conformal Hubble rate is $\HH = a'/a$.

Finally, the electromagnetic
Lorentz four-force is described in \Sec{four_lorentz},
\begin{equation}
    \tilde f_\Lor^0 = \tEE \cdot \tJJ\,, \qquad
    \tilde f^i_\Lor = \tilde J^0 \tEE + \tJJ 
    \times \tBB\,. \label{fLor_summary}
\end{equation}
The four-current $\tilde J^\mu$ couples the equations of motion of the fluid with
the electric and magnetic fields,
and can be computed using 
the generalized Ohm's law
(see \Sec{cov_ohms}),
\begin{equation}
    \tilde J^0 = \gamma (\tilde \rho_e + 
    \tilde \sigma \uu \cdot \tEE)\,, \qquad
    \tilde J^i = \gamma (\tilde \rho_e \uu +
    \tilde \sigma [\tEE + \uu \times \tBB])\,.
\end{equation}
The electromagnetic fields are then evolved together
with the equations of motion of the fluid using Maxwell
equations.

One last step is still required to evolve \Eq{conservation_summary},
since $\tilde T^{ij}_{\rm pf}$ needs to be expressed in terms of the
dynamical components $\tilde T^{0\mu}_{\rm pf}$ to close the system.
Furthermore, we need to reconstruct the peculiar velocity $\uu$ and energy density
$\tilde \rho$ from the dynamical variables $\tilde T^{0\mu}_{\rm pf}$ to express the imperfect (viscous)
and Hubble four-forces.
This procedure is described in \Sec{conservation_rel_perf}.
In first place, for a constant $\cs^2$, the Lorentz factor can be computed in the following way,
\begin{equation}
\gamma^2 = \frac{1}{2(1 - r^2)} \left[ 1 - 2 r^2 \frac{\cs^2}{1 + \cs^2} +
\sqrt{1 - 4 r^2 \frac{\cs^2}{(1 + \cs^2)^2}}
    \ \right] \,, \quad {\rm where \ \ }
    r^2 = \frac{\tilde T^{0i}_{\rm pf} \tilde T^{0i}_{\rm pf}}{({\tilde T^{00}_{\rm pf}})^2} \,.
\end{equation}
Then, once $\gamma^2$ is computed, the stress tensor can be reconstructed using
\begin{equation}
    \tilde T^{ij}_{\rm pf} = [(1 + \cs^2) \gamma^2 u^i u^j + \cs^2 \delta^{ij}]
    \, \tilde  \rho = \frac{\tilde T^{0i}_{\rm pf} \, \tilde T^{0j}_{\rm pf}}{(1 + \cs^2) \tilde \rho \gamma^2 } + \cs^2 \, \tilde \rho \, \delta^{ij}\,,
\end{equation}
where the energy density $\tilde \rho$ and the peculiar velocity $\uu$ can be
reconstructed from $\tilde T^{0\mu}_{\rm pf}$ as
\begin{equation}
    \tilde \rho = \frac{\tilde T_{\rm pf}^{00}}{(1 + \cs^2) \gamma^2 - \cs^2}\,, \qquad
    u^i = \frac{\tilde T^{0i}_{\rm pf}}
    {(1 + \cs^2) \tilde \rho \gamma^2}\,.
\end{equation}

An alternative procedure is described in \Sec{conservation_MHD} where,
instead of evolving dynamically the stress-energy tensor
components of the perfect fluid, $\tilde T^{0\mu}_{\rm pf}$, one
can evolve the combined components of the perfect fluid
and the electromagnetic fields,
$\tilde T^{0\mu} = \tilde T^{0\mu}_{\rm pf} + \tilde T^{0\mu}_{\rm EM}$.

\subsection*{Subrelativistic limit}

The subrelativistic limit of the conservation form of the MHD equations has been
considered in \Sec{conservation_rel_perf}, where it is shown that terms of order
$r^2 \sim {\cal O} (u^2)$ need to be kept in the time derivatives to lead to the correct limit
in the subrelativistic regime $u^2 \ll 1$.
The same procedure as the one described in the relativistic conservation form
is taken. In this case, the stress tensor $\tilde T^{ij}_{\rm pf}$ is related
to the dynamical variables in the following way,
\begin{equation}
    \lim_{u^2 \ll 1} \tilde T^{ij}_{\rm pf} = \frac{1}{1 + \cs^2} \frac{\tilde T^{0i}_{\rm pf}
    \, \tilde T^{0j}_{\rm pf}}{\tilde T^{00}_{\rm pf}} \biggl(1 + \frac{r^2 \cs^2}{(1 + \cs^2)^2}
    \biggr) 
    + \cs^2 \, \tilde T^{00}_{\rm pf} \biggl(1 - \frac{r^2}
    {1 + \cs^2} \biggr)\, \delta^{ij}\,.
\end{equation}
In this regime, Ohm's law becomes
\begin{equation} \label{ohms_subrel}
    \lim_{u^2 \ll 1} \tilde J^0 = \tilde \rho_e + \tilde \sigma \, \uu \cdot \tEE \,, \qquad
    \lim_{u^2 \ll 1} \tilde J^i = \tilde \rho_e \uu + \tilde \sigma
     (\tEE + \uu \times \tBB)\,.
\end{equation}

\subsection{Relativistic MHD equations in the non-conservation form}
\label{MHD_summary_nonconservation}

The fully relativistic fluid equations of motion
in the non-conservation form are presented, up to our knowledge,
for the first time, in \Sec{nonconservation_MHD}
[cf.~\Eqq{eqs_relativistic_mhd} and discussion in \Sec{intro} around \Eqq{summary_rel}],
\begin{subequations}
\label{eqs_summary_rel_1}
\begin{align}
    \partial_{\tau} \ln \tilde \rho \, = &\, -
    \frac{1 + \cs^2}{1 - \cs^2 u^2}
    \nab \cdot \uu
    -
    \frac{1 - \cs^2}{1 - \cs^2 u^2} \, (\uu \cdot \nab) \ln \tilde \rho
     + 
    \Bigl[(\beta - 4) + \frac{1 + u^2}{1 - \cs^2 u^2} (1 - 3 \cs^2) \Bigr] \, \HH
    \nonumber \\
    &\, + 
    \frac{1}{1 - \cs^2 u^2} \frac{1}{\tilde \rho} \bigl[(\tilde f_\ipf^0 + \tilde f_\Lor^0)
    (1 + u^2) - 2 \, \uu \cdot (\tff_\ipf + \tff_\Lor) \bigr]
    \,, \\
    D_{\tau} \uu  = &\,
    \frac{\uu}{(1 - \cs^2 u^2)\gamma^2} \biggl[
    \cs^2 \, \nab \cdot \uu + 
    \cs^2 \frac{1 - \cs^2}{1 + \cs^2} (\uu \cdot \nab) \ln \tilde \rho -
    \frac{1}{\tilde \rho} \biggl(\tilde f_\ipf^0 + \tilde f_\Lor^0 -
    \frac{2 \cs^2}{1 + \cs^2} \uu \cdot [\tff_\ipf + \tff_\Lor] \biggr)
    \biggr]  \nonumber \\ 
    &\,  -  \frac{\cs^2}{1 + \cs^2}  \frac{\nab \ln \tilde \rho}{\gamma^2}
     + \frac{1}{1 + \cs^2} \,
    \frac{\tff_\ipf + \tff_\Lor}{\tilde \rho \gamma^2} +
    \frac{3 \cs^2 - 1}{1 - \cs^2 u^2} \frac{\uu \, \HH}{\gamma^2}\,,
\end{align}
\end{subequations}
where $D_\tau = \partial_\tau + \uu \cdot \nab$, and $\tilde f_\ipf^\mu$ and $\tilde f_\Lor^\mu$
are the imperfect (viscous) and Lorentz four-forces.

\subsection*{Subrelativistic limit}

In the subrelativistic limit, \Eqq{eqs_summary_rel_1} reduce to [cf.~\Eqq{subrel_eqs}]
\begin{subequations}
\begin{align}
    \hspace{-20mm}
    \lim_{u^2\ll 1}\partial_\tau \ln \tilde \rho =
    & - (1 + \cs^2) \,\nab \cdot \uu -
    (1 - \cs^2) (\uu \cdot \nab) \ln \tilde \rho
    \nonumber \\ &\, + \frac{1}{\tilde \rho} 
    \bigl[\tilde f^0_\ipf + \tilde f^0_\Lor - 2 \, \uu \cdot (\tff_\ipf + \tff_\Lor)\bigr]
    + \bigl[\beta - 3 (1 + \cs^2)\bigr]
    \,, \label{cont_summary_nonrel_1}   
\end{align}
\begin{align}
    \lim_{u^2\ll 1} D_\tau \uu = &\, \uu \, \cs^2 \biggl[\nab \cdot \uu +
    \frac{1 - \cs^2}{1 + \cs^2} (\uu \cdot \nab) \ln \tilde \rho \biggr]  - \frac{\uu}{\tilde \rho}
    \biggl[\tilde f^0_\ipf + \tilde f^0_\Lor  + \frac{2 \cs^2}
    {1 + \cs^2} \uu \cdot (\tff_\ipf + \tff_\Lor) \biggr]
    \nonumber \\ &\, 
    - \frac{\cs^2}{1 + \cs^2} \nab \ln \tilde \rho + \frac{1}{1 + \cs^2} \frac{\tff_\ipf + \tff_\Lor}{\tilde \rho}
    + (3 \cs^2 - 1) \, \uu \, \HH\,.
    \label{mom_summary_nonrel_1}
\end{align}
\end{subequations}
These equations present corrections due to the subrelativistic limit
of $D_\tau \ln \gamma^2$ that, up to our knowledge, have not been
taken into account in previous work (see discussion in \Sec{intro}
and the explicit corrections in \Sec{MHD_eom}, cf.~\Eqq{corr_eqs_MHD} for $\cs^2 = \third$,
and in \Eqq{eqs_cont2} for a generic constant $\cs^2$).
The subrelativistic heat flux, $\tilde f_\ipf^0$, and viscous forces, $\tilde f_\ipf^i$,
are presented in \Eq{visc_forces_sum}. The Lorentz four-force components
are given in \Eqs{fLor_summary}{ohms_subrel},
and
the work against viscous forces can be expressed as [cf.~\Eq{work_visc}],
\begin{equation}
    \frac{\uu \cdot \tff_\ipf}{\tilde p + \tilde \rho}
    \simeq
    - 2 \, \tilde\nu \, \tilde
    S^{ij} \, \tilde S_{ij} +
    \bigl(\twothird \tilde \nu - \tilde \xi\bigr)
    \, \tilde \theta^2\,,
\end{equation}
for homogeneous $\tilde \nu$ and $\tilde \xi$.

In the MHD limit, i.e., when the conductivity is large and the
displacement current can be neglected, the current density that
appears in the Lorentz force is $\tJJ = \nab \times \tBB$ and
the electric field is computed from Ohm's law [cf.~\Eq{E_Ohms}].
In this limit, when the Alfv\'en speed can become locally large, one
can include the Boris correction (see \Sec{boris_correction}) and modify \Eq{mom_summary_nonrel_1}
to the following
\begin{equation}
    \partial_\tau u^i =
    {\rm RHS}^i ({\uu}) - \vA^2 \gamma_A (\delta^{ij} - \hat b^i \hat b^j) \,
    {\rm RHS}_j
    ({\uu}) \,,
\end{equation}
where RHS$(\uu)$ is the right-hand-side of \Eq{mom_summary_nonrel_1}.
This correction allows to always maintain subluminal Alfv\'en speeds,
where
\begin{equation}
    \vA^2 = \frac{\tBB^2}{(1 + \cs^2) \tilde \rho}\,, \qquad
    \gamma_A = \frac{1}{1 + \vA^2}\,.
\end{equation}

\subsection{Maxwell equations}
\label{maxwell_summary}

Maxwell equations describe the dynamics of the electric and
magnetic fields (see \Sec{maxwell_eqs}),
\begin{equation}
    \partial_\tau \tEE = \nab \times \tBB - \tJJ\,, \qquad
    \partial_\tau \tBB = - \nab \times \tEE\,, \qquad
    \nab \cdot \tEE = \tilde J^0\,, \qquad \nab \cdot \tBB = 0\,,
    \label{Maxwell_summary}
\end{equation}
where the four-current $\tilde J^\mu$ is given by the generalized
Ohm's law.
In terms of the gauge field $A_\mu = (\chi, A_i)$, Maxwell equations
are
\begin{equation}
    \partial_\tau \tEE = - \nab^2 \AAA + \nab (\nab \cdot \AAA) - \tJJ\,, \qquad \partial_\tau \AAA = \nab \chi - \tEE\,, \qquad
    \nab \cdot \tEE = \tilde J^0\,, 
\end{equation}
where $\tBB = \nab \times \tAA$ and a particular gauge can be chosen
(e.g., $\chi = 0$ in the temporal gauge).
In general, this system of equations (in terms of $\tEE$ and $\tBB$
or in terms of $\tEE$ and $\AAA$) can be evolved together with the
equations of motion of the fluid, representing a closed system
of equations.

In the MHD limit of large conductivity, the displacement current can be neglected, $\partial_\tau \tEE = 0$.
Hence, Faraday's law becomes a constraint equation
$\tJJ = \nab \times \tBB$,
and Maxwell equations reduce to one
dynamical variable, $\tBB$,
described by the magnetic induction equation
(see \Sec{induction_sec})
\begin{equation}
    D_\tau \tBB = (\tBB \cdot \nab) \uu - \tBB (\nab \cdot \uu) 
    + \frac{\tilde \eta}{\gamma}
    \nab^2 \tBB + \tilde \eta \tilde \rho_e \nab \times \uu
    - \tilde \eta \,
    (1 - u^2) (\tJJ \times \nab \gamma)\,, \label{induction}
\end{equation}
where $\tEE$ has been substituted by Ohm's law [cf.~\Eq{E_Ohms}],
\begin{equation}
    \tEE = \frac{\tilde \eta}{\gamma} \tJJ - \tilde \eta \, \tilde \rho_e \, \uu -
    \uu \times \tBB\,,
\end{equation}
and
we have assumed homogeneous magnetic diffusivity $\tilde \eta = 1/\tilde \sigma$ and charge density $\tilde \rho_e$.
The last term in \Eq{induction} can explicitly be expressed as
\begin{equation}
    \tJJ \times \nab \gamma = (\nab \gamma \cdot \nab) \tBB - (\tBB \cdot \nab) \gamma\,.
\end{equation}
In terms of the vector potential, the induction equation is
\begin{equation}
    D_\tau \AAA = \nab \chi + \uu \cdot (\nab \AAA) + \frac{\tilde \eta}{\gamma}
    \, \bigl[\nab^2 \AAA - \nab (\nab \cdot \AAA)\bigr] + \tilde \eta \, \tilde \rho_e
    \uu \,.
\end{equation}
In the subrelativistic limit ($u^2 \ll 1$),
the induction equations for $\tBB$ and $\AAA$ simplify to
\begin{subequations}
\begin{align} 
    \lim_{u^2 \ll 1} D_\tau \tBB =& (\tBB \cdot \nab) \uu - \tBB (\nab \cdot \uu) +
    \tilde \eta \, \nab^2 \tBB + \tilde \eta \tilde \rho_e \nab \times \uu \,,  \\ 
    \lim_{u^2 \ll 1} D_\tau \AAA = & \nab \chi + \uu \cdot (\nab \AAA) + \tilde \eta\bigl[
    \nab^2 \AAA - \nab (\nab \cdot \AAA)\bigr] + \tilde \eta \tilde \rho_e \uu\,.
\end{align}
\end{subequations}

Finally, the Lorentz force that appears in the fluid equations of
motion [c.f.~\Eq{fLor_summary}] is obtained using $\tJJ = \nab \times \tBB$,
$\tBB = \nab \times \AAA$, and taking $\tEE$ from Ohm's law [cf.~\Eqs{f0_MHD_rel}{fi_MHD_rel}],
\begin{equation}
    \tilde f_\Lor^0 = \frac{\tilde \eta}{\gamma} \, \tJJ^2 - \tilde \eta
    \,
    \tilde \rho_e \, \uu \cdot \tJJ + \uu \cdot (\tJJ \times \tBB)\,, \qquad
    \tff_\Lor = \biggl(\frac{\tilde \rho_e}{\gamma} + \uu \cdot 
    \tJJ \biggr) \biggl(\frac{\tilde \eta}{\gamma} \tJJ - \tilde \eta 
    \tilde \rho_e \uu - \uu \times \tBB \biggr) + \tJJ \times \tBB\,.
\end{equation}

\acknowledgments

We are grateful to Jennifer Schober, Axel Brandenburg, Kenneth Marschall, Daniel G. Figueroa, Chiara Caprini,  Tanmay Vachaspati, Deepen Garg, Simona Procacci, and Ramkishor Sharma for useful
and stimulating discussions.
We thank the great support and hospitality provided by the Bernoulli Center in
Lausanne to fund the
``Generation, evolution, and observations of
cosmological magnetic fields'' program.
During this program, co-organized by ARP, part of this work was developed and lectured
by ARP, as part of the minicourse ``Simulations of Early Universe Magnetohydrodynamics'' lectured together with J. Schober at EPFL.
ARP is grateful to the students who participated in the lectures, raising questions,
concerns, and providing feedback that has been enormously beneficial for this work.
This work is supported by the Swiss National
Science Foundation (SNSF Ambizione grant \href{https://data.snf.ch/grants/grant/208807}{208807}).

\newpage

\appendix

\section{Vorticity production}
\label{vorticity}

In this appendix, we focus on the vorticity production in the case
of a relativistic fluid with $\cs^2 \sim {\cal O}(1)$.
Let us consider the relativistic momentum equation \Eq{mom_relativistic_mhd} for an equation of state $\tilde p = \cs^2 \tilde \rho$, where the squared speed of sound $\cs^2$
is a constant (e.g., $\cs^2 = \third$ for radiation-dominated fluids),
\begin{equation}
    D_\tau \uu =  
    \frac{\uu \Psi}{(1 - \cs^2 u^2)\gamma^2}
    - \frac{1}{1 + \cs^2} \frac{\nab \tilde p}{\tilde \rho \gamma^2} +  \frac{1}{1 + \cs^2} \frac{\tff_\tot}{\tilde \rho \gamma^2} \,, \label{mom_vort}
\end{equation}
where $\tff_\tot$ refers to any external
forces, e.g., viscous and Lorentz forces
$\tff_\tot = \tff_\ipf + \tff_\Lor$.
We omit the Hubble friction in the momentum equation, $\tff_H$, since it vanishes for the choice $\beta = 4$.
For compactness, we have defined the scalar $\Psi$ as
\begin{equation}
    \Psi = \cs^2 \, \biggl[\nab \cdot \uu + 
    \frac{1 - \cs^2}{1 + \cs^2} (\uu \cdot \nab) \ln \tilde \rho \biggr]
    - \frac{1}{\tilde \rho} \biggl[\tilde f^0_{\rm tot}
    + \frac{2 \, \cs^2}{1 + \cs^2}
    \uu \cdot \tff_\tot \biggr] +
    (3\cs^2 - 1) \HH \,,
    \label{Psi}
\end{equation}
where $\tilde f_\tot^0 = \tilde f_\ipf^0 + \tilde f_\Lor^0$ are
the viscous and Lorentz dissipative forces.
We can then take the curl of \Eq{mom_vort} to find a conservation equation for the
vorticity $\oom = \nab \times \uu$,
\begin{align}
    D_\tau \oom + 
    \oom \, \nab \cdot \uu -   (\oom \cdot \nab) \uu = &\,
    \frac{\oom \Psi}{(1 - \cs^2 u^2)\gamma^2}
    + \uu \times \nab \biggl[\frac{\Psi}{(1 - \cs^2u^2)\gamma^2}\biggr] \nonumber \\ &\, +
    \frac{1}{1 + \cs^2} \frac{(\nab \tilde p - \tff_\tot) \times \nab (\tilde \rho \gamma^2)}{\tilde \rho^2 \gamma^4} + \frac{1}{1 + \cs^2} \frac{\nab \times \tff_\tot}{\tilde \rho \gamma^2}\,. \label{vort_prodc}
\end{align}
To find the curl of the convective derivative, we
first take into account that
\begin{equation}
    \oom \times \uu = \bigl[(\uu \cdot \nab) \uu - \half \nab \uu^2\bigr]\,,
\end{equation}
such that $\nab \times [(\uu \cdot \nab) \uu] = \nab \times (\oom \times \uu)$. Then we apply the following identity,
\begin{equation}
    \nab \times (\oom \times \uu) = 
    \oom (\nab \cdot \uu) + (\uu \cdot \nab) \oom - (\oom \cdot \nab) \uu\,,
\end{equation}
where we have used $\nab \cdot \oom = 0$.
In \Eq{vort_prodc},
we have also used the following identity
\begin{equation}
    \nab \times (\uu \psi) = \psi \oom + \uu \times \nab \psi\,,
\end{equation}
which applies
for any scalar $\psi$.

Let us now focus on each of the
last three terms of \Eq{vort_prodc}, which
can lead to the production of vorticity even if $\oom$ is
zero initially,
\begin{equation}
    \lim_{|\oom| \to 0} D_\tau \oom = \uu \times \nab \biggl[
    \frac{\Psi}{(1 - \cs^2 u^2) \gamma^2}\biggr] +
    \frac{1}{1 + \cs^2} \frac{(\nab \tilde p - \tff_\tot) \times \nab (\tilde \rho \gamma^2)}{\tilde \rho^2 \gamma^4} + \frac{1}{1 + \cs^2} \frac{\nab \times \tff_\tot}{\tilde \rho \gamma^2} \label{vort_prod2}\,.
\end{equation}
The first of these terms is proportional to $\Psi$, which
vanishes when $\cs^2 \to 0$ (i.e., matter-dominated fluid),
in the absence of Hubble friction, and in the subrelativistic limit,
such that $\tilde f_\tot^0$ and $\uu \cdot \tff_\tot$ can be neglected.
However, when one of these terms is non-zero, vorticity can be
produced from an initial purely compressional (irrotational)
velocity field.
For example, the term arising from $\cs^2 = \third$ had been pointed
out and studied in \cite{Dahl:2021wyk,Dahl:2024eup}, while all the
subrelativistic terms are considered in \cite{vorticity}.
The remaining two terms in \Eq{vort_prod2}
correspond to the generalization of the
baroclinic term $\nab \tilde p \times \nab \tilde \rho$
and the vorticity production due to external forces (magnetic fields
or viscosity).
In general, unlike for inviscid
matter-dominated subrelativistic fluids in
flat space-time, where vorticity is a topological invariant described
by Euler equations, vorticity can be produced when one includes
a relativistic equation of state or the expansion of the
Universe, even for a perfect fluid
with no dissipation and neglecting external forces.
In the following, we briefly describe each of the potential
terms that lead to the production of vorticity.

\subsection*{A.1 \ Baroclinic contribution}

The baroclinic term contributing to the production of vorticity is 
the following,
\begin{equation}
    \frac{\nab \tilde p \times \nab (\tilde \rho \gamma^2)}{\tilde \rho^2 \gamma^4}
    = 
    \frac{\nab \tilde p \times \nab \tilde \rho}{\tilde \rho^2 \gamma^2} + \frac{\nab \tilde p \times \nab u^2}{\tilde \rho}\,,
\end{equation}
where we have used $\nab \gamma^2 = \gamma^4 \nab u^2$.
The first term is the usual subrelativistic baroclinic term
and it becomes zero whenever the pressure is only a function of $\tilde \rho$,
while
the second contribution is a relativistic term that can lead to the production
of vorticity even when the first term is zero, as it is proportional to the
misalignment between the pressure and the speed gradients.

\subsection*{A.2 \ Vorticity production when $\cs^2 \neq 0$ and $\tilde f^\mu_\tot = \tilde f_H^\mu = 0$}

We now consider the term that can lead to the production of vorticity
when $\cs^2 \neq 0$, in the absence of external forces $\tilde f^\mu_\tot = 0$ and Hubble friction,
\begin{equation}
    \uu \times \nab \biggl[
    \frac{\Psi}{(1 - \cs^2 u^2) \gamma^2}\biggr] = 
    \frac{\uu \times \nab \Psi}{(1 - \cs^2 u^2)\gamma^2}
    - \frac{\cs^2 - \gamma^2}{1 - \cs^2 u^2} \, \Psi \, \uu \times \nab u^2\,.
\end{equation}
The second term is a relativistic term that leads to the production of vorticity
with a strength proportional to the misalignment of $\uu$ and the gradient
of $u^2$.
The first contribution is proportional to
\begin{align}
    \uu \times \nab \Psi = &\, \cs^2 \uu \times [\nab (\nab \cdot \uu)] +
    \cs^2 \frac{1 - \cs^2}{1 + \cs^2}  \uu \times [(\uu \cdot \nab)     \ln \tilde \rho ]\,,
\end{align}
which is, in general, different than zero, so it can lead to the production
of vorticity in the subrelativistic limit.

\subsection*{A.3 \ Vorticity production when $\cs^2 = 0$ and $\tilde f_\tot^\mu \neq 0$}

When $\cs^2 = 0$, the term proportional to $\Psi$ is zero.
The remaining term that can produce vorticity can be split
in the following way,
\begin{equation}
    \frac{\nab \times \tff_\tot}{\tilde \rho \gamma^2}
    - \frac{\tff_\tot \times \nab (\tilde \rho \gamma^2)}{\tilde \rho^2 \gamma^4}  =
    \frac{\nab \times \tff_\tot}{\tilde \rho \gamma^2} -
    \frac{\tff_\tot \times \nab \tilde \rho}{\tilde \rho^2 \gamma^2} -
    \frac{\tff_\tot \times \nab u^2}{\tilde \rho} \,.
\end{equation}
The first two terms describe the usual subrelativistic
production of vorticity in the presence
of external forces, while the third term corresponds to a relativistic
term producing vorticity that will be non-zero when the external
forces and gradients of $u^2$ are misaligned.

\newpage

\bibliographystyle{JCAP}
\bibliography{references}

@article{Caprini:2024gyk,
    author = "Caprini, Chiara and Jinno, Ryusuke and Konstandin, Thomas and Roper Pol, Alberto and Rubira, Henrique and Stomberg, Isak",
    title = "{Gravitational waves from first-order phase transitions: from weak to strong}",
    eprint = "2409.03651",
    archivePrefix = "arXiv",
    primaryClass = "gr-qc",
    doi = "10.1007/JHEP07(2025)217",
    journal = "JHEP",
    volume = "07",
    pages = "217",
    year = "2025"
}

@article{MaxwellJuttner,
    author = {J\"uttner, Ferencz},
    title = "{Das Maxwellsche gesetz der geschwindigkeitsverteilung in der relativtheorie}",
    journal = "{Annalen der Physik}",
    volume = 339,
    pages = {856},
    year = {1911}
}

@article{Baym:1997gq,
    author = "Baym, Gordon and Heiselberg, Henning",
    title = "{The Electrical conductivity in the early universe}",
    eprint = "astro-ph/9704214",
    archivePrefix = "arXiv",
    doi = "10.1103/PhysRevD.56.5254",
    journal = "Phys. Rev. D",
    volume = "56",
    pages = "5254--5259",
    year = "1997"
}

@article{Brandenburg:1996sa,
    author = "Brandenburg, Axel and Enqvist, Kari and Olesen, Poul",
    title = "{The Effect of Silk damping on primordial magnetic fields}",
    eprint = "hep-ph/9608422",
    archivePrefix = "arXiv",
    reportNumber = "HU-TFT-96-35",
    doi = "10.1016/S0370-2693(96)01566-3",
    journal = "Phys. Lett. B",
    volume = "392",
    pages = "395--402",
    year = "1997"
}

@article{Navier,
    author = {Navier, C. L.},
    title = "{Sur les lois des mouvements des fluides, en ayant \'egard \`a l’adh\'esion des molecules}",
    journal = "{Annales de Chimie et de Physique}",
    pages = {244},
    year = {1821}
}

@article{Stokes,
    author = {Stokes, G. G.},
    title = "{On the Theories of the Internal Friction of Fluids in Motion, and of the Equilibrium and
Motion of Elastic Solid}",
    journal = "{Cambridge
University Press}",
    pages = {75},
    year = {1880}
}

@article{Uchida:2022vue,
    author = "Uchida, Fumio and Fujiwara, Motoko and Kamada, Kohei and Yokoyama, Jun'ichi",
    title = "{New description of the scaling evolution of the cosmological magneto-hydrodynamic system}",
    eprint = "2212.14355",
    archivePrefix = "arXiv",
    primaryClass = "astro-ph.CO",
    reportNumber = "RESCEU-9/22",
    doi = "10.1016/j.physletb.2023.138002",
    journal = "Phys. Lett. B",
    volume = "843",
    pages = "138002",
    year = "2023"
}

@article{Uchida:2024ude,
    author = "Uchida, Fumio and Fujiwara, Motoko and Kamada, Kohei and Yokoyama, Jun'ichi",
    title = "{New comprehensive description of the scaling evolution of the cosmological magneto-hydrodynamic system}",
    eprint = "2405.06194",
    archivePrefix = "arXiv",
    primaryClass = "astro-ph.CO",
    reportNumber = "RESCEU-10/24, KEK-TH-2620",
    doi = "10.1088/1475-7516/2025/08/017",
    journal = "JCAP",
    volume = "08",
    pages = "017",
    year = "2025"
}

@ARTICLE{1977ApJ...211..361L,
       author = {{Liang}, E.~P.~T.},
        title = "{Relativistic simple waves: shock damping and entropy production.}",
      journal = {Astrophys.\,Journal},
         year = 1977,
        month = jan,
       volume = {211},
        pages = {361-376},
          doi = {10.1086/154942},
       adsurl = {https://ui.adsabs.harvard.edu/abs/1977ApJ...211..361L}
}

@article{Parker:1968mv,
    author = "Parker, L.",
    title = "{Particle creation in expanding universes}",
    doi = "10.1103/PhysRevLett.21.562",
    journal = "Phys. Rev. Lett.",
    volume = "21",
    pages = "562--564",
    year = "1968"
}

@article{Arnold:2000dr,
    author = "Arnold, Peter Brockway and Moore, Guy D. and Yaffe, Laurence G.",
    title = "{Transport coefficients in high temperature gauge theories. 1. Leading log results}",
    eprint = "hep-ph/0010177",
    archivePrefix = "arXiv",
    reportNumber = "UW-PT-00-15",
    doi = "10.1088/1126-6708/2000/11/001",
    journal = "JHEP",
    volume = "11",
    pages = "001",
    year = "2000"
}

@article{Durrer:2013pga,
    author = "Durrer, Ruth and Neronov, Andrii",
    title = "{Cosmological Magnetic Fields: Their Generation, Evolution and Observation}",
    eprint = "1303.7121",
    archivePrefix = "arXiv",
    primaryClass = "astro-ph.CO",
    doi = "10.1007/s00159-013-0062-7",
    journal = "Astron. Astrophys. Rev.",
    volume = "21",
    pages = "62",
    year = "2013"
}

@article{Dahl:2024eup,
    author = "Dahl, Jani and Hindmarsh, Mark and Rummukainen, Kari and Weir, David J.",
    title = "{Primordial acoustic turbulence: Three-dimensional simulations and gravitational wave predictions}",
    eprint = "2407.05826",
    archivePrefix = "arXiv",
    primaryClass = "gr-qc",
    reportNumber = "HIP-2024-16/TH",
    doi = "10.1103/PhysRevD.110.103512",
    journal = "Phys. Rev. D",
    volume = "110",
    number = "10",
    pages = "103512",
    year = "2024"
}

@article{Banerjee:2003xk,
    author = "Banerjee, Robi and Jedamzik, Karsten",
    title = "{Are cluster magnetic fields primordial?}",
    eprint = "astro-ph/0306211",
    archivePrefix = "arXiv",
    doi = "10.1103/PhysRevLett.93.179901",
    journal = "Phys. Rev. Lett.",
    volume = "91",
    pages = "251301",
    year = "2003",
    note = "[Erratum: Phys.Rev.Lett. 93, 179901 (2004)]"
}

@book{Weinberg:1972kfs,
    author = "Weinberg, Steven",
    title = "{Gravitation and Cosmology}: {Principles and Applications of the General Theory of Relativity}",
    isbn = "978-0-471-92567-5, 978-0-471-92567-5",
    publisher = "John Wiley and Sons",
    address = "New York",
    year = "1972"
}

@article{Kahniashvili:2012vt,
    author = "Kahniashvili, Tina and Brandenburg, Axel and Campanelli, Leonardo and Ratra, Bharat and Tevzadze, Alexander G.",
    title = "{Evolution of inflation-generated magnetic field through phase transitions}",
    eprint = "1206.2428",
    archivePrefix = "arXiv",
    primaryClass = "astro-ph.CO",
    reportNumber = "NORDITA-2012-45, KSUPT-12-2",
    doi = "10.1103/PhysRevD.86.103005",
    journal = "Phys. Rev. D",
    volume = "86",
    pages = "103005",
    year = "2012"
}

@article{Brandenburg:2014mwa,
    author = "Brandenburg, Axel and Kahniashvili, Tina and Tevzadze, Alexander G.",
    title = "{Nonhelical inverse transfer of a decaying turbulent magnetic field}",
    eprint = "1404.2238",
    archivePrefix = "arXiv",
    primaryClass = "astro-ph.CO",
    reportNumber = "NORDITA-2014-42",
    doi = "10.1103/PhysRevLett.114.075001",
    journal = "Phys. Rev. Lett.",
    volume = "114",
    number = "7",
    pages = "075001",
    year = "2015"
}

@article{Kahniashvili:2012uj,
    author = "Kahniashvili, Tina and Tevzadze, Alexander G. and Brandenburg, Axel and Neronov, Andrii",
    title = "{Evolution of Primordial Magnetic Fields from Phase Transitions}",
    eprint = "1212.0596",
    archivePrefix = "arXiv",
    primaryClass = "astro-ph.CO",
    reportNumber = "NORDITA-2012-96",
    doi = "10.1103/PhysRevD.87.083007",
    journal = "Phys. Rev. D",
    volume = "87",
    number = "8",
    pages = "083007",
    year = "2013"
}

@article{Christensson:2002xu,
    author = "Christensson, Mattias and Hindmarsh, Mark and Brandenburg, Axel",
    title = "{Scaling laws in decaying helical 3-D magnetohydrodynamic turbulence}",
    eprint = "astro-ph/0209119",
    archivePrefix = "arXiv",
    reportNumber = "SUSX-TH-02-014, NORDITA-2002-43-AP",
    doi = "10.1002/asna.200510365",
    journal = "Astron. Nachr.",
    volume = "326",
    pages = "393--399",
    year = "2005"
}

@article{jay_p__boris_1970,
        title={ A Physically Motivated Solution of the Alfv\'en Problem },
        author={ Jay P. Boris },
        year = 1970,
        journal ={ NRL Memorandum Report, },
        doi={ 10.21236/AD0715774 },  
      }

@article{Caprini:2024hue,
    author = "Caprini, Chiara and Jinno, Ryusuke and Lewicki, Marek and Madge, Eric and Merchand, Marco and Nardini, Germano and Pieroni, Mauro and Roper Pol, Alberto and Vaskonen, Ville",
    collaboration = "LISA Cosmology Working Group",
    title = "{Gravitational waves from first-order phase transitions in LISA: reconstruction pipeline and physics interpretation}",
    eprint = "2403.03723",
    archivePrefix = "arXiv",
    primaryClass = "astro-ph.CO",
    reportNumber = "LISA-COSWG-24-01, CERN-TH-2024-029",
    doi = "10.1088/1475-7516/2024/10/020",
    journal = "JCAP",
    volume = "10",
    pages = "020",
    year = "2024"
}

@article{EPTA:2023xxk,
    author = "Antoniadis, J. and others",
    collaboration = "EPTA, InPTA",
    title = "{The second data release from the European Pulsar Timing Array - IV. Implications for massive black holes, dark matter, and the early Universe}",
    eprint = "2306.16227",
    archivePrefix = "arXiv",
    primaryClass = "astro-ph.CO",
    doi = "10.1051/0004-6361/202347433",
    journal = "Astron. Astrophys.",
    volume = "685",
    pages = "A94",
    year = "2024"
}

@article{RoperPol:2023bqa,
    author = "Roper Pol, A. and Neronov, A. and Caprini, C. and Boyer, T. and Semikoz, D.",
    title = "{LISA and $\gamma$-ray telescopes as multi-messenger probes of a first-order cosmological phase transition}",
    eprint = "2307.10744",
    archivePrefix = "arXiv",
    primaryClass = "astro-ph.CO",
    doi = "10.1051/0004-6361/202658936",
    journal = "Astron. Astrophys.",
    volume = "708",
    pages = "A337",
    year = "2026"
}

@ARTICLE{2002JCoPh.177..176G,
       author = {{Gombosi}, Tamas I. and {T{\'o}th}, G{\'a}bor and {De Zeeuw}, Darren L. and {Hansen}, Kenneth C. and {Kabin}, Konstantin and {Powell}, Kenneth G.},
        title = "{Semirelativistic Magnetohydrodynamics and Physics-Based Convergence Acceleration}",
      journal = {Journal of Computational Physics},
         year = 2002,
        month = mar,
       volume = {177},
       number = {1},
        pages = {176-205},
          doi = {10.1006/jcph.2002.7009},
       adsurl = {https://ui.adsabs.harvard.edu/abs/2002JCoPh.177..176G}
}

@article{2018arXiv180608166C,
author = {Piyali Chatterjee},
title = {Testing Alfvén wave propagation in a “realistic” set-up of the solar atmosphere},
journal = {Geophys. Astrophys. Fluid Dynamics},
volume = {114},
number = {1-2},
pages = {213--234},
year = {2020},
publisher = {Taylor \& Francis},
doi = {10.1080/03091929.2019.1672676},
eprint = {1806.08166}
}

@article{Brandenburg:2004jv,
    author = "Brandenburg, Axel and Subramanian, Kandaswamy",
    title = "{Astrophysical magnetic fields and nonlinear dynamo theory}",
    eprint = "astro-ph/0405052",
    archivePrefix = "arXiv",
    reportNumber = "NORDITA-2004-37",
    doi = "10.1016/j.physrep.2005.06.005",
    journal = "Phys. Rept.",
    volume = "417",
    pages = "1--209",
    year = "2005"
}

@book{Rezzolla:2013dea,
    author = "Rezzolla, Luciano and Zanotti, Olindo",
    title = "{Relativistic Hydrodynamics}",
    doi = "10.1093/acprof:oso/9780198528906.001.0001",
    isbn = "978-0-19-174650-5, 978-0-19-852890-6",
    publisher = "Oxford University Press",
    month = "9",
    year = "2013"
}

@article{PC_relativistic,
  title = "{Simulations of relativistic MHD in Pencil Code}",
  author = {Brandenburg, A. and Roper Pol, A.},
  journal = "{{\em in preparation\!\!}}",
}

@article{CosmoLattice_MHD,
  title = "{The art of simulating the early Universe.
Part III. Scalar-Gauge-Fluid Dynamics}",
  author = {Figueroa, D.~G. and Marschall, K. and Midiri, A.~S. and Roper Pol, A.},
  journal = "{{\em in preparation\!\!}}",
}

@article{vorticity,
  author = {Caprini, C. and Garg, D. and Roper Pol, A.},
  title = {Vorticity generation in relativistic perfect fluids with
  application to first-order phase transitions},
  journal = "{{\em in preparation\!\!}}",
}

@article{Dahl:2021wyk,
    author = "Dahl, Jani and Hindmarsh, Mark and Rummukainen, Kari and Weir, David J.",
    title = "{Decay of acoustic turbulence in two dimensions and implications for cosmological gravitational waves}",
    eprint = "2112.12013",
    archivePrefix = "arXiv",
    primaryClass = "gr-qc",
    reportNumber = "HIP-2021-29/TH",
    doi = "10.1103/PhysRevD.106.063511",
    journal = "Phys. Rev. D",
    volume = "106",
    number = "6",
    pages = "063511",
    year = "2022"
}

@article{Ghosh:2025bqp,
    author = "Ghosh, Oindrila and Brandenburg, Axel and Caprini, Chiara and Neronov, Andrii and Vazza, Franco",
    title = "{Can galactic magnetic fields diffuse into the voids?}",
    eprint = "2510.26918",
    archivePrefix = "arXiv",
    primaryClass = "astro-ph.CO",
    reportNumber = "NORDITA-2025-055, CERN-TH-2025-219",
    doi = "10.1103/4114-mgsp",
    journal = "Phys. Rev. D",
    volume = "113",
    number = "2",
    pages = "023523",
    year = "2026"
}

@book{Cercignani2002RelativisticBoltzmann,
  author    = {Carlo Cercignani and Gilberto Medeiros Kremer},
  title     = {The Relativistic Boltzmann Equation: Theory and Applications},
  publisher = {Birkh{\"a}user Basel},
  year      = {2002},
  series    = {Progress in Mathematical Physics},
  volume    = {22},
  isbn      = {978-3-7643-6693-3},
  doi       = {10.1007/978-3-0348-8165-4}
}

@book{Wald:1984rg,
    author = "Wald, Robert M.",
    title = "{General Relativity}",
    doi = "10.7208/chicago/9780226870373.001.0001",
    publisher = "Chicago Univ. Pr.",
    address = "Chicago, USA",
    year = "1984"
}

@book{Anderson:1992,
    author = "Anderson Jr., John D.",
    title = "{Computational Fluid Dynamics. The Basics with Applications}",
    publisher = "McGraw Hill",
    address = "New York, USA",
    year = "1992"
}

@book{Carroll:2004st,
    author = "Carroll, Sean M.",
    title = "{Spacetime and Geometry}: {An Introduction to General Relativity}",
    doi = "10.1017/9781108770385",
    isbn = "978-0-8053-8732-2, 978-1-108-48839-6, 978-1-108-77555-7",
    publisher = "Cambridge University Press",
    month = "7",
    year = "2019"
}

@ARTICLE{Subramanian1997,
   author = {{Subramanian}, K.},
    title = "{Dynamics of fluctuating magnetic fields in turbulent dynamos incorporating ambipolar drifts}",
eprint = "astro-ph/9708216",
    archivePrefix = "arXiv",
     year = 1997,
   adsurl = {http://adsabs.harvard.edu/abs/1997astro.ph..8216S}
}

@BOOK{Biskamp2003,
       author = {{Biskamp}, Dieter},
        title = "{Magnetohydrodynamic Turbulence}",
         year = 2003,
       adsurl = {https://ui.adsabs.harvard.edu/abs/2003matu.book.....B},
      place={Cambridge},
      publisher={Cambridge University Press}
}

@book{Misner:1973prb,
    author = "Misner, Charles W. and Thorne, K. S. and Wheeler, J. A.",
    title = "{Gravitation}",
    isbn = "978-0-7167-0344-0, 978-0-691-17779-3",
    publisher = "W. H. Freeman",
    address = "San Francisco",
    year = "1973"
}

@article{Font:2008fka,
    author = "Font, Jose A.",
    title = "{Numerical Hydrodynamics and Magnetohydrodynamics in General Relativity}",
    doi = "10.12942/lrr-2008-7",
    journal = "Living Rev. Rel.",
    volume = "11",
    pages = "7",
    year = "2008"
}

@article{Subramanian:1997gi,
    author = "Subramanian, Kandaswamy and Barrow, John D.",
    title = "{Magnetohydrodynamics in the early universe and the damping of nonlinear Alfven waves}",
    eprint = "astro-ph/9712083",
    archivePrefix = "arXiv",
    doi = "10.1103/PhysRevD.58.083502",
    journal = "Phys. Rev. D",
    volume = "58",
    pages = "083502",
    year = "1998"
}

@BOOK{2008Kundu,
       author = {{Kundu}, Pijush K. and {Cohen}, Ira M.},
        title = "{Fluid Mechanics: Fourth Edition}",
         year = 2008,
       adsurl = {https://ui.adsabs.harvard.edu/abs/2008flme.book.....K}
}

@BOOK{1970mtnu.book.....C,
       author = {{Chapman}, Sydeny and {Cowling}, T.~G.},
        title = "{The mathematical theory of non-uniform gases. an account of the kinetic theory of viscosity, thermal conduction and diffusion in gases}",
         year = 1970,
       adsurl = {https://ui.adsabs.harvard.edu/abs/1970mtnu.book.....C}
}

@article{Romatschke:2009im,
    author = "Romatschke, Paul",
    title = "{New Developments in Relativistic Viscous Hydrodynamics}",
    eprint = "0902.3663",
    archivePrefix = "arXiv",
    primaryClass = "hep-ph",
    reportNumber = "INT-PUB-09-010",
    doi = "10.1142/S0218301310014613",
    journal = "Int. J. Mod. Phys. E",
    volume = "19",
    pages = "1--53",
    year = "2010"
}

@article{RoperPol:2022iel,
    author = "Roper Pol, Alberto and Caprini, Chiara and Neronov, Andrii and Semikoz, Dmitri",
    title = "{Gravitational wave signal from primordial magnetic fields in the Pulsar Timing Array frequency band}",
    eprint = "2201.05630",
    archivePrefix = "arXiv",
    primaryClass = "astro-ph.CO",
    doi = "10.1103/PhysRevD.105.123502",
    journal = "Phys. Rev. D",
    volume = "105",
    number = "12",
    pages = "123502",
    year = "2022"
}

@article{Figueroa:2020rrl,
    author = "Figueroa, Daniel G. and Florio, Adrien and Torrenti, Francisco and Valkenburg, Wessel",
    title = "{The art of simulating the early Universe -- Part I}",
    eprint = "2006.15122",
    archivePrefix = "arXiv",
    primaryClass = "astro-ph.CO",
    doi = "10.1088/1475-7516/2021/04/035",
    journal = "JCAP",
    volume = "04",
    pages = "035",
    year = "2021"
}

@book{Shukurov_Subramanian_2021, place={Cambridge}, series={Cambridge Astrophysics}, title={Astrophysical Magnetic Fields: From Galaxies to the Early Universe}, publisher={Cambridge University Press}, author={Shukurov, Anvar and Subramanian, Kandaswamy}, year={2021}, collection={Cambridge Astrophysics}}

@inproceedings{RoperPol:2021gjc,
    author = "Roper Pol, Alberto",
    title = "{Gravitational radiation from MHD turbulence in the early universe}",
    booktitle = "{55th Rencontres de Moriond on Gravitation}",
    eprint = "2105.08287",
    archivePrefix = "arXiv",
    primaryClass = "gr-qc",
    month = "5",
    year = "2021"
}

@inproceedings{RoperPol:2022hxn,
    author = "Roper Pol, Alberto",
    title = "{Gravitational waves from MHD turbulence at the QCD phase transition as a source for Pulsar Timing Arrays}",
    booktitle = "{56th Rencontres de Moriond on Gravitation}",
    eprint = "2205.09261",
    archivePrefix = "arXiv",
    primaryClass = "gr-qc",
    month = "5",
    year = "2022"
}

@article{Banerjee:2004df,
    author = "Banerjee, Robi and Jedamzik, Karsten",
    title = "{The Evolution of cosmic magnetic fields: From the very early universe, to recombination, to the present}",
    eprint = "astro-ph/0410032",
    archivePrefix = "arXiv",
    doi = "10.1103/PhysRevD.70.123003",
    journal = "Phys. Rev. D",
    volume = "70",
    pages = "123003",
    year = "2004"
}

@article{Christensson:2000sp,
    author = "Christensson, Mattias and Hindmarsh, Mark and Brandenburg, Axel",
    title = "{Inverse cascade in decaying 3-D magnetohydrodynamic turbulence}",
    eprint = "astro-ph/0011321",
    archivePrefix = "arXiv",
    doi = "10.1103/PhysRevE.64.056405",
    journal = "Phys. Rev. E",
    volume = "64",
    pages = "056405",
    year = "2001"
}

@article{Figueroa:2021yhd,
    author = "Figueroa, Daniel G. and Florio, Adrien and Torrenti, Francisco and Valkenburg, Wessel",
    title = "{${\cal C}$osmo${\cal L}$attice: A modern code for lattice simulations of scalar and gauge field dynamics in an expanding universe}",
    eprint = "2102.01031",
    archivePrefix = "arXiv",
    primaryClass = "astro-ph.CO",
    doi = "10.1016/j.cpc.2022.108586",
    journal = "Comput. Phys. Commun.",
    volume = "283",
    pages = "108586",
    year = "2023"
}

@article{Figueroa:2023xmq,
    author = "Figueroa, Daniel G. and Florio, Adrien and Torrenti, Francisco",
    title = "{Present and future of ${\mathcal{C}}$osmo${\mathcal{L}}$attice}",
    eprint = "2312.15056",
    archivePrefix = "arXiv",
    primaryClass = "astro-ph.CO",
    doi = "10.1088/1361-6633/ad616a",
    journal = "Rept. Prog. Phys.",
    volume = "87",
    number = "9",
    pages = "094901",
    year = "2024"
}

@article{Brandenburg:2017rnt,
    author = "Brandenburg, Axel and Kahniashvili, Tina and Mandal, Sayan and Roper Pol, Alberto and Tevzadze, Alexander G. and Vachaspati, Tanmay",
    title = "{The dynamo effect in decaying helical turbulence}",
    eprint = "1710.01628",
    archivePrefix = "arXiv",
    primaryClass = "physics.flu-dyn",
    reportNumber = "NORDITA-2017-099",
    doi = "10.1103/PhysRevFluids.4.024608",
    journal = "Phys. Rev. Fluids.",
    volume = "4",
    pages = "024608",
    year = "2019"
}

@article{Brandenburg:2017neh,
    author = "Brandenburg, Axel and Kahniashvili, Tina and Mandal, Sayan and Roper Pol, Alberto and Tevzadze, Alexander G. and Vachaspati, Tanmay",
    title = "{Evolution of hydromagnetic turbulence from the electroweak phase transition}",
    eprint = "1711.03804",
    archivePrefix = "arXiv",
    primaryClass = "astro-ph.CO",
    reportNumber = "NORDITA-2017-116",
    doi = "10.1103/PhysRevD.96.123528",
    journal = "Phys. Rev. D",
    volume = "96",
    number = "12",
    pages = "123528",
    year = "2017"
}

@article{Brandenburg:1996fc,
    author = "Brandenburg, Axel and Enqvist, Kari and Olesen, Poul",
    title = "{Large scale magnetic fields from hydromagnetic turbulence in the very early universe}",
    eprint = "astro-ph/9602031",
    archivePrefix = "arXiv",
    reportNumber = "NORDITA-96-6-A",
    doi = "10.1103/PhysRevD.54.1291",
    journal = "Phys. Rev. D",
    volume = "54",
    pages = "1291--1300",
    year = "1996"
}

@article{Brandenburg:2020vwp,
    author = "Brandenburg, Axel and Durrer, Ruth and Huang, Yiwen and Kahniashvili, Tina and Mandal, Sayan and Mukohyama, Shinji",
    title = "{Primordial magnetic helicity evolution with a homogeneous magnetic field from inflation}",
    eprint = "2005.06449",
    archivePrefix = "arXiv",
    primaryClass = "astro-ph.CO",
    reportNumber = "NORDITA-2020-044, YITP-20-47, IPMU20-0042",
    doi = "10.1103/PhysRevD.102.023536",
    journal = "Phys. Rev. D",
    volume = "102",
    number = "2",
    pages = "023536",
    year = "2020"
}

@article{Brandenburg:2018ptt,
    author = "Brandenburg, Axel and Durrer, Ruth and Kahniashvili, Tina and Mandal, Sayan and Yin, Weichen Winston",
    title = "{Statistical Properties of Scale-Invariant Helical Magnetic Fields and Applications to Cosmology}",
    eprint = "1804.01177",
    archivePrefix = "arXiv",
    primaryClass = "astro-ph.CO",
    reportNumber = "NORDITA-2018-024",
    doi = "10.1088/1475-7516/2018/08/034",
    journal = "JCAP",
    volume = "08",
    pages = "034",
    year = "2018"
}

@article{Jinno:2020eqg,
    author = "Jinno, Ryusuke and Konstandin, Thomas and Rubira, Henrique",
    title = "{A hybrid simulation of gravitational wave production in first-order phase transitions}",
    eprint = "2010.00971",
    archivePrefix = "arXiv",
    primaryClass = "astro-ph.CO",
    reportNumber = "DESY-20-170, DESY 20-170",
    doi = "10.1088/1475-7516/2021/04/014",
    journal = "JCAP",
    volume = "04",
    pages = "014",
    year = "2021"
}

@article{Brandenburg:2021tmp,
    author = "Brandenburg, Axel and Clarke, Emma and He, Yutong and Kahniashvili, Tina",
    title = "{Can we observe the QCD phase transition-generated gravitational waves through pulsar timing arrays?}",
    eprint = "2102.12428",
    archivePrefix = "arXiv",
    primaryClass = "astro-ph.CO",
    reportNumber = "NORDITA-2021-016",
    doi = "10.1103/PhysRevD.104.043513",
    journal = "Phys. Rev. D",
    volume = "104",
    number = "4",
    pages = "043513",
    year = "2021"
}

@article{Brandenburg:2021bvg,
    author = "Brandenburg, Axel and Gogoberidze, Grigol and Kahniashvili, Tina and Mandal, Sayan and Roper Pol, Alberto and Shenoy, Nakul",
    title = "{The scalar, vector, and tensor modes in gravitational wave turbulence simulations}",
    eprint = "2103.01140",
    archivePrefix = "arXiv",
    primaryClass = "gr-qc",
    reportNumber = "NORDITA-2021-019",
    doi = "10.1088/1361-6382/ac011c",
    journal = "Class. Quant. Grav.",
    volume = "38",
    number = "14",
    pages = "145002",
    year = "2021"
}

@article{Kahniashvili:2020jgm,
    author = "Kahniashvili, Tina and Brandenburg, Axel and Gogoberidze, Grigol and Mandal, Sayan and Roper Pol, Alberto",
    title = "{Circular polarization of gravitational waves from early-Universe helical turbulence}",
    eprint = "2011.05556",
    archivePrefix = "arXiv",
    primaryClass = "astro-ph.CO",
    reportNumber = "NORDITA-2020-102",
    doi = "10.1103/PhysRevResearch.3.013193",
    journal = "Phys. Rev. Res.",
    volume = "3",
    number = "1",
    pages = "013193",
    year = "2021"
}

@article{RoperPol:2019wvy,
    author = "Roper Pol, Alberto and Mandal, Sayan and Brandenburg, Axel and Kahniashvili, Tina and Kosowsky, Arthur",
    title = "{Numerical simulations of gravitational waves from early-universe turbulence}",
    eprint = "1903.08585",
    archivePrefix = "arXiv",
    primaryClass = "astro-ph.CO",
    reportNumber = "NORDITA-2019-024",
    doi = "10.1103/PhysRevD.102.083512",
    journal = "Phys. Rev. D",
    volume = "102",
    number = "8",
    pages = "083512",
    year = "2020"
}

@article{RoperPol:2018sap,
    author = "Roper Pol, Alberto and Brandenburg, Axel and Kahniashvili, Tina and Kosowsky, Arthur and Mandal, Sayan",
    title = "{The timestep constraint in solving the gravitational wave equations sourced by hydromagnetic turbulence}",
    eprint = "1807.05479",
    archivePrefix = "arXiv",
    primaryClass = "physics.flu-dyn",
    reportNumber = "NORDITA-2018-054",
    doi = "10.1080/03091929.2019.1653460",
    journal = "Geophys. Astrophys. Fluid Dynamics",
    volume = "114",
    number = "1-2",
    pages = "130--161",
    year = "2020"
}

@article{Vachaspati:2020blt,
    author = "Vachaspati, Tanmay",
    title = "{Progress on cosmological magnetic fields}",
    eprint = "2010.10525",
    archivePrefix = "arXiv",
    primaryClass = "astro-ph.CO",
    doi = "10.1088/1361-6633/ac03a9",
    journal = "Rept. Prog. Phys.",
    volume = "84",
    number = "7",
    pages = "074901",
    year = "2021"
}

@article{Subramanian:2015lua,
    author = "Subramanian, Kandaswamy",
    title = "{The origin, evolution and signatures of primordial magnetic fields}",
    eprint = "1504.02311",
    archivePrefix = "arXiv",
    primaryClass = "astro-ph.CO",
    doi = "10.1088/0034-4885/79/7/076901",
    journal = "Rept. Prog. Phys.",
    volume = "79",
    number = "7",
    pages = "076901",
    year = "2016"
}

@article{Joyce:1997uy,
    author = "Joyce, M. and Shaposhnikov, Mikhail E.",
    title = "{Primordial magnetic fields, right-handed electrons, and the Abelian anomaly}",
    eprint = "astro-ph/9703005",
    archivePrefix = "arXiv",
    reportNumber = "CERN-TH-97-31",
    doi = "10.1103/PhysRevLett.79.1193",
    journal = "Phys. Rev. Lett.",
    volume = "79",
    pages = "1193--1196",
    year = "1997"
}

@article{Caprini:2009yp,
    author = "Caprini, Chiara and Durrer, Ruth and Servant, Geraldine",
    title = "{The stochastic gravitational wave background from turbulence and magnetic fields generated by a first-order phase transition}",
    eprint = "0909.0622",
    archivePrefix = "arXiv",
    primaryClass = "astro-ph.CO",
    doi = "10.1088/1475-7516/2009/12/024",
    journal = "JCAP",
    volume = "12",
    pages = "024",
    year = "2009"
}

@article{Jedamzik:2020krr,
    author = "Jedamzik, Karsten and Pogosian, Levon",
    title = "{Relieving the Hubble tension with primordial magnetic fields}",
    eprint = "2004.09487",
    archivePrefix = "arXiv",
    primaryClass = "astro-ph.CO",
    doi = "10.1103/PhysRevLett.125.181302",
    journal = "Phys. Rev. Lett.",
    volume = "125",
    number = "18",
    pages = "181302",
    year = "2020"
}

@article{Jedamzik:2018itu,
    author = "Jedamzik, Karsten and Saveliev, Andrey",
    title = "{Stringent Limit on Primordial Magnetic Fields from the Cosmic Microwave Background Radiation}",
    eprint = "1804.06115",
    archivePrefix = "arXiv",
    primaryClass = "astro-ph.CO",
    doi = "10.1103/PhysRevLett.123.021301",
    journal = "Phys. Rev. Lett.",
    volume = "123",
    number = "2",
    pages = "021301",
    year = "2019"
}

@article{Kahniashvili:2018mzl,
    author = "Kahniashvili, Tina and Brandenburg, Axel and Kosowsky, Arthur and Mandal, Sayan and Roper Pol, Alberto",
    title = "{Magnetism in the Early Universe}",
    eprint = "1810.11876",
    archivePrefix = "arXiv",
    primaryClass = "astro-ph.CO",
    reportNumber = "NORDITA-2018-100",
    doi = "10.1017/S1743921319004447",
    journal = "IAU Symp. A",
    volume = "30",
    pages = "295--298",
    year = "2020"
}

@article{Schober:2020ogz,
    author = "Schober, Jennifer and Schober, Jennifer and Fujita, Tomohiro and Fujita, Tomohiro and Durrer, Ruth and Durrer, Ruth",
    title = "{Generation of chiral asymmetry via helical magnetic fields}",
    eprint = "2002.09501",
    archivePrefix = "arXiv",
    primaryClass = "physics.plasm-ph",
    doi = "10.1103/PhysRevD.101.103028",
    journal = "Phys. Rev. D",
    volume = "101",
    number = "10",
    pages = "103028",
    year = "2020",
    note = "[Erratum: Phys.Rev.D 105, 069901 (2022)]"
}

@article{Perrone:2021srr,
    author = "Perrone, L. M. and Gregori, G. and Reville, B. and Silva, L. O. and Bingham, R.",
    title = "{Neutrino-electron magnetohydrodynamics in an expanding universe}",
    eprint = "2106.14892",
    archivePrefix = "arXiv",
    primaryClass = "astro-ph.CO",
    doi = "10.1103/PhysRevD.104.123013",
    journal = "Phys. Rev. D",
    volume = "104",
    number = "12",
    pages = "123013",
    year = "2021"
}

@article{Dvornikov:2022cyz,
    author = "Dvornikov, Maxim",
    title = "{Interaction of inhomogeneous axions with magnetic fields in the early universe}",
    eprint = "2201.10586",
    archivePrefix = "arXiv",
    primaryClass = "hep-ph",
    doi = "10.1016/j.physletb.2022.137039",
    journal = "Phys. Lett. B",
    volume = "829",
    pages = "137039",
    year = "2022"
}

@article{Armua:2022rvx,
    author = "Armua, Andres and Berera, Arjun and Figueroa, Jaime Calderon",
    title = "{Parameter study of decaying magnetohydrodynamic turbulence}",
    eprint = "2212.02418",
    archivePrefix = "arXiv",
    primaryClass = "physics.plasm-ph",
    doi = "10.1103/PhysRevE.107.055206",
    journal = "Phys. Rev. E",
    volume = "107",
    number = "5",
    pages = "055206",
    year = "2023"
}

@article{Jedamzik:2023rfd,
    author = "Jedamzik, Karsten and Abel, Tom and Ali-Haimoud, Yacine",
    title = "{Cosmic recombination in the presence of primordial magnetic fields}",
    eprint = "2312.11448",
    archivePrefix = "arXiv",
    primaryClass = "astro-ph.CO",
    doi = "10.1088/1475-7516/2025/03/012",
    journal = "JCAP",
    volume = "03",
    pages = "012",
    year = "2025"
}

@article{Trivedi:2018ejz,
    author = "Trivedi, Pranjal and Reppin, Johannes and Chluba, Jens and Banerjee, Robi",
    title = "{Magnetic heating across the cosmological recombination era: Results from 3D MHD simulations}",
    eprint = "1805.05315",
    archivePrefix = "arXiv",
    primaryClass = "astro-ph.CO",
    doi = "10.1093/mnras/sty1757",
    journal = "Mon. Not. Roy. Astron. Soc.",
    volume = "481",
    number = "3",
    pages = "3401--3422",
    year = "2018"
}

@article{Zhou:2022xhk,
    author = "Zhou, Hongzhe and Sharma, Ramkishor and Brandenburg, Axel",
    title = "{Scaling of the Hosking integral in decaying magnetically dominated turbulence}",
    eprint = "2206.07513",
    archivePrefix = "arXiv",
    primaryClass = "physics.plasm-ph",
    reportNumber = "NORDITA 2022-040",
    doi = "10.1017/S002237782200109X",
    journal = "J. Plasma Phys.",
    volume = "88",
    number = "6",
    pages = "905880602",
    year = "2022"
}

@article{Schober:2017cdw,
    author = {Schober, Jennifer and Rogachevskii, Igor and Brandenburg, Axel and Boyarsky, Alexey and Fr\"ohlich, J\"urg and Ruchayskiy, Oleg and Kleeorin, Nathan},
    title = "{Laminar and turbulent dynamos in chiral magnetohydrodynamics. II. Simulations}",
    eprint = "1711.09733",
    archivePrefix = "arXiv",
    primaryClass = "physics.flu-dyn",
    reportNumber = "NORDITA-2017-118",
    doi = "10.3847/1538-4357/aaba75",
    journal = "Astrophys. J.",
    volume = "858",
    number = "2",
    pages = "124",
    year = "2018"
}

@article{Brandenburg:2021aln,
    author = "Brandenburg, Axel and He, Yutong and Kahniashvili, Tina and Rheinhardt, Matthias and Schober, Jennifer",
    title = "{Relic gravitational waves from the chiral magnetic effect}",
    eprint = "2101.08178",
    archivePrefix = "arXiv",
    primaryClass = "astro-ph.CO",
    reportNumber = "NORDITA-2021-005",
    doi = "10.3847/1538-4357/abe4d7",
    journal = "Astrophys. J.",
    volume = "911",
    number = "2",
    pages = "110",
    year = "2021"
}

@book{Kolb:1990vq,
    author = "Kolb, Edward W. and Turner, Michael S.",
    title = "{The Early Universe}",
    reportNumber = "FERMILAB-BOOK-1990-01",
    doi = "10.1201/9780429492860",
    isbn = "978-0-201-62674-2",
    volume = "69",
    year = "1990"
}

@article{Jinno:2022mie,
    author = "Jinno, Ryusuke and Konstandin, Thomas and Rubira, Henrique and Stomberg, Isak",
    title = "{Higgsless simulations of cosmological phase transitions and gravitational waves}",
    eprint = "2209.04369",
    archivePrefix = "arXiv",
    primaryClass = "astro-ph.CO",
    reportNumber = "DESY 22-148, IFT-UAM/CSIC-22-100, TUM-HEP-1416/22",
    doi = "10.1088/1475-7516/2023/02/011",
    journal = "JCAP",
    volume = "02",
    pages = "011",
    year = "2023"
}

@article{Rogachevskii:2017uyc,
    author = {Rogachevskii, Igor and Ruchayskiy, Oleg and Boyarsky, Alexey and Fr\"ohlich, J\"urg and Kleeorin, Nathan and Brandenburg, Axel and Schober, Jennifer},
    title = "{Laminar and turbulent dynamos in chiral magnetohydrodynamics-I: Theory}",
    eprint = "1705.00378",
    archivePrefix = "arXiv",
    primaryClass = "physics.plasm-ph",
    reportNumber = "PREPRINT-NORDITA-2017-39",
    doi = "10.3847/1538-4357/aa886b",
    journal = "Astrophys. J.",
    volume = "846",
    number = "2",
    pages = "153",
    year = "2017"
}

@book{Schutz:1985jx,
    author = "Schutz, Bernard F.",
    title = "{A first course in General Relativity}",
    doi = "10.1017/CBO9780511984181",
    isbn = "978-0-511-98418-1",
    publisher = "Cambridge Univ. Pr.",
    address = "Cambridge, UK",
    year = "1985"
}

@book{Jackson:1998nia,
    author = "Jackson, John David",
    title = "{Classical Electrodynamics}",
    isbn = "978-0-471-30932-1",
    publisher = "Wiley",
    year = "1998"
}

@book{Baumann_2022, place={Cambridge}, title={Cosmology}, publisher={Cambridge University Press}, author={Baumann, Daniel}, year={2022}}

@article{Schober:2024vtv,
    author = "Schober, Jennifer and Rogachevskii, Igor and Brandenburg, Axel",
    title = "{Efficiency of dynamos from an autonomous generation of chiral asymmetry}",
    eprint = "2404.07845",
    archivePrefix = "arXiv",
    primaryClass = "physics.plasm-ph",
    reportNumber = "NORDITA-2024-008",
    doi = "10.1103/PhysRevD.110.043515",
    journal = "Phys. Rev. D",
    volume = "110",
    number = "4",
    pages = "043515",
    year = "2024"
}

@article{Brandenburg:2024tyi,
    author = "Brandenburg, A. and Neronov, A. and Vazza, F.",
    title = "{Resistively controlled primordial magnetic turbulence decay}",
    eprint = "2401.08569",
    archivePrefix = "arXiv",
    primaryClass = "astro-ph.CO",
    reportNumber = "NORDITA-2024-001",
    doi = "10.1051/0004-6361/202449267",
    journal = "Astron. Astrophys.",
    volume = "687",
    pages = "A186",
    year = "2024"
}

@article{Brandenburg:2023imm,
    author = "Brandenburg, Axel and Clarke, Emma and Kahniashvili, Tina and Long, Andrew J. and Sun, Guotong",
    title = "{Relic gravitational waves from the chiral plasma instability in the standard cosmological model}",
    eprint = "2307.09385",
    archivePrefix = "arXiv",
    primaryClass = "astro-ph.CO",
    reportNumber = "NORDITA-2023-034",
    doi = "10.1103/PhysRevD.109.043534",
    journal = "Phys. Rev. D",
    volume = "109",
    number = "4",
    pages = "043534",
    year = "2024"
}

@article{Brandenburg:2023rrd,
    author = "Brandenburg, Axel and Sharma, Ramkishor and Vachaspati, Tanmay",
    title = "{Inverse cascading for initial magnetohydrodynamic turbulence spectra between Saffman and Batchelor}",
    eprint = "2307.04602",
    archivePrefix = "arXiv",
    primaryClass = "physics.plasm-ph",
    reportNumber = "NORDITA-2023-035",
    doi = "10.1017/S0022377823001253",
    journal = "J. Plasma Phys.",
    volume = "89",
    number = "6",
    pages = "905890606",
    year = "2023"
}

@article{Brandenburg:2023aco,
    author = "Brandenburg, Axel and Kamada, Kohei and Mukaida, Kyohei and Schmitz, Kai and Schober, Jennifer",
    title = "{Chiral magnetohydrodynamics with zero total chirality}",
    eprint = "2304.06612",
    archivePrefix = "arXiv",
    primaryClass = "hep-ph",
    reportNumber = "NORDITA-2023-014, RESCEU-5/23, KEK-TH-2504, MS-TP-23-13",
    doi = "10.1103/PhysRevD.108.063529",
    journal = "Phys. Rev. D",
    volume = "108",
    number = "6",
    pages = "063529",
    year = "2023"
}

@article{Brandenburg:2023rul,
    author = "Brandenburg, Axel and Kamada, Kohei and Schober, Jennifer",
    title = "{Decay law of magnetic turbulence with helicity balanced by chiral fermions}",
    eprint = "2302.00512",
    archivePrefix = "arXiv",
    primaryClass = "physics.plasm-ph",
    reportNumber = "NORDITA-2023-001, RESCEU-1/23",
    doi = "10.1103/PhysRevResearch.5.L022028",
    journal = "Phys. Rev. Res.",
    volume = "5",
    number = "2",
    pages = "L022028",
    year = "2023"
}

@article{Sharma:2022ysf,
    author = "Sharma, Ramkishor and Brandenburg, Axel",
    title = "{Low frequency tail of gravitational wave spectra from hydromagnetic turbulence}",
    eprint = "2206.00055",
    archivePrefix = "arXiv",
    primaryClass = "astro-ph.CO",
    reportNumber = "NORDITA 2022-036",
    doi = "10.1103/PhysRevD.106.103536",
    journal = "Phys. Rev. D",
    volume = "106",
    number = "10",
    pages = "103536",
    year = "2022"
}

@article{Cattaneo:1948,
    author = "Cattaneo, C.",
    title = "{Sulla conduzione del calore}",
    journal = "Atti. Mat. Fis. Univ. Modena.",
    number = "3",
    pages = "83",
    year = "1948"
}

@article{Israel:1976tn,
    author = "Israel, W.",
    title = "{Nonstationary irreversible thermodynamics: A Causal relativistic theory}",
    doi = "10.1016/0003-4916(76)90064-6",
    journal = "Annals Phys.",
    volume = "100",
    pages = "310--331",
    year = "1976"
}

@article{Israel:1976efz,
    author = "Israel, W. and Stewart, J. M.",
    title = "{Thermodynamics of nonstationary and transient effects in a relativistic gas}",
    doi = "10.1016/0375-9601(76)90075-X",
    journal = "Phys. Lett. A",
    volume = "58",
    number = "4",
    pages = "213--215",
    year = "1976"
}

@book{Landau1987Fluid,
  author = {Landau, L. D. and Lifshitz, E. M.},
  day = 15,
  edition = 2,
  howpublished = {Paperback},
  isbn = {0750627670},
  month = jan,
  publisher = {Butterworth-Heinemann},
  title = {Fluid Mechanics, Second Edition: Volume 6 (Course of Theoretical Physics)},
  url = {http://www.worldcat.org/isbn/0750627670},
  year = 1987
}

@article{Weinberg:1971mx,
    author = "Weinberg, Steven",
    title = "{Entropy generation and the survival of protogalaxies in an expanding universe}",
    doi = "10.1086/151073",
    journal = "Astrophys. J.",
    volume = "168",
    pages = "175",
    year = "1971"
}

@article{Jedamzik:1996wp,
    author = "Jedamzik, Karsten and Katalinic, Visnja and Olinto, Angela V.",
    title = "{Damping of cosmic magnetic fields}",
    eprint = "astro-ph/9606080",
    archivePrefix = "arXiv",
    doi = "10.1103/PhysRevD.57.3264",
    journal = "Phys. Rev. D",
    volume = "57",
    pages = "3264--3284",
    year = "1998"
}

@article{Arnold:2003zc,
    author = "Arnold, Peter Brockway and Moore, Guy D and Yaffe, Laurence G.",
    title = "{Transport coefficients in high temperature gauge theories. 2. Beyond leading log}",
    eprint = "hep-ph/0302165",
    archivePrefix = "arXiv",
    doi = "10.1088/1126-6708/2003/05/051",
    journal = "JHEP",
    volume = "05",
    pages = "051",
    year = "2003"
}

@article{Arnold:2006fz,
    author = "Arnold, Peter Brockway and Dogan, Caglar and Moore, Guy D.",
    title = "{The Bulk Viscosity of High-Temperature QCD}",
    eprint = "hep-ph/0608012",
    archivePrefix = "arXiv",
    doi = "10.1103/PhysRevD.74.085021",
    journal = "Phys. Rev. D",
    volume = "74",
    pages = "085021",
    year = "2006"
}

@article{Ahonen:1998iz,
    author = "Ahonen, Jarkko",
    title = "{Transport coefficients in the early universe}",
    eprint = "hep-ph/9801434",
    archivePrefix = "arXiv",
    doi = "10.1103/PhysRevD.59.023004",
    journal = "Phys. Rev. D",
    volume = "59",
    pages = "023004",
    year = "1999"
}

@article{Kahniashvili:2010gp,
    author = "Kahniashvili, Tina and Brandenburg, Axel and Tevzadze, Alexander G. and Ratra, Bharat",
    title = "{Numerical simulations of the decay of primordial magnetic turbulence}",
    eprint = "1004.3084",
    archivePrefix = "arXiv",
    primaryClass = "astro-ph.CO",
    reportNumber = "NORDITA-PREPRINT-NORDITA-2010-20",
    doi = "10.1103/PhysRevD.81.123002",
    journal = "Phys. Rev. D",
    volume = "81",
    pages = "123002",
    year = "2010"
}

@article{RoperPol:2021xnd,
    author = "Roper Pol, Alberto and Mandal, Sayan and Brandenburg, Axel and Kahniashvili, Tina",
    title = "{Polarization of gravitational waves from helical MHD turbulent sources}",
    eprint = "2107.05356",
    archivePrefix = "arXiv",
    primaryClass = "gr-qc",
    reportNumber = "NORDITA-2021-062",
    doi = "10.1088/1475-7516/2022/04/019",
    journal = "JCAP",
    volume = "04",
    number = "04",
    pages = "019",
    year = "2022"
}

@article{Planck:2014ylh,
    author = "Ade, P. A. R. and others",
    collaboration = "Planck",
    title = "{Planck intermediate results - XXIV. Constraints on variations in fundamental constants}",
    eprint = "1406.7482",
    archivePrefix = "arXiv",
    primaryClass = "astro-ph.CO",
    doi = "10.1051/0004-6361/201424496",
    journal = "Astron. Astrophys.",
    volume = "580",
    pages = "A22",
    year = "2015"
}

@book{Weinberg:2008zzc,
    author = "Weinberg, Steven",
    title = "{Cosmology}",
    isbn = "978-0-19-852682-7",
    year = "2008",
    publisher = {Oxford University Press}
}

@article{Planck:2013pxb,
    author = "Ade, P. A. R. and others",
    collaboration = "Planck",
    title = "{Planck 2013 results. XVI. Cosmological parameters}",
    eprint = "1303.5076",
    archivePrefix = "arXiv",
    primaryClass = "astro-ph.CO",
    reportNumber = "CERN-PH-TH-2013-129",
    doi = "10.1051/0004-6361/201321591",
    journal = "Astron. Astrophys.",
    volume = "571",
    pages = "A16",
    year = "2014"
}

@article{Brandenburg:2016odr,
    author = "Brandenburg, Axel and Kahniashvili, Tina",
    title = "{Classes of hydrodynamic and magnetohydrodynamic turbulent decay}",
    eprint = "1607.01360",
    archivePrefix = "arXiv",
    primaryClass = "physics.flu-dyn",
    reportNumber = "NORDITA-2016-82",
    doi = "10.1103/PhysRevLett.118.055102",
    journal = "Phys. Rev. Lett.",
    volume = "118",
    number = "5",
    pages = "055102",
    year = "2017"
}

@article{Brandenburg:2017rcb,
    author = "Brandenburg, Axel and Schober, Jennifer and Rogachevskii, Igor and Kahniashvili, Tina and Boyarsky, Alexey and Frohlich, Jurg and Ruchayskiy, Oleg and Kleeorin, Nathan",
    title = "{The turbulent chiral-magnetic cascade in the early universe}",
    eprint = "1707.03385",
    archivePrefix = "arXiv",
    primaryClass = "astro-ph.CO",
    reportNumber = "NORDITA-2017-037",
    doi = "10.3847/2041-8213/aa855d",
    journal = "Astrophys. J. Lett.",
    volume = "845",
    number = "2",
    pages = "L21",
    year = "2017"
}

@article{Durrer:2003ja,
    author = "Durrer, Ruth and Caprini, Chiara",
    title = "{Primordial magnetic fields and causality}",
    eprint = "astro-ph/0305059",
    archivePrefix = "arXiv",
    doi = "10.1088/1475-7516/2003/11/010",
    journal = "JCAP",
    volume = "11",
    pages = "010",
    year = "2003"
}

@article{Kahniashvili:2015msa,
    author = "Kahniashvili, Tina and Brandenburg, Axel and Tevzadze, Alexander G.",
    title = "{The evolution of primordial magnetic field since its generation}",
    eprint = "1507.00510",
    archivePrefix = "arXiv",
    primaryClass = "astro-ph.CO",
    reportNumber = "NORDITA-2015-83",
    doi = "10.1088/0031-8949/91/10/104008",
    journal = "Phys. Scripta",
    volume = "91",
    number = "10",
    pages = "104008",
    year = "2016"
}

@article{Hosking:2022umv,
    author = "Hosking, David N. and Schekochihin, Alexander A.",
    title = "{Cosmic-void observations reconciled with primordial magnetogenesis}",
    eprint = "2203.03573",
    archivePrefix = "arXiv",
    primaryClass = "astro-ph.CO",
    doi = "10.1038/s41467-023-43258-3",
    journal = "Nature Commun.",
    volume = "14",
    number = "1",
    pages = "7523",
    year = "2023"
}

@article{Schober:2018wlo,
    author = "Schober, J. and Brandenburg, A. and Rogachevskii, I.",
    title = "{Chiral fermion asymmetry in high-energy plasma simulations}",
    eprint = "1808.06624",
    archivePrefix = "arXiv",
    primaryClass = "physics.plasm-ph",
    reportNumber = "NORDITA-2018-076",
    doi = "10.1080/03091929.2019.1591393",
    journal = "Geophys. Astrophys. Fluid Dynamics",
    volume = "114",
    number = "1-2",
    pages = "106--129",
    year = "2020"
}

@article{PencilCode:2020eyn,
    author = "Brandenburg, A. and others",
    collaboration = "Pencil Code",
    title = "{The Pencil Code, a modular MPI code for partial differential equations and particles: multipurpose and multiuser-maintained}",
    eprint = "2009.08231",
    archivePrefix = "arXiv",
    primaryClass = "astro-ph.IM",
    reportNumber = "NORDITA-2020-087",
    doi = "10.21105/joss.02807",
    journal = "J. Open Source Softw.",
    volume = "6",
    number = "58",
    pages = "2807",
    year = "2021"
}

@article{Kahniashvili:2021gym,
    author = "Kahniashvili, Tina and Clarke, Emma and Stepp, Jonathan and Brandenburg, Axel",
    title = "{Big Bang Nucleosynthesis Limits and Relic Gravitational-Wave Detection Prospects}",
    eprint = "2111.09541",
    archivePrefix = "arXiv",
    primaryClass = "astro-ph.CO",
    reportNumber = "NORDITA-2021-089",
    doi = "10.1103/PhysRevLett.128.221301",
    journal = "Phys. Rev. Lett.",
    volume = "128",
    number = "22",
    pages = "221301",
    year = "2022"
}

@article{Subramanian:2009fu,
    author = "Subramanian, Kandaswamy",
    title = "{Magnetic fields in the early universe}",
    eprint = "0911.4771",
    archivePrefix = "arXiv",
    primaryClass = "astro-ph.CO",
    doi = "10.1002/asna.200911312",
    journal = "Astron. Nachr.",
    volume = "331",
    pages = "110--120",
    year = "2010"
}

@article{Heckler:1993nc,
    author = "Heckler, A. and Hogan, C. J.",
    title = "{Neutrino heat conduction and inhomogeneities in the early universe}",
    doi = "10.1103/PhysRevD.47.4256",
    journal = "Phys. Rev. D",
    volume = "47",
    pages = "4256--4260",
    year = "1993"
}

@article{PhysRev.58.919,
  title = {The Thermodynamics of Irreversible Processes. {III. R}elativistic Theory of the Simple Fluid},
  author = {Eckart, Carl},
  journal = {Phys. Rev.},
  volume = {58},
  issue = {10},
  pages = {919--924},
  numpages = {0},
  year = {1940},
  month = {Nov},
  publisher = {American Physical Society},
  doi = {10.1103/PhysRev.58.919},
  url = {https://link.aps.org/doi/10.1103/PhysRev.58.919}
}

@book{Durrer:2008eom,
    author = "Durrer, Ruth",
    title = "{The Cosmic Microwave Background}",
    doi = "10.1017/CBO9780511817205",
    isbn = "978-0-511-81720-5",
    publisher = "Cambridge University Press",
    address = "Cambridge",
    year = "2008"
}

@article{Planck:2018vyg,
    author = "Aghanim, N. and others",
    collaboration = "Planck",
    title = "{Planck 2018 results. VI. Cosmological parameters}",
    eprint = "1807.06209",
    archivePrefix = "arXiv",
    primaryClass = "astro-ph.CO",
    doi = "10.1051/0004-6361/201833910",
    journal = "Astron. Astrophys.",
    volume = "641",
    pages = "A6",
    year = "2020",
    note = "[Erratum: Astron.Astrophys. 652, C4 (2021)]"
}

@article{Martel:1997hk,
    author = "Martel, Hugo and Shapiro, Paul R.",
    title = "{A convenient set of comoving cosmological variables and their application}",
    eprint = "astro-ph/9710119",
    archivePrefix = "arXiv",
    doi = "10.1046/j.1365-8711.1998.01497.x",
    journal = "Mon. Not. Roy. Astron. Soc.",
    volume = "297",
    pages = "467",
    year = "1998"
}

@article{Giombi:2025tkv,
    author = "Giombi, Lorenzo and Dahl, Jani and Hindmarsh, Mark",
    title = "{Acoustic gravitational waves beyond leading order in bubble over Hubble radius}",
    eprint = "2504.08037",
    archivePrefix = "arXiv",
    primaryClass = "gr-qc",
    reportNumber = "HIP-2025-13/TH",
    month = "4",
    year = "2025"
}

@article{Jedamzik:2025cax,
    author = "Jedamzik, Karsten and Pogosian, Levon and Abel, Tom",
    title = "{Hints of Primordial Magnetic Fields at Recombination and Implications for the Hubble Tension}",
    eprint = "2503.09599",
    archivePrefix = "arXiv",
    primaryClass = "astro-ph.CO",
    reportNumber = "SCG-2025-01",
    doi = "10.1038/s41550-025-02737-x",
    month = "3",
    year = "2025"
}

@ARTICLE{1980Afz....16..769S,
       author = {{Shandarin}, S.~F.},
        title = "{Evolution of perturbations in Friedmann models of the universe}",
      journal = {Astrofizika},
         year = 1980,
        month = oct,
       volume = {16},
        pages = {769-779},
       adsurl = {https://ui.adsabs.harvard.edu/abs/1980Afz....16..769S}
}

@article{Giombi:2024kju,
    author = "Giombi, Lorenzo and Dahl, Jani and Hindmarsh, Mark",
    title = "{Signatures of the speed of sound on the gravitational wave power spectrum from sound waves}",
    eprint = "2409.01426",
    archivePrefix = "arXiv",
    primaryClass = "gr-qc",
    reportNumber = "HIP-2024-19/TH",
    doi = "10.1088/1475-7516/2025/01/100",
    journal = "JCAP",
    volume = "01",
    pages = "100",
    year = "2025"
}

@book{Weinberg:1995mt,
    author = "Weinberg, Steven",
    title = "{The Quantum theory of fields. Vol. 1: Foundations}",
    doi = "10.1017/CBO9781139644167",
    isbn = "978-0-521-67053-1, 978-0-511-25204-4",
    publisher = "Cambridge University Press",
    month = "6",
    year = "2005"
}

\label{RealLastPage}

\end{document}